\documentclass[11pt,reqno]{amsart}
\usepackage{amsmath}
\usepackage{amssymb}
\usepackage{graphicx}

\usepackage{helvet}

\theoremstyle{definition}

\numberwithin{equation}{section}

\newcommand{\bx}{{\mathbf x}}
\newcommand{\bs}{{\mathbf s}}
\newcommand{\by}{{\mathbf y}}
\newcommand{\bu}{{\mathbf u}}
\newcommand{\bv}{{\mathbf v}}
\newcommand{\bz}{{\mathbf z}}

\newcommand{\R}{{\mathbb R}}

\newcommand{\la}{\lambda}

\newcommand{\di}{{\displaystyle}}

\title[A Thouless-like effect ]{A Thouless-Like Effect in the Dyson 
Hierarchical Model with Continuous Symmetry}

\author[Pavel Bleher]{Pavel Bleher}

\address[Pavel Bleher]{University -- Purdue University at Indianapolis, 
402 N. Blackford Street, Indianapolis, Indiana 46202, USA}
\email{{\tt pbleher@iupui.edu}}

\author[P\'eter  Major]{P\'eter Major}

\address[P\'eter Major]{Alfr\'ed R\'enyi Institute of Mathematics, 
Budapest, P.O.B. 127 H--1364, Hungary} 
\email{{\tt major@renyi.hu}}

\keywords{Dyson's hierarchical model, continuous symmetry, Thouless'
effect, renormalization transformation, limit distribution of the
average spin, super-polynomial critical asymptotics}

\begin{document}

\begin{abstract}
In this paper we study Dyson's classical $r$-component
hierarchical model with a Hamiltonian function which has a continuous
$O(r)$-symmetry, $r\ge 2$.
 This is a one-dimensional ferromagnetic model
with a long range interaction potential $U(i,j)=-l(d(i,j)) d^{-2}(i,j)$,
where $d(i,j)$ denotes the hierarchical distance. We are interested in
the case when $l_n=l(2^n)$, $n=1,2,\ldots$,  is an increasing sequence,
with a sub-exponential growth as $n\to\infty$. For a class of free 
measures, we prove a conjecture of Dyson. This conjecture states that 
the convergence of the series $l_1^{-1}+l_2^{-1}+\dots$ is a necessary 
and sufficient condition of the existence of phase transition in the 
model under consideration, and the spontaneous magnetization vanishes 
at the critical point, i.e., there is no Thouless' effect. We find, 
however, that the distribution of the normalized mean spin at the 
critical temperature $T_c$ tends to the uniform distribution on the 
unit sphere in ${\mathbb R}^r$ as the volume tends to infinity, a 
phenomenon which resembles the Thouless effect. We prove that the 
limit distribution of the normalized  mean spin is Gaussian for $T>T_c$, 
and it is non-Gaussian for $T\le T_c$. We also show that the density of 
the limit distribution of the  normalized mean spin for $T\le T_c$ is 
a nice analytic function which can be found from the unique solution 
of a nonlinear fixed point integral
equation. Finally, we  determine some critical asymptotics and show that
the divergence of the correlation length and magnetic susceptibility is
super-polynomial as $T\to T_c$.
\end{abstract}

\maketitle


\noindent
{\bf Contents}

\bigskip
\halign{#&\hskip10pt #\hfill\cr
1.& Introduction. Formulation of the Main Results.\cr
2.& Analytic Reformulation of the Problem. Strategy of the Proof.\cr
3.& Formulation of Auxiliary Theorems.\cr
4.& Basic Estimates in the Low Temperature Region.\cr
5.& Estimates in the Intermediate Region. The proof of Theorem 3.1.\cr
6.& Estimates in the High Temperature Region. The proof of Theorem 3.3.\cr
7.& Estimates in the Low Temperature Region. The proof of Theorem 3.2.\cr
8.& Estimates Near the Critical Point. The proof of Theorems 3.4, 1.3 
    and 1.5.\cr
  & Appendices A and B\cr
  & References\cr}
 
\section{Introduction. Formulation of the Main Results.}

In this paper we investigate Dyson's hierarchical vector-valued model
with {\it continuous symmetry}. The model consists of spin variables
$\sigma(j)\in{\mathbb R}^r,\; j\in{\mathbb N}=\{1,2,\dots\}$, where 
$r\ge2$. We define the {\it hierarchical distance} $d(\cdot,\cdot)$ on
${\mathbb N}$ as
$$
d(j,k)=2^{n(j,k)-1}\quad \textrm{for } j\not= k 
$$
with
\begin{eqnarray*}
n(j,k)&=&\{\min n\colon  \textrm{ there is an integer }l 
\textrm{ such that } (l-1)2^n<j,k\le l2^n\} \\ 
&&\qquad\qquad\qquad\qquad\qquad\qquad\qquad\qquad\qquad 
\textrm{if}\quad  j\not= k, 
\end{eqnarray*}
and  $d(j,j)=0$. The Hamiltonian of the ferromagnetic Dyson's 
hierarchical $r$-component model in the volume 
$V_n=\{1,2,\dots, 2^n\}$ is
\begin{equation}
\mathcal{H}_n(\sigma)=-\sum_{1\le j<k\le 2^n}
\frac{l(d(j,k))}{ d^2(j,k)}\,\sigma(j)\sigma(k) ,
\label{1.1}
\end{equation}
where $\sigma(j)\sigma(k)$ denotes a scalar product in ${\mathbb R}^r$, and
$l(t)$ is a positive function. In this paper we will be interested in
the case when $l(t)$ is a positive increasing function such that
$$
\lim_{t\to\infty} l(t)=\infty;\qquad
\lim_{t\to\infty} \frac{l(t)}{ t^\varepsilon}=0,\quad 
\textrm{for all }\varepsilon>0.
$$
Since the hierarchical distance $d(j,k)$ for $j\not =k$ takes the 
values $2^n$, $n=0,1,2,\ldots$, only,
we consider the function $l(t)$ for $t=2^n$ only and define
$$
l_n=l(2^n).
$$
Let $\nu(d\bx)$ be a probability measure on ${\mathbb R}^r$. Then 
the Gibbs measure in $V_n$ at a temperature $T>0$ with free boundary 
conditions and the free measure $\nu(d\bx)$ is defined as
$$
\mu_n(d\bx;T)=Z^{-1}_n(T)\exp\left\{-\beta \mathcal{H}_n(\bx)\right\}
\prod_{j=1}^{2^n}\nu(d\bx_j), \quad \beta=T^{-1}.
$$
We will assume that the free measure $\nu(d\bx)$ is invariant with
respect to the group $O(r)$ of orthogonal transformations, i.e.,
$\nu(UA)=\nu(A)$ for all $U\in O(r)$ and all Borel sets $A\in
B({\mathbb R}^r)$. Then
the Gibbs measure $\mu_n(d\bx;T)$ is $O(r)$-invariant as well,
\begin{eqnarray*}
\mu_n(UA_1,\dots, UA_{2^n};T)&=&\mu_n(A_1,\dots,A_{2^n};T),
\quad \textrm{for all } U\in O(r),\\
&&\quad A_j\in B({\mathbb R}^r),\; j=1,\dots, 2^n.
\end{eqnarray*}
In \cite{Dys2}, Dyson proved the following theorem (see also \cite{Dys3}). 
Assume that $r=3$, and $\nu(d\bx)$ is a uniform measure on the unit sphere
in ${\mathbb R}^3$. This is the classical Heisenberg hierarchical model.

\medskip\noindent 
{\bf Theorem 1.1.} {\rm (see \cite{Dys2}).} {\it The classical
Heisenberg hierarchical model has a phase transition if
\begin{equation}
B=\sum_{n=1}^\infty l_n^{-1}<\infty.\label{1.3}
\end{equation}
It has a long-range order so long as $\beta>B$.}

\medskip
Dyson also formulated the following conjecture (see \cite{Dys2}):
``It also seems likely that for sequences $l_n$ which are positive
and increasing with $n$ the condition (\ref{1.3}) is necessary for a phase
transition in Heisenberg hierarchical models.''
The goal of this paper is to {\it prove} Dyson's conjecture for a class of
hierarchical models and to study the {\it limit distribution of the normalized
mean spin} both below and above the critical temperature if condition
(\ref{1.3}) holds. Dyson's proof is a clever application of {\it correlation
inequalities}. Our approach is based on an analytical study
of the {\it renormalization group transformation} for the hierarchical models.

The renormalization group (RG) approach to the Dyson hierarchical models
was initiated in the works of Bleher and Sinai \cite{BS1}--\cite{BS3} 
(see also 
the monograph \cite{Sin} and the review \cite{Ble}, and references therein). 
The Dyson hierarchical models are of a great interest because for this model 
the RG transformation reduces to a nonlinear integral equation, and this 
allows a study of critical phenomena unavailable in other models. 
The works of Bleher and Sinai were concerned with the critical phenomena 
and phase transitions in the scalar Dyson hierarchical models. They 
were extended to the study of critical phenomena and phase transitions 
in the vector Dyson hierarchical models with continuous symmetry in 
the works of Bleher and Major \cite{BM1}--\cite{BM5}. The present paper is
 a continuation of the works \cite{BM1}--\cite{BM5}.

We apply a perturbation technique which works if the free measure
$\nu(d\bx)$ is a {\it small perturbation} of the Gaussian measure.
Hence, we cannot treat the case when $\nu(d\bx)$ is a uniform measure 
on the unit sphere. On the other hand, we will consider {\it arbitrary} 
spin dimension $r\ge 2$. We will focus on free measures $\nu(d\bx)$, 
which have a density function $p(\bx)$ on ${\mathbb R}^r$ such that 
$p(\bx)$ is close, in an appropriate sense, to the density function
\begin{equation}
p_0(\bx)
=C(\kappa)\exp\left\{-\frac{|\bx|^2}2-\kappa\, \frac{|\bx|^4}4\right\} 
\label{1.4}
\end{equation}
with a sufficiently small parameter $\kappa>0$. Precise conditions on
$p(\bx)$ are given below. We also will assume some
regularity conditions about the sequence $l_n=l(2^n)$ (see
below).

We are investigating the following question. Let $p_n(\bx,T)$ denote
the density function of the mean spin
$2^{-n}\sum\limits_{j=1}^{2^n}\sigma(j)$, where
$(\sigma(1),\dots,\sigma(2^n))$ is a $\mu_n(T)$-distributed random
vector. Because of the rotational invariance of the model, the
function $p_n(\bx,T)$ is a function of $|\bx|$. We are interested in 
the limit behaviour of the function $p_n(\bx,T)$ as $n\to\infty$, with 
an appropriate normalization. In our papers \cite{BM1}--\cite{BM5}
this problem was considered for the Hamiltonian 
$$
\mathcal{H}_n(\sigma)=-\sum_{1\le j<k\le 2^n}
\frac1{d^{\alpha}(j,k)}\,\sigma(j)\sigma(k),
$$
where $1<\alpha<2$. Observe that if $\alpha\le 1$ then the 
thermodynamic limit of the model does not exist,
and if $\alpha\ge 2$ then there is no phase transition, 
hence the range $1<\alpha<2$ is natural.
We distinguished in  \cite{BM1}--\cite{BM5} the three cases for~$\alpha$:

\medskip
(i) $1<\alpha<3/2$, \ (ii) $\alpha=3/2$, and (iii) $3/2<\alpha<2$.

\smallskip
The difference between these cases appears in the asymptotic behavior
of $p_n(\bx,T)$ at small $T$. When $T$ is small the spontaneous
magnetization $M(T)$ is positive, and the function $p_n(\bx,T)$ is 
concentrated in a narrow spherical shell near the sphere 
$|\bx|=M(T)$. The question is what the width of this shell is and 
what the limiting shape of $p_n(\bx,T)$ is like  along the radius after
an appropriate rescaling. In case (i), the width is of the order of
$2^{-n/2}$ and the limit shape of $p_n(\bx,T)$ is Gaussian
(see \cite{BM1}). In case (ii), there is a logarithmic correction 
in the asymptotics of the width, but the limit shape is still 
Gaussian  (see \cite{BM4}). In case (iii), the width of the shell 
has a nonstandard asymptotics of the order of $2^{-n(2-\alpha) }$, and the 
limit shape of $p_n(\bx,T)$ along the radius (after a rescaling) is 
a non-Gaussian function which is a solution of a nonlinear integral 
equation (see \cite{BM3} and the review \cite{BM2}). In the present 
paper we are interested in the marginal potential $l(d(j,k))/ d^2(j,k)$,
with an extra factor  $l(t)$ of a sub-polynomial growth.

Before formulating the main results we would like to discuss the
importance of Dyson's condition (\ref{1.3}). In the case of the Ising
hierarchical model ($r=1$), Dyson proved in \cite{Dys2} that there exists
a ``weakest'' interaction function $l(t)$ for which the hierarchical 
model (\ref{1.1}) has a phase transition. This function is 
$l(t)=\log\log t$, which corresponds to $l_n=\log n$. Dyson has 
proved that if
$$
\lim_{n\to\infty} \frac{l_n}{\log n}=0,
$$
then the spontaneous magnetization is equal to zero for all
temperatures $T>0$. On the other hand, if
$$
\frac{l_n}{\log n}>\varepsilon \quad\textrm{for all } n>0\textrm{ with
some }\varepsilon>0,
$$
then the spontaneous magnetization is positive at sufficiently low
temperatures $T>0$. In the borderline model, when
$$
l_n=J\log n,\qquad J>0,
$$
Dyson proved that the spontaneous magnetization $M(T)$ has a jump at
the critical temperature $T_c$. The existence of the jump for the
1D Ising model with long-range interaction was first predicted by
Thouless (see \cite{Tho}, and also the works \cite{YA} of
Anderson, Yuval and \cite{Ham} of Hamann and references therein) 
for the translationally invariant Ising model with the interaction
\begin{equation}
H(\sigma)=-\sum_{j,k} \frac{\sigma(j)\sigma(k)}{(j-k)^2}.
\label{1.7}
\end{equation}
This phenomenon (the jump of $M(T)$ at $T=T_c$) is called the 
{\it Thouless effect}. The existence of a phase transition in the 
ferromagnetic one-dimensional Ising model with $1/(j-k)^2$ 
interaction energy was proved by Fr\"ohlich and Spencer in \cite{FS}.
A rigorous proof of the existence of the Thouless effect in the 
Ising model with the inverse square  interaction (\ref{1.7}) was 
given by Aizenman, J.~Chayes, L.~Chayes, and Newman \cite{ACCN}.  
Simon proved in \cite{Sim} the absence of continuous symmetry 
breaking in the one-dimensional $r$-component Heisenberg model 
with the interaction (\ref{1.7}), in the case when $r\ge 2$.

Dyson formulated a general heuristic principle in \cite{Dys2} which
tells us when one should expect the Thouless effect
 in a 1D long-range ferromagnetic model: It should
occur for the ``weakest'' interaction (if it exists)
for which a phase transition appears.
Dyson wrote that in the hierarchical model ``in the Ising case, there
exists a borderline model $l_n=\log n$ which is the `weakest'
ferromagnet for which a transition occurs, and this borderline model
shows a Thouless effect. In the Heisenberg case there exists no
borderline model, since there is no `most slowly converging' series
(\ref{1.3}). Thus we do not expect to find a Thouless effect in any
one-dimensional Heisenberg hierarchical ferromagnet.'' This conjecture
of Dyson, about the absence of a Thouless effect in the Heisenberg
case, plays a very essential role in our investigation. We show that
in the class of the $r$-component hierarchical models under
consideration, the spontaneous magnetization $M(T)$ approaches 
zero as $T$ approaches the critical temperature, i.e., there is 
no Thouless effect.  On the other hand, we observe a phenomenon 
which resembles the Thouless effect: at $T=T_c$ the
rescaled distribution
$$
\bar M_n^r(T_c) p_n(M_n(T_c) \bx,T_c)\,d\bx,\qquad 
\bar M_n(T)=\left(\int_{{\mathbb R}^r} |\bx|^2 p_n(\bx,T)\,d\bx\right)^{1/2},
$$
approaches, as $n\to\infty$, a uniform measure on the unit sphere in
${\mathbb R}^r,\;r\ge 2$. Thus, although the spontaneous magnetization
$M(T_c)=\lim\limits_{n\to\infty}  \bar M_n(T_c)$ is equal to zero at
the critical point, the distribution of the normalized mean spin
converges to a uniform measure on the unit sphere. This is a
``remnant'' of the spontaneous magnetization at the critical
temperature $T_c$.

To formulate our results we will need some conditions on the sequence
$l_n=l(2^n)$. We need different conditions on $l_n$ in different
theorems. We formulate the conditions we shall later apply.

\medskip\noindent
{\bf Conditions on the sequence $l_n$, $n=0,1,2,\dots$}. Let us
introduce the notation
$$
c_n=\frac{l_n}{l_{n-1}}, \quad n=0,1,\dots,\textrm{ with}\quad l_{-1}=1.
$$

\medskip\noindent 
{\bf Condition 1.}
\begin{equation}
l_0=1;\quad 1\le c_n\le 1.01\quad\textrm{ for all } n;
\qquad \lim_{n\to\infty} c_n=1. \label{1.10}
\end{equation}

\medskip\noindent
{\it Remark.}\/ The requirement $l_0=1$ is not a real condition, it can be
reached by a rescaling of the temperature. We use it just for
a normalization.
 
\medskip\noindent
{\bf Condition 2.} {\it
$$
\lim_{n\to\infty}l_n\sum_{j=n}^\infty l_j^{-1}=\infty.
$$
Moreover, the above condition is uniform in the following sense: For all
$\varepsilon>0$ there are some numbers 
$K(\varepsilon)>0$ and $L(\varepsilon)>0$ such that
$$
l_n\sum_{j=n}^{n+K(\varepsilon)}l_j^{-1}\ge \varepsilon^{-1}
$$
for all $n>L(\varepsilon)$.}

\medskip\noindent
{\bf Condition 3.} 
$$
\sup_{1<n<\infty}\sum_{k=1}^n
\left(l_k\sum\limits_{j=k}^n l_j^{-1}\right)^{-2}<\infty.
$$

\medskip\noindent 
{\bf Condition 4}. 
$$
\sum_{n=1}^\infty l_j^{-1} > 400\,\kappa^{-1}.
$$

\medskip\noindent 
{\bf Condition 5.} 
$$
\frac{l_n}{l_{n+k}}>\bar\eta\quad\textrm{for all }\;n=0,1,2,\dots,
\quad \textrm{and all } k=1,\dots, L.
$$

The numbers $\kappa,\bar\eta>0$, and $L\in{\mathbb N}$ in these 
conditions will be
chosen later.
An example of sequences $l_n$ satisfying Conditions 1--5 is given in
the following proposition.

\medskip\noindent
{\bf Proposition 1.2.} {\it The sequence
\begin{equation}
l_n=(1+an)^\lambda,\qquad a>0,\;\lambda>1,
\label{1.16}
\end{equation}
satisfies Conditions 2 and 3 for all $a>0$ and $\lambda>1$.
There exists a number $a_0=a_0(\lambda)>0$ such that this sequence
satisfies Condition 1 for all $0<a<a_0$, a number
$a_1=a_1(\kappa,\lambda)>0$ such that this sequence
satisfies Condition 4 for all $0<a<a_1$,
and finally there exists a number
$a_2=a_2(\bar\eta, L)>0$ such that this sequence satisfies Condition 5
for all $0<a<a_2$.}

\medskip 
Thus, for all $\lambda>1$ there exists a number
$$
a_3=a_3(\lambda,\kappa,\bar\eta,L)=
\min\{a_0(\lambda),a_1(\kappa,\lambda),a_2(\bar\eta,L)\}>0
$$
such that for all $0<a<a_3$, the sequence (\ref{1.16}) satisfy 
Conditions 1---5.
We prove Proposition 1.2 in Appendix B below.
Now we describe the class of initial densities we shall consider.

\medskip\noindent
{\bf Class of initial densities.}\/ We say that a probability density
$p(\bx)$ on ${\mathbb R}^r$ belongs to the class $\mathcal{P}_\kappa$ if
\begin{equation}
p(\bx)=C(1+\varepsilon(|\bx|^2))\exp 
\left(-\frac{|\bx|^2}2-\kappa\,\frac{|\bx|^4}4\right),
\label{1.17}
\end{equation}
where $C>0$ is a norming factor, and
\begin{equation}
\| \varepsilon(t)\|_{C^4({\mathbb R}^1)}<0.01.
\label{1.18}
\end{equation}

Now we formulate our main results. We denote by $p_n(\bx,T)$ the
distribution of the mean spin $2^{-n}[\sigma(1)+\dots+\sigma(2^n)]$
with respect to the Gibbs measure $\mu_n(d\bx;T)$ and put
\begin{equation}
\bar M_n(T)=\left(\int_{{\mathbb R}^r}|\bx|^2p_n(\bx,T)\,d\bx\right)^{1/2}.
\label{1.19}
\end{equation}
By $\tilde p_n(\bx,T)$ we denote the rescaled density function
\begin{equation}
\tilde p_n(\bx,T)=\bar M_n^r(T)p_n(\bar M_n(T)\bx,T)
\label{1.19a}
\end{equation}
and by $\tilde\nu_{n,T}(d\bx)$ the corresponding probability distribution
\begin{equation}
\tilde\nu_{n,T}(d\bx)=\tilde p_n(\bx,T)\,d\bx.
\label{1.19b}
\end{equation}

\medskip\noindent
{\bf Formulation of the main results.} We fix a sufficiently small 
positive number $\eta$ which will be 
the same through the whole paper. For instance, $\eta=10^{-100}$ is 
a good choice. Define the following number $N=N(\eta)$:
\begin{equation}
N=\min\{ n\colon  l_n>\eta^{-1}\}.\label{1.22}
\end{equation}
Assume that an arbitrary number $\bar\eta$ in the interval
$0<\bar\eta\le\eta$ is fixed. (The number $\bar\eta$ 
appears in Condition 5). 

\medskip\noindent
{\bf Theorem 1.3.} {\rm (Necessity of Dyson's condition).}
 {\it Let us consider the case when
$$
\sum_{n=1}^\infty l_n^{-1}=\infty. 
$$
Then there exists a number $\kappa_0=\kappa_0(N)$ such that for all 
$0<\kappa<\kappa_0$ the following statements hold.

Assume that the density $p(\bx)=\di\frac{\nu(d\bx)}{d\bx}$ belongs 
to the class $\mathcal{P}_\kappa$ and the sequence $\{l_n,\;n\ge 0\}$ 
satisfies Conditions 1--3.
Then there exists a constant $L=L(\bar\eta,\kappa)$ such that if
the sequence $\{l_n,\;n\ge 0\}$ satisfies Condition 5,
then for all $T>0$, there exists the limit
\begin{equation}
\lim_{n\to\infty} 2^n \bar M_n^2(T)=\chi(T)>0.
\label{1.24}
\end{equation}
In particular, the spontaneous magnetization satisfies the relation
$$
M(T)=\lim_{n\to\infty} \bar M_n(T)=0.
$$
In addition, the distribution $\tilde\nu_{n,T}(d\bx)$ tends weakly
to the $r$-dimensional standard normal distribution as $n\to\infty$.}

\medskip
To formulate our results for the case when the Dyson condition 
(\ref{1.3}) holds, we define a function $\bar p_n(t,T)$ by the formula
\begin{equation}
p_n(\bx,T)=C_n(T)^{-1}\bar p_n(|\bx|,T), \label{1.26}
\end{equation}
for $t=|\bx|>0$ and $\bar p_n(t,T)=0$ for $t<0$. The norming constant 
$C_n(T)$ is chosen in such a way that $\bar p_n(t,T)$ is a 
probability density function, i.e. 
\[
\di\int_0^\infty \bar p_n(t,T)\,dt=1.
\]
We will call $\bar p_n(t,T)$ the {\it probability density of 
the mean spin distribution 
along the radius}. 

In Parts 2 and 3 we will describe the limit behaviour of an
appropriate rescaling of the probability density $\bar p_n(t,T)$ for $T=T_c$
and $T<T_c$. Then we will formulate a Corollary which gives a 
good asymptotics for the norming constants $C_n(T)$ in (\ref{1.26}).
In such a way we get a good asymptotics for the probability
density functions $p_n(\bx,T)$ for $T\le T_c$.  
To do this we introduce the notations
\begin{equation}
\begin{aligned}
&\hat M_n(T)=\int_{-\infty}^\infty t\bar p_n(t,T)\,dt ,\\ 
&V_n(T)=\left(\int_{-\infty}^\infty 
(t-\hat M_n(T))^2\bar p_n(t,T)\,dt\right)^{1/2}, 
\end{aligned}
\label{1.27}
\end{equation}
and the {\it rescaled} probability density
\begin{equation}\label{1.28}
\pi_n(t,T)=V_n(T)\bar p_n\left(\hat M_n(T)+V_n(T)\,t,T\right) 
\end{equation}
which can be rewritte in an equivalent form as
\begin{equation}
\bar p_n(t,T)=\frac1{V_n(T)}
\pi_n\left(\frac{t-\hat M_n(T)}{V_n(T)},\,T\right). \label{1.29}
\end{equation}
Observe that, in general, $\hat M_n(T)$ and $\bar M_n(T)$, which 
is defined in \eqref{1.19}, are different,
but as we will see later, 
\[
\lim_{n\to\infty} [\hat M_n(T)-\bar M_n(T)]=0.
\] 

Our aim is to prove that in the case when the Dyson condition (\ref{1.3})
holds, there exists a critical temperature $T_c$ such that the
spontaneous  magnetization $M(T)=\lim\limits_{n\to\infty}\hat M_n(T)$
is positive for $T<T_c$ and it is zero for $T\ge T_c$. For 
$T<T_c$ the density function $\bar p_n(t,T)$ is concentrated near 
the point $t=\hat M_n(T)$, and the function $\pi_n(t,T)$ 
represents a rescaled distribution of $\bar p_n(t,T)$ near this point. 
 We want to prove that $\pi_n(t,T)$ tends to a 
limit $\pi(t)$ as $n\to\infty$. It turns out that this limit does 
exist, and the limit function $\pi(t)$ is a nice analytic 
function, although it is non-Gaussian. The function $\pi(t)$ is 
expressed in terms of a solution of a nonlinear fixed point 
equation, and the next  proposition concerns the existence of 
such a solution. Introduce the space of probability densities 
$p(t)$ on the line
\begin{equation*} 
\mathcal A=\left\{p(t)\colon\; \int_{-\infty}^\infty 
e^{\varepsilon |t|}p(t)\,dt<\infty 
\;\textrm{for some}\; \varepsilon=\varepsilon(p(t))>0\right\}.
\end{equation*}
Consider also the subspace $\mathcal A_0\subset \mathcal A$,
\begin{equation*}  
\mathcal A_0=\left\{p(t)\colon\; p(t)\in\mathcal A, \; \;
\int_{-\infty}^\infty tp(t)\,dt=0\right\}.
\end{equation*}

\medskip\noindent
{\bf Proposition 1.4.}  {\it There exists a unique probability density
function $g\in\mathcal A_0$ which satisfies the following 
fixed point equation:
\begin{equation} \label{1.30}
\begin{aligned}
g(t)=\frac {2}{\pi^{\frac{r-1}{2}}}
&\int_{u\in {\mathbb R}^1,\, \bv\in  {\mathbb R}^{r-1}} e^{-|\bv|^2}  
g\left(t-\frac{r-1}{4}-u+\frac{|\bv|^2}2\right)  \\
&  \times g\left(t-\frac{r-1}{4}+u+\frac{|\bv|^2}2\right)\,du\,d\bv. 
\end{aligned}
\end{equation} 
The density $g(t)$ can be extended to an entire  function on the
complex plane, and for real $t$ it satisfies the estimate
\begin{equation}
0< g(t)< C_\varepsilon\exp \{-(2-\varepsilon) |t|\},\quad 
\textrm{for all }\varepsilon>0.
\label{1.30a}
\end{equation}}

\medskip
For a proof of Proposition 1.4 see the proof of Lemmas~12 and 13 in
\cite{BM3}. It is worth noticing that the Fourier transform of $g$,
\[
\tilde g(\xi)=\int_{-\infty}^\infty e^{i\xi t} g(t)\,dt,
\]
solves the equation
\begin{equation}\label{1.30aa}
\tilde g(\xi)=\frac{e^{\frac{i\xi(r-1)}{4}}
\tilde g^2(\frac{\xi}{2})}{\left(1+\frac{i\xi}{2}\right)^{\frac{r-1}{2}}}.
\end{equation}

Using the probability density $g(t)$ of Proposition 1.4, we
introduce a probability density $\pi(t)$ on the line of the form
\begin{equation} 
\pi(t)=c e^{-2bt/3}g(bt-a), \label{1.30b}
\end{equation}
where the numbers $b>0$, $c>0$, and $a$ are chosen in such a way that
\begin{equation} 
\int_{-\infty}^\infty \pi(t)\,dt=1, \quad \int_{-\infty}^\infty t\,\pi(t)\,dt=0, 
\quad \int_{-\infty}^\infty t^2\pi(t)\,dt=1. \label{1.30c}
\end{equation}
Observe that such $a,b,c$ exist and are unique. Indeed, after 
the change of variable $u=bt-a$,
the second equation in \eqref{1.30c} gives $a$ as
\begin{equation} \label{1.30ca}
a=-\frac{\di\int_{-\infty}^\infty ue^{-2u/3}g(u)\,du}{\di\int_{-\infty}^\infty e^{-2u/3}g(u)\,du}.
\end{equation}
Then the first and third equations determine $b$ and $c$
 uniquely from the system,
\begin{equation}\label{1.30cb}
\left\{
\begin{aligned}
&\frac{c}{b}\int_{\R} e^{-2(u+a)/3} g(u)\,du=1, \\
&\frac{c}{b^3}\int_{\R} (u+a)^2 e^{-2(u+a)/3} g(u)\,du=1. 
\end{aligned}
\right.
\end{equation}
Estimate \eqref{1.30a} secures the convergence of the integrals in 
\eqref{1.30ca} and \eqref{1.30cb}.

Now we formulate

\medskip\noindent
{\bf Theorem 1.5.} {\it Assume that
\begin{equation}
\sum_{n=1}^\infty l_n^{-1}<\infty. \label{1.31a}
\end{equation}
Then there exists a number $\kappa_0=\kappa_0(N)$, where $N$ is defined 
in \eqref{1.22}, such that for all 
$0<\kappa<\kappa_0$ the following statements hold.

Assume that the density $p(\bx)=\frac{\nu(d\bx)}{d\bx}$ belongs 
to the class $\mathcal{P}_\kappa$, and the sequence $\{l_n,\;n\ge 0\}$ 
satisfies Conditions 1--4. Then there exists a constant 
$L=L(\bar\eta,\kappa)$ such that if the sequence $\{l_n,\;n\ge 0\}$ 
satisfies Condition 5, then there exists a critical temperature 
$T_c>0$ with the following properties:

\medskip\noindent
\begin{enumerate}
\item If $T>T_c$ then
\begin{equation}
\lim_{n\to\infty} 2^n \bar M_n^2(T)=\chi(T)>0, \label{1.32}
\end{equation} 
and the distribution $\tilde\nu_{n,T}(d\bx)$ tends weakly, as
$n\to\infty$, to the $r$-dimensional standard normal distribution. 
The function $\chi(T)$ in (\ref{1.32}) satisfies the following 
estimates near the critical point: 
There exists a temperature $T_0>T_c$ and numbers $C_2>C_1>0$
such that for all $T_0>T>T_c$ there
exists a number $\bar n(T)$ such that
\begin{equation}
\begin{aligned}
&C_1\sum_{k=\bar n(T)}^\infty l_k^{-1}< T-T_c\le
C_2 \sum_{k=\bar n(T)}^\infty l_k^{-1},\\
&C_1\frac{2^{\bar n(T)}}{l_{\bar n(T)}}<\chi(T)< C_2\frac{2^{\bar n(T)}}
{l_{\bar n(T)}}.
\end{aligned} \label{1.33}
\end{equation}
(The number $\xi(T)=2^{\bar n(T)}$ is the correlation length.)

\item At $T=T_c$, $\lim\limits_{n\to\infty} M_n(T_c)=0$ 
(there is no Thouless' effect), and moreover
\begin{equation}\label{1.34}
\lim_{n\to\infty} L_n^{-1} M_n(T_c)=1, 
\end{equation}
where $M_n(T)$ is defined in (\ref{2.15}), and
\begin{equation}\label{1.34a}
L_n=\left(\frac{r-1}6\sum_{j=n}^\infty l_j^{-1}\right)^{1/2}.
\end{equation}
(Condition (\ref{1.31a}) implies that $\lim\limits_{n\to\infty} L_n=0$.)

Let us define the rescaled version $\rho_n(t)$
of the probability density function $\bar p_n(t,T_c)$ as
\begin{equation}\label{1.34b}
\rho_n(t)=\frac{\hat M_n(T_c)}{d_n}
\bar p_n \left( \hat M_n(T_c)
\left(1+\frac{t}{d_n}\right),T_c\right), 
\end{equation}
where $\hat M_n(T)$ is defined in (\ref{1.27}), and
\begin{equation}
d_n= \frac{(r-1)l_n}{2b}\sum_{k=n}^\infty l_k^{-1}.
\label{1.34c}
\end{equation}
(Observe that $\lim\limits_{n\to\infty}d_n=\infty$  by Condition~2 on $\{l_n\}$.)
The function $\rho_n(t)$ is defined on the half-line $[-d_n,\infty)$.
Then
\begin{equation} 
\lim_{n\to\infty}\|\rho_n(t)-\pi(t)\|=0,  \label {1.35a}
\end{equation}
where the probability density  $\pi(t)$ is defined 
in equations  (\ref{1.30b}), (\ref{1.30c}) and
\begin{equation}
\|f(t)\|=\sum_{j=0}^2\sup_{t\ge -d_n} 
\left\{e^{|t|/3}\left|\frac{d^jf(t)}{dt^j}\right|\; \right\}
\label{1.40}
\end{equation}

\medskip\noindent
\item If $T<T_c$, then the numbers $\hat M_n(T)$ and $V_n(T)$ 
defined in formula (\ref{1.27}) satisfy the following relations:
The limit
\begin{equation}
\lim_{n\to\infty} \hat M_n(T)=M(T)>0  \label{1.36}
\end{equation}
exists, and  
\begin{equation}
C_1|T-T_c|^{1/2}<M(T)< C_2|T-T_c|^{1/2}.
\label{1.37}
\end{equation}
In addition,
\begin{equation}
\lim_{n\to\infty} l_nV_n(T)=\gamma(T)= \frac{b T}{3M(T)}>0
\label{1.38}
\end{equation}
with the number $b$ appeared in formula (\ref{1.30b}), and 
\begin{equation} 
\lim_{n\to\infty} \|\pi_n(t,T)-\pi(t)\|=0 \label{1.39}
\end{equation}
where the  probability densities $\pi_n(t,T)$ and $\pi(t)$ are defined 
in equations (\ref{1.28}) and (\ref{1.30b}), (\ref{1.30c}), 
respectively, and $\|f(t)\|$ is defined in (\ref{1.40}), with  
$d_n=\frac{\hat M_n(T)}{V_n(T)}$.  
\end{enumerate}}

\medskip
Theorems 1.3 and 1.5 are the central results of the present paper.
Let us make some remarks about Theorem 1.5.
Relations (\ref{1.32}) and (\ref{1.34}) imply that
$$
M(T)=\lim_{n\to\infty} \hat M_n(T)=0,\quad \textrm{for all } T\ge T_c,
$$
i.e., the spontaneous magnetization $M(T)$ vanishes at $T\ge T_c$. 
Relation \ref{1.37}) implies that
$$
\lim_{T\to T_c^-}M(T)=0,
$$
with the classical critical exponent $1/2$ for the magnetization.

The number $\bar n(T)$ in (\ref{1.33}) is very important for our 
investigation in the subsequent sections. It shows how many 
iterations of the recursive equation (renormalization group 
transformation) is needed to reach the ``high temperature region'' 
(see Section 3 below for precise definitions). The quantity
$\xi(T)=2^{\bar n(T)}$ is the {\it correlation length.} Usually the
correlation length has a power-like asymptotics $\xi(T)\asymp
|T-T_c|^{-\nu}$ as $T\to T_c$ where $\nu$ is the critical exponent of
the correlation length (see, e.g., \cite{Fish} or \cite{WK}). It 
follows from (\ref{1.33}) that in the case under consideration, 
$\xi(T)$ grows super-polynomially as $T\to T_c^+$. For instance, 
if $l_n$ is a sequence determined by equation (\ref{1.16}) then
$\xi(T)$ grows like $\exp \left[C_0(T-T_c)^{1/(\lambda-1)}\right]$. 
Similarly, (\ref{1.33}) implies that the magnetic susceptibility 
$\chi(T)$ diverges super-polynomially as $T\to T_c^+$.

Relation (\ref{1.38}) shows that the mean square deviation of the 
mean spin along the radius behaves, when $n\to\infty$, as
$$
V_n(T)\sim \frac{bT}{3M(T)l_n}, \qquad T<T_c,
$$
so that it goes to zero very slowly as $n\to \infty$ (comparing 
with the standard behavior of $C2^{-n/2}$). In fact, it goes to zero
sub-polynomially with respect to the number of spins $2^n$.
And according to \eqref{1.34c}, at $T=T_c$ the scaled mean square deviation 
 of the mean spin along the radius, $d_n^{-1}$, goes to zero 
even slower, than at $T<T_c$, namely,
$$
d_n^{-1}\sim \frac{2b}{r-1}\left(l_n \sum_{k=n}^\infty l_k^{-1}\right)^{-1},\quad T=T_c.
$$
On the other hand, observe that by (\ref{1.35a}) and (\ref{1.39}), the limit 
distribution density $\pi(t)$ of the normalized mean spin along the 
radius is the same for all $T<T_c$ and for $T=T_c$ as well.

Let us say some words about our methods.
The questions we investigate in this paper lead to a problem of the
following type:  We have a starting probability density
function $p_0(\bx,T)$ which depends
on a parameter $T$, the temperature, and we apply the powers of an
appropriately defined nonlinear
operator $\mathbf Q$ to it. This operator $\mathbf Q$
is the renormalization group operator. We want to describe
the behavior of the sequence of functions
$p_n(\bx,T)={\mathbf Q}^n p_0(\bx,T)$, $n=1,2,\dots$. In particular, we
want to understand how the behavior of this sequence of functions
$p_n(\bx,T)$, $n=1,2,\dots$, depends on the parameter~$T$. Our
investigation shows that if the function $p_n(\bx,T)$ is essentially
concentrated around the origin, then a negligible error is committed
when $p_{n+1}(\bx,T)={\mathbf Q} p_n(\bx,T)$ is replaced by the
convolution of the function $p_n(\bx,T)$ with itself, and this is
the case for all $n$ if the parameter $T$ is large. The replacement
of the operator $\mathbf Q$ by the convolution is called the 
{\it high temperature approximation}. 

On the other hand, if the function $p_n(\bx,T)$ is essentially 
concentrated in a narrow shell far from the origin, and this is 
the case for all $n$ if the parameter $T$ is small, then another 
good approximation of the  function 
$p_{n+1}(\bx,T)={\mathbf Q}_n p_n(\bx,T)$ is possible. This is 
called the  {\it low temperature approximation}. The high temperature 
approximation  actually means the application of the standard methods 
of classical probability theory. The low temperature approximation 
applied in this paper is a natural modification of the methods in 
our paper \cite{BM3} where a similar problem was investigated. But in 
the present paper we have to make a more careful and detailed
analysis. The reason for it is that  while in \cite{BM3} it was enough 
to investigate only very low temperatures $T$, now we have to follow
carefully when the high and when the low temperature approximation is
applicable. Moreover, --- and this is a most important part of this
paper, ---  to describe the behavior of the functions $p_n(\cdot,T)$
for all temperatures~$T$ we have to follow the behavior of these
functions also in the case when neither the high nor the low
temperature approximation is applicable. This is the so called
{\it intermediate region}.\/ (See Section~3 for precise definitions.)

We study the intermediate region in Section~5. There we show that if 
the function $p_n(\bx,T)$ ``is not very far from the origin", namely, 
the low temperature approximation is not applicable for it, then 
the functions $p_{n+k}(\bx,T)$ are getting closer and closer to the 
origin as the index $n+k$ is increasing. Moreover, after 
finitely many steps $k$ the high temperature approximation is 
already applicable, and the number of steps $k$ we need to get into 
this situation can be bounded by a constant independent of the 
parameter~$T$. The proof given in Section~5 contains arguments 
essentially different from  the rest of the paper. Here we 
heavily exploit that the numbers $c_n=\frac{l_n}{l_{n-1}}$
are very close to one. Informally speaking,
the sequence of numbers $c_n-1$ behaves like a small parameter, and
this ``small parameter" enables us to handle our model near the
critical temperature.

The setup of the rest of the paper is the following. In
Section 2 we give an analytic reformulation of the problem and
connect Dyson's condition (\ref{1.3}) with an approximate
recursive formula for some quantities  $M_n(T)$ related to the
spontaneous magnetization (see (\ref{2.28}) below). In Section 3 
we introduce a notion of low and high temperature regions together 
with an intermediate region. Then we formulate the basic auxiliary theorems
about the characterization of these regions. In Sections~4,~5, and~6
we prove the main estimates concerning the low temperature region,
the intermediate region, and the high temperature region,
respectively. In Section~7 we prove the convergence of the recursive
iterations to the fixed point for all $T<T_c$. Finally, in Section~8 
we prove Theorem 3.4 concerning some asymptotics near the critical 
point $T_c$ and derive Theorems 1.3 and 1.5 from the auxiliary theorems.


\section{Analytic Reformulation of the Problem. Strategy of the Proof.}

The hierarchical structure of the Hamiltonian (\ref{1.1}) leads to the
following {\it recursive equation} for the density functions $p_n(\bx,T)$
(see, e.g.,\ Appendix A to the paper \cite{BM3}):
\begin{equation}
p_{n+1}(\bx,T)=C_n(T)\int_{{\mathbb R}^r} 
\exp\left(\frac{l_n}T(\bx^2-\bu^2)\right)p_n(\bx-\bu,T)p_n(\bx+\bu,T)\,d\bu,
\label{2.1}
\end{equation}
for $n\ge0$, where $p_0(\bx,T)=p_0(\bx)$ is defined in~(\ref{1.17}),
$$
l_n=l(2^n),           
$$
and $C_n(T)$ is an
appropriate norming constant which turns $p_{n+1}(\bx,T)$ into a density
function. We are interested in the asymptotic behaviour of the functions
$p_n(\bx,T)$ as $n\to\infty$. For the sake of simplicity we will assume
that $\varepsilon(t)=0$ in (\ref{1.17}), so that $p_0(\bx)$ coincides with 
(\ref{1.4}). All the proofs below are easily extended to the case of 
nonzero $\varepsilon(t)$ satisfying estimate (\ref{1.18}).

Define
\begin{equation}
c_n=\frac{ l_n}{l_{n-1}},\quad n=0,1,\dots
\textrm{ with }\quad l_{-1}=1, \label{2.3}
\end{equation}
\begin{equation}
A_n=1+\sum_{j=1}^\infty \frac{
c_{n+1}}2\cdots\frac{ c_{n+j}}2
=1+l_n^{-1}\sum_{j=1}^\infty 2^{-j}l_{n+j},\qquad n=0,1,\dots.
\label{2.4}
\end{equation}
%
Then 
\begin{equation}
l_n=\prod\limits_{j=0}^n c_j, \quad n\ge 0,
\label{2.6}
\end{equation}
and
\begin{equation}
l_nA_n=l_n+\frac{l_{n+1}A_{n+1}}2.
\label{2.7}
\end{equation}
Indeed, by (\ref{2.4}),
$$
l_nA_n=l_n+\sum_{j=1}^\infty 2^{-j}l_{n+j}=\sum_{j=0}^\infty 2^{-j}l_{n+j},
$$
hence
\begin{equation}
l_nA_n-l_n=\sum_{j=1}^\infty 2^{-j}l_{n+j}=\frac{1}{2}
\sum_{j=0}^\infty 2^{-j}l_{n+1+j}=\frac{l_{n+1}A_{n+1}} 2,
\label{2.9}
\end{equation}
and (\ref{2.7}) follows.

Define 
\begin{equation}
q_n(\bx,T)=\Lambda_n(T)^{-1}\exp\left(\frac{A_nl_n\bx^2}2\right)
p_n(\sqrt T\, \bx,T),
\label{2.10}
\end{equation}
where $\Lambda_n(T)>0$ is a norming constant such that
$$
\int_{{\mathbb R}^r}q_n(\bx,T)\,d\bx=1.
$$
Let
\begin{equation}
c^{(n)}=(1+A_{n})\,l_n,\qquad n=0,1,2,\dots \label{2.11}
\end{equation}
Then it follows from equations (\ref{2.1}) and \eqref{2.7} that
\begin{equation}
q_{n+1}(\bx,T)=\frac1{Z_n(T)}\int_{{\mathbb R}^r} e^{-c^{(n)} \bu^2}
q_n(\bx-\bu,T)q_n(\bx+\bu,T)\,d\bu. \label{2.12}
\end{equation}
Also, by (\ref{1.4}),
\begin{equation}
q_0(\bx,T)=\frac1{Z_0(T)}\exp\left\{( c_0A_0-T)\frac
{|\bx|^2}2-\kappa T^2\frac{|\bx|^4}4\right\}.  \label{2.13}
\end{equation}
The norming constants $Z_n(T)$ in the previous formulas 
are determined by the condition that
$$
\int_{{\mathbb R}^r}q_n(\bx,T)\,d\bx=1.
$$
Thus, the functions $q_n(\bx,T)$ are defined recursively
by formulas (\ref{2.12}) and (\ref{2.13}). Our goal is to derive an
asymptotics of the functions $q_n(\bx,T)$ as $n\to\infty$.
Then the asymptotics of the functions $p_n(\bx,T)$ can be found
by means of formula (\ref{2.10}). The advantage of the functions 
$q_n(\bx,T)$ is that their recursive equation \eqref{2.12} does not 
depend on $T$.

The method of paper~\cite{BM3} can be adapted in the study of the low
temperature approximation. We shall follow this approach.
Due to the rotational symmetry of the Hamiltonian (\ref{1.1}),
the function $q_n(\bx,T)$ depends only on $|\bx|$.
Define the function $\bar q_n(t,T)$, $t\in{\mathbb R}^1$,
$n=0,1,2,\dots$, such that
\begin{equation}
q_n(\bx,T)= C_n(T) \,\bar q_n(|\bx|,T), \label{2.14}
\end{equation}
with a norming constant $C_n(T)$ such that 
$$
\int_0^\infty \bar q_n(t,T)\,dt=1.
$$
 We will define
$$
\bar q_n(t,T)=0 \quad \textrm{ for}\quad t<0.
$$
 Put also
\begin{equation}
M_n(T)=\int_0^{\infty} t\, \bar q_n(t,T)\,dt,\quad n=0,1,\dots,
\label{2.15}
\end{equation}
and define the rescaled probability density functions
\begin{equation}
f_n(t,T)=\frac1{c^{(n)}}\bar q_n\left(M_n(T)+\frac t{c^{(n)}},T\right),
\quad t\in {\mathbb R}^1, \quad
n=0,1,\dots. \label{2.16}
\end{equation}
Then
\begin{equation}
\bar q_n(t,T)=c^{(n)} f_n\left(c^{(n)}(t-M_n(T)),T\right),
\label{2.17}
\end{equation}
and
$$
\int_{-\infty}^\infty f_n(t,T)\,dt=1, \quad
\int_{-\infty}^\infty t f_n(t,T)\,dt=0.
$$
The order parameter $M_n(T)$ in \eqref{2.15} is very convenient 
for the asymptotic recursive analysis. Later we will relate it to
the parameters $\bar M_n(T)$ and $\hat M_n(T)$ introduced in formulae
\eqref{1.19} and \eqref{1.27}, respectively.

A low temperature approximation can be applied in the case when
$M_n(T)$ is relatively large, comparing with the size of the
neighborhood of $M_n(T)$ in which the function $f_n(t,T)$ is essentially
concentrated. In this case we
follow the behaviour of the pair $(f_n(t,T), M_n(T))$. To describe this
procedure introduce the notation ${\mathbf c}=\{c^{(n)},\;n=0,1,\dots\}$. 
The rotational invariance of the function $q_n(\cdot,T)$ suggests the
definition of the operator
\[
\begin{aligned}
\bar  {\mathbf Q}_{n,M}^{\mathbf c}f(t)&=
\int_{u\in{\mathbb R}^1,\,\bv\in{\mathbb R}^{r-1}}
\exp\left\{-\frac{u^2}{c^{(n)}}-\bv^2\right\}  \\
& \times f\left(c^{(n)}\left(\sqrt{\left(M+\frac t{c^{(n+1)}}+\frac
u{c^{(n)}}\right)^2+\frac{\bv^2}{c^{(n)}}}-M\right)\right) \\
&\times f\left(c^{(n)}\left(\sqrt{\left( M+
\frac t{c^{(n+1)}}-\frac u{c^{(n)}}\right)^2+\frac{\bv^2}
{c^{(n)}}}-M\right)\right) \,du\,d\bv. 
\end{aligned}
\]
Formula (\ref{2.12}) together with the definition of the 
function $f_n(t,T)$ yields that
$$
\bar q_{n+1}\left(M_n(T)+\frac t{c^{(n+1)}},T\right)=\frac{c^{(n+1)}}{Z_n(T)}
\bar{\mathbf Q}_{n,M_n(T)}^{\mathbf c}f_n(t,T)
$$
with
$$
Z_n(T)=\int_{-c^{(n+1)}M_n(T)}^\infty
\bar {\mathbf Q}_{n,M_n(T)}^{\mathbf c}f_n(t,T)\,dt.
$$
The norming constant $Z_n(T)$ is determined by the condition
$$
\int_0^\infty \bar q_{n+1}(t,T)\,dt=1.
$$
 Define also
\begin{equation}
m_n(T)=m_n(f_n(t,T))=\frac1{Z_n(T)}\int_{-c^{(n+1)}M_n(T)}^\infty
t\,\bar {\mathbf Q}_{n,M_n(T)}^{\mathbf c}f_n(t,T)\,dt     \label{2.22}
\end{equation}
and
$$
{\mathbf Q}_{n,M_n(T)}^{\mathbf c}f_n(t,T)=\frac1{Z_n(T)}
\bar {\mathbf Q}_{n,M_n(T)}^{\mathbf c}f_n(t+m_n(T),T).
$$
Then
\begin{equation}
f_{n+1}(t,T)={\mathbf Q}_{n,M_n(T)}^{{\mathbf c}}f_n(t,T) 
\quad\textrm{and} \quad
M_{n+1}(T)=M_n(T)+\frac{m_{n}(T)}{c^{(n+1)}}. \label{2.24}
\end{equation}

To formulate a good approximation of the operator 
${\mathbf Q}_{n,M_n(T)}^{\mathbf c}$, let us introduce the numbers
\begin{equation}
\bar c_n=
\frac{c^{(n)}}{c^{(n-1)}}=\frac{(1+A_n)l_n}{(1+A_{n-1})l_{n-1}}, 
\quad n=1,2,\dots. \label{2.24a}
\end{equation}
The arguments of the function $f$ in the definition of the operator
$  \bar {\mathbf Q}_{n,M}^{\mathbf c}$,
\begin{equation}
\ell_{n,M}^{{\mathbf  c},\pm}(t,u,\bv)=c^{(n)}
\left(\sqrt{\left( M+\frac{t}{c^{(n+1)}}
\pm\frac u{c^{(n)}}\right)^2+\frac{\bv^2}{c^{(n)}}}-M\right), \label{2.25}
\end{equation}
can be well approximated by a simpler expression because of the
estimate
$$
\left|\ell_{n,M}^{{\mathbf c},\pm}(t,u,\bv) -\left(\frac{t}{\bar c_{n+1}}
\pm u+\frac{\bv^2}{2M}\right)\right|\le
100\left(\frac{|\bv|^4}{c^{(n)}M^3}+\frac{t^2+u^2} {c^{(n)}M}\right) 
$$
which holds for $|t|<\frac14c^{(n+1)}M$, \ $|u|<\frac14c^{(n)}M$ and 
$\bv^2<c^{(n)}M^2$. This estimate suggests that for low temperatures~$T$, 
when $M_n(T)$ is not small, the operator $\bar{\mathbf Q}_{n,M_n(T)}^{\mathbf c}$ 
can be well approximated by the operator 
$\bar{\mathbf T}_{n,M_n(T)}^{\mathbf c}$ defined as
\begin{equation}
\begin{aligned}
\bar{\mathbf T}_{n,M_n(T)}^{\mathbf c} f(t,T)&=
\int_{u\in{\mathbb R}^1,\,\bv\in{\mathbb R}^{r-1}} e^{-\bv^2}
f\left(\frac{t}{\bar c_{n+1}}+u+\frac{\bv^2}{2M_n(T)},T\right)  \\
&\times f\left(\frac t{\bar c_{n+1}}-u+\frac{\bv^2}
{2M_n(T)},T\right) \,du\,d\bv.
\end{aligned}  \label{2.27}
\end{equation}
The elaboration of the above indicated method will be called the 
{\it low temperature approximation}. It works well when $M_n(T)$ 
is much larger than the range where the function $f_n(t,T)$ is 
essentially concentrated. For $n=0$ the starting value $M_0(T)$ 
at very low temperatures $T>0$ is very large. In this case the low 
temperature expansion can be applied. As we shall see later, 
the approximation of $\bar {\mathbf Q}_{n,M_n(T)}^{\mathbf c}$ by 
$\mathbf {\mathbf T}_{n,M_n(T)}^{\bar c}$ yields that
\begin{equation}
M_{n+1}(T)\sim M_n(T)-\frac{r-1}{4c^{(n)}M_n(T)},
 \label{2.28}
\end{equation}
which, in turn, implies that
\begin{equation}
M_{n+1}^2(T)\sim M_n^2(T)-\frac{r-1}{2c^{(n)}}.
\label{2.29}
\end{equation}
It follows from (\ref{2.4}) and (\ref{1.10}) that
\begin{equation}
2\le A_n\le 2.03,\qquad \lim_{n\to\infty}A_n=2,
\label{2.30}
\end{equation}
hence if Condition 1 is satisfied, then not only 
$\lim\limits_{n\to\infty}c_n=1$, 
but also $\lim\limits_{n\to\infty}\bar c_n=1$, and by (\ref{2.11}),
\begin{equation}
3\le\frac{c^{(n)}}{l_n}\le 3.03,\qquad
\lim_{n\to\infty} \frac {c^{(n)}}{l_n}=3.
\label{2.31}
\end{equation}
This allows us to rewrite (\ref{2.29}) as
\begin{equation}
M_{n+1}^2(T)\sim M_n^2(T)-\frac{r-1}{6l_n}
\label{2.32}
\end{equation}
This formula underlines the importance of the Dyson condition
(\ref{1.3}). \hfill\break 
Namely, if the series
\begin{equation}
B=\sum_{n=1}^\infty l_n^{-1}
\label{2.33}
\end{equation}
converges then $M_n(T)$ remains large for all $n$ if $T>0$ is
small. Indeed, assume that $T<c_0A_0/2$. Then it follows
from (\ref{2.13}) that $M^2_0(T)>C(\kappa T^2)^{-1}$, hence by 
(\ref{2.32}), neglecting the error term,
$$
M_n^2(T)\ge M^2_0(T)-\frac{r-1}6\sum_{n=0}^\infty l_n^{-1}\ge
C(\kappa T^2)^{-1}-C_1\gg 1
$$
for all $n$ if $T>0$ is small, which was stated. On the other hand, if
the series (\ref{2.33}) diverges, then for some $n$, $M_n(T)$ 
becomes small, and the approximation (\ref{2.28}) becomes inapplicable.

The low temperature approximation can be applied when $M_n(T)$ is
not small. When $M_n(T)$ is small
a different approximation is natural. If the function
$q_n(\bx,T)$ is essentially concentrated in a ball whose radius is 
much less than $\left(c^{(n)}\right)^{-1/2}$, then a small error is 
committed if the kernel function $e^{-c^{(n)}\bu^2}$ in formula 
(\ref{2.12}) is omitted. This means that the formula expressing 
$q_{n+1}(\bx)$ by $q_n(\bx)$ can be well approximated through the 
convolution $q_{n+1}(\bx)=q_n*q_n(\bx)$.
This approximation will be called the {\it high temperature
approximation}. If the high temperature approximation can be applied 
for $q_n(\bx,T)$, then the function $q_{n+1}(\bx,T)$ is even more 
strongly concentrated around zero. Hence, as a detailed analysis 
will show, if at a temperature $T$ it can be applied for a certain 
$n_0$, then it can be applied for all $n\ge n_0$.

Finally, there are such pairs $(n,T)$ for which the function
$q_n(\bx,T)$ can be studied neither by the low nor by the high 
temperature approximation. We call the set of such pairs an 
{\it intermediate region.} We shall prove that if the sequence $c^{(n)}$
sufficiently slowly tends to infinity and the function $q_n(\bx,T)$ is
out of the region where the low temperature approximation is
applicable, then the density function $q_{n+1}(\bx,T)$ will be more
strongly concentrated around zero than the function $q_n(\bx,T)$.
Moreover, in {\it finitely many}\/ steps the function $q_{n+k}(\bx,T)$
will be so
strongly concentrated around zero that after this step the high
temperature approximation is applicable. It is important that the
number of steps $k$ needed to get into the high temperature
region can be bounded independently of the parameter $T$.
 
The main part of the paper consists of an elaboration of the 
above heuristic argument.

\section{Formulation of Auxiliary Theorems.}

To describe the region where the low temperature approximation will be
applied we define some sequences $\beta_n(T)$ which depend on the
temperature~$T$. Define recursively,
\begin{equation}
\begin{aligned}
\beta_N(T)&=\frac{\left( c^{(N)}\right)^2}{2^N},\\
\beta_{n+1}(T)&=\left(\frac{\bar c_{n+1}^2}2+\sqrt{\frac{\beta_n(T)}
{c^{(n)}}}\right)
\beta_n(T) +\frac{10}{M_n^2(T)}\quad \textrm{ for }\;n\ge N,
\end{aligned} \label{3.1}
\end{equation}
where the number $N$ is defined in (\ref{1.22}), $\bar c_n$ in 
(\ref{2.24a}) and $M_n(T)$ in (\ref{2.15}). As it will be seen 
later, these numbers measure how strongly the functions $f_n(x,T)$ 
are concentrated around zero. We define the low temperature region, 
where low temperature approximation will be applied.

\medskip\noindent
{\bf Definition of the low temperature region.} {\it A pair $(n,T)$ 
is in the low temperature region if the following properties (1) 
and (2) hold.
 
\begin{enumerate}
\item $0<T\le c_0 A_0/2$, where $A_0$ was defined in (\ref{2.4}). 
\item{} Either $0\le n\le N$ with
the number $N$ introduced in (\ref{1.22}) or $n>N$ and
$\frac{\beta_{n-1}(T)}{c^{(n-1)}}\le\eta$ with the number $\eta$ appearing
also in (\ref{1.22}). 
\end{enumerate}
The temperature $T$ is in the low temperature region if the pair 
$(n,T)$ is in the low temperature region for all numbers~$n$.} 
\medskip
Let us remark that by (\ref{2.6}) and (\ref{1.10})
$$
1\le l_n=\prod_{j=1}^n c_j\le 1.01^n,
$$
hence by (\ref{2.31}),
\begin{equation}
3\le c^{(n)}\le 3.03\cdot 1.01^n.
\label{3.3}
\end{equation}
Therefore, by (\ref{3.1}),
\begin{equation}
\frac{\beta_N(T)}{c^{(N)}}=\frac{c^{(N)}}{2^N}\le \frac1{c^{(N)}}\le\eta
\label{3.4}
\end{equation}
hence the pair $(N+1,T)$ is in the low temperature region if 
$T\le c_0 A_0/2$. Since $\beta_{n+1}(T)\ge\frac{10}{M_n^2(T)}$ the 
pair $(n,T)$ can get out of the low temperature region only if 
$M_n(T)$ becomes very small.

To define the high temperature region introduce the notations
\begin{equation}
\begin{aligned}
h_n(\bx,T)&=\left(c^{(n)}\right)^{-r/2}q_n\left(\frac
{\bx}{\sqrt{c^{(n)}}},T\right),\\
D_n^2(T)&=\int_{{\mathbb R}^r}\bx^2h_n(\bx,T)\,d\bx.
\end{aligned}\label{3.5}
\end{equation}
where the function $q_n(\bx,T)$ is defined in (\ref{2.10}). 
Let us also introduce the probability measure $H_{n,T}$,

\begin{equation}
H_{n,T}(\mathbf A)=\int_{\mathbf A} h_n(\bx,T)\,d\bx,\quad {\mathbf A}
\subset {\mathbb R}^r, \label{3.6}
\end{equation}
on ${\mathbb R}^r$.

\medskip\noindent
{\bf Definition of the high temperature region.} {\it
A pair $(n,T)$ is in the high temperature region if
$D^2_n(T)<e^{-1/\eta^2}$  with the number $\eta$ in formula (\ref{1.22}), 
where $D_n^2(T)$ is defined in (\ref{3.5}). The temperature $T$ is 
in the high temperature region if there exists a threshold index 
$n_0(T)$ such that $(n,T)$ is in the high temperature region for 
all $n\ge n_0(T)$.}

\medskip\noindent
It may happen that a pair $(n,T)$ belongs neither to the low nor to the
high temperature region. Then we say that $(n,T)$ belongs to the {\it
intermediate region}. Let us remark that we  introduced two numbers $N$ 
and $\eta$ in formula (\ref{1.22}), and in the formulation of the 
subsequent results $N$ and $\eta$ will denote these numbers.  
The following result is very important for us.

\medskip\noindent
{\bf Theorem 3.1.} {\it There exists a number $\kappa_0=\kappa_0(N)$
such that for all $0<\kappa<\kappa_0$ (where $\kappa$ appears in 
formula (\ref{1.4})) and $0<\bar\eta<\eta$ there is a number 
$L=L(\bar\eta,\kappa)$ for which the following is true. Assume that 
Conditions~1 and~5 (with $\bar\eta$ and this number 
$L=L(\bar\eta,\kappa)$) hold. We consider such temperatures $T$ for 
which there are numbers $n$  such that the pair $(n,T)$ does not 
belong to the low temperature region. Let $\bar n(T)\ge 0$ be the 
smallest number $n$ with this property. 

If the pair $(\bar n(T),T)$ does not belong to the high temperature 
region (which means that $(\bar n(T),T)$ is in the intermediate 
region), then there exist some numbers $K=K(\bar\eta,\kappa)>0$, 
$\tilde\eta=\tilde\eta(\bar\eta,\kappa)>0$, and 
$k=k(\bar\eta,\kappa)\in{\mathbb N}$ such that
$$
D^2_{\bar n(T)}(T)<K,\qquad \tilde\eta<D^2_{\bar n(T)+k}(T)<e^{-1/\eta^2}.
$$
This implies in particular that the pair $(\bar n(T)+k,T)$ with 
this index $k$ belongs to the high temperature region.}

\medskip
We shall also prove the following corollary of Theorem~3.1. 
(See the Remark after the proof of Lemma~6.1.) 

\medskip\noindent 
{\bf Corollary.} {\it Under the conditions of Theorem~3.1 all
temperatures $T>0$ belong either to the low or to the high 
temperature region. If the Dyson condition (\ref{1.3}) holds, 
then all sufficiently low temperatures belong to the low and all 
sufficiently high temperatures to the high temperature region. 
If the Dyson condition (\ref{1.3}) is violated, then all
temperatures~$T>0$ belong to the high temperature region.}

\medskip 
The next theorem concerns the {\it low temperature region.}

\medskip\noindent
{\bf Theorem 3.2.} {\it There exists a number $\kappa_0=\kappa_0(N)$ 
such that for all $0<\kappa<\kappa_0$ the following is true. Assume 
that the Dyson condition (\ref{1.3}) and Conditions 1 and 2 hold.
Assume that  the temperature $T$ is in the low temperature region. 
Then the numbers $M_n(T)$ defined in (\ref{2.15}) have a
limit, 
\begin{equation}
\lim_{n\to\infty} M_n(T)=M_\infty(T), \label{3.8}
\end{equation}
and
\begin{equation}
\lim_{n\to\infty}\frac{M^2_n(T)-M^2_\infty(T)}{\frac{r-1}2
\sum\limits_{k=n}^\infty \frac1{c^{(k)}}}=1. 
\label{3.9}
\end{equation}
In addition,
\begin{equation}
\lim_{n\to\infty}
\left\|\frac1{M_n(T)}
f_n\left(\frac t{M_n(T)},T\right)-
g\left(t\right)\right\|=0,
 \label{3.10}
\end{equation}
where
\begin{equation}
\| f(t) \|=\sum_{j=0}^2\,\sup_{t\ge -c^{(n)}M_n(T)}e^{|t|}
\left| \frac{d^jf(t)}{d\,t^j} \right|,
\label{3.11}
\end{equation}
$f_n(t,T)$ is introduced in (\ref{2.16}), and the probability 
density $g(t)$ is defined as a solution of the fixed point 
equation (\ref{1.30}).}

\medskip 
Part (3) of Theorem 1.5,  with the exception of estimate (\ref{1.37}),
follows from Theorem~3.2 and the additional relation $M_\infty(T)>0$ 
if $T<T_c$ which follows from the results in Theorem~3.4 formulated 
at the end of this section.  Indeed, we can express the function 
$p_n(\bx,T)$ in terms of $f_n(t,T)$. Namely, by (\ref{2.10}), 
(\ref{2.14}), and (\ref{2.17})
\begin{equation}
\begin{aligned}
p_n(\bx,T)&=L_n^{-1}(T)\, \exp\left(- \frac{A_nl_n|\bx|^2}{2T}\right)\\
&\qquad \times f_n\left(\frac{c^{(n)}}{\sqrt T}\left(|\bx|
- \sqrt T\,M_n(T)\right),T\right)
\end{aligned}
\label{3.12}
\end{equation}
with an approriate norming constant $L_n(T)$. Let us write that
$|\bx|^2=\left(\sqrt T\,M_n(T)+|\bx|-\sqrt T\,M_n(T)\right)^2$, hence
\begin{eqnarray*}
&&\exp\left(- \frac{A_nl_n|\bx|^2}{2T}\right)=
\exp\biggl\{- \frac{A_nl_n}{2T}\bigl[TM_n^2(T) \\
&&\qquad +2\sqrt T\,M_n(T)(|\bx|-\sqrt T\,M_n(T))
+(|\bx|-\sqrt T\,M_n(T))^2\bigr]\biggr\}, 
\end{eqnarray*}
and substitute it into (\ref{3.12}). This leads to the equation
\begin{equation}
p_n(\bx,T)=\tilde L_n^{-1}(T)\tilde f_n\left(
\frac{|\bx|-\tilde M_n(T)}{\tilde V_n(T)},\,T\right) \label{3.12a}
\end{equation}
with an appropriate norming constant $\tilde L_n(T)$, where
\begin{eqnarray*}
\tilde M_n(T)&=&\sqrt T\, M_n(T),\qquad\tilde V_n(T)=\frac{\sqrt T}
{c^{(n)}M_n(T)}, \\
\tilde f_n(t,T)&=&f_n\left( \frac{t}{M_n(T)},T\right)\exp\left(-\frac{A_nl_nt}
{c^{(n)}}-\varepsilon_n(t,T)\right), \\
\varepsilon_n(t,T)&=&\frac{A_nl_nt^2}{2(c^{(n)})^2M^2_n(T)}.
\end{eqnarray*}
Observe that by (\ref{2.30}) and (\ref{2.31})
$$
\lim_{n\to\infty} \frac{A_nl_n}{c^{(n)}}=\frac23,
\qquad \lim_{n\to\infty}\frac{A_nl_n}{2(c^{(n)})^2M^2_n(T)}=0,
$$
hence  (\ref{3.10}) implies that there is some $C_0>0$ such that
\begin{equation*}
\lim_{n\to\infty}\left\|\frac1{M_n(T)}\tilde f_n(t,T)-
C_0 g\left(t\right) e^{-2t/3}\right\|'=0, 
\end{equation*}
where
$$
\| f(t) \|'=\sum_{j=0}^2\,\sup_{t\ge -c^{(n)}M^2_n(T)}e^{|t|/3}
\left| \frac{d^jf(t)}{d\,t^j} \right|.
$$
This also implies that there exist some real number $a$ and $C'>0$ 
such that
\begin{equation}
\lim_{n\to\infty}\left\|\frac1{M_n(T)}\tilde f_n(t-a,T)-
C' g\left(t-a\right) e^{-2t/3}\right\|'=0, \label{3.12b} 
\end{equation}
with such numbers $a$ and $C'>0$ for which the relations  
$$
\int_\infty^\infty C' g(t-a) e^{-2t/3}\,dt=1  \textrm{ and }
\int_\infty^\infty C' tg(t-a) e^{-2t/3}\,dt=0
$$  
hold.

Let us define for all $b>0$ the function $\pi(t|b)=
C'b g(bt-a) e^{-2bt/3}\,dt$. These functions satisy the
relations 
$$
\int_\infty^\infty C'b \pi(t|b)\,dt=1 \textrm{ and } 
\int_\infty^\infty C'b t\pi(t|b)\,dt=0.
$$
Moreover, the number $b>0$
can be chosen in such a way that the identity 
\begin{eqnarray*}
\int_\infty^\infty C'b t^2\pi(t|b)\,dt
&=&b^{-2}\int_\infty^\infty C'b (bt)^2\pi(t|b)\,dt  \\
&=&b^{-2}\int_\infty^\infty C' t^2 g(t-a)e^{2t/3}\,dt=1
\end{eqnarray*}
also holds. Let us define the function $\pi(t)=\pi(t|b)$ with this
parameter $b$. In such a way we constructed a function $\pi(t)$
that satisfies relations (\ref{1.30b}) and (\ref{1.30c}).
Moreover, if we define the functions
\begin{equation}
\tilde \pi_n(t,T)=C_n(T) \tilde f_n(bt-a,T)
\label{3.12d}
\end{equation}
with these numbers $a$ and $b$ and with such a norming constant 
$C_n(T)$ for which
$$
\int_{-\infty}^\infty\tilde \pi_n(t,T)\,dt=1,
$$
then these functions satisfy the relation
\begin{equation}
\lim_{n\to\infty}\left\|\tilde\pi_n(t)-\pi(t)\right\|'=0 \label{3.12c} 
\end{equation}
because of relations (\ref{3.12b}). Because of (\ref{3.12c}) we
also have $\int_{-\infty}^\infty \tilde \pi_n(t,T)\,dt=1$,
$\lim\limits_{n\to\infty}\int_{-\infty}^\infty t\tilde \pi_n(t,T)\,dt=0$,
$\lim\limits_{n\to\infty}\int_{-\infty}^\infty t^2\tilde \pi_n(t,T)\,dt=1$, 
and because of (\ref{3.12a}) and (\ref{3.12d})
\begin{eqnarray*}
\tilde \pi_n(t,T)&=&C'_n(T)^{-1}\bar p_n
((bt-a)\tilde V_N(T)+\tilde M_n(T),T) \\ 
&=& b{\tilde V_n(T)}\bar p_n (bt-a)\tilde V_N(T)+\tilde M_n(T),T).
\end{eqnarray*} 
(The normalization constant in the second identity of the last 
formula is determined by the fact that both $\bar p_n(t,T)$ and 
$\tilde \pi_nt,T)$ are probability density functions).

Hence
\begin{eqnarray}
\int_{-\infty}^\infty t\tilde \pi_n(t,T)\,dt
&=& b\tilde V_n(T)\int_{-\infty}^\infty t
\bar p_n (bt-a)\tilde V_N(T)+\tilde M_n(T),T)  \,dt \nonumber \\
&=&\int_{-\infty}^\infty 
\frac{t- \tilde M_n(T)+a\tilde V_n(T)}{b\tilde V_n(T)} \bar p_n(t,T)\,dt 
\label{3.13}\\
&=&\frac{\bar M_n(T) -\tilde M_n(T)+a\tilde V_n(T)}{b\tilde V_n(T)}
\to0 \quad\textrm{as }n\to\infty, \nonumber
\end{eqnarray}
and
\begin{eqnarray}
&&  \!\!\! \!\!\!\!\!
 \int_{-\infty}^\infty t^2\tilde \pi_n(t,T)\,dt
= b\tilde V_n(T)\int_{-\infty}^\infty t^2
\bar p_n (bt-a)\tilde V_N(T)+\tilde M_n(T),T)\,dt \nonumber \\
&& \!\!=\int_{-\infty}^\infty
\left(\frac{(t-\bar M_n(T))+(\bar M_n(T)- \tilde M_n(T)+a\tilde V_n(T))}
{b\tilde V_n(T)}\right)^2 \bar p_n(t,T)\,dt \nonumber \\
&& \!\!=\frac{V_n^2(T)
+\left((\bar M_n(T)-\tilde M_n(T))+a\tilde V_n(T)\right)^2}
{b^2\tilde V^2_n(T)}\to1 \quad\textrm{as }n\to\infty. \label{3.14}
\end{eqnarray}

Relations (\ref{3.13}) and (\ref{3.14}) together with Theorem 3.2, 
the inequqailty $M_\infty(T)>0$ and the definition of the quantities 
$\tilde M_n(T)$ and $\tilde V_n(T)$ imply relations (\ref{1.36}) and 
(\ref{1.38}). Indeed, by Theorem~3.2 
$\lim\limits_{n\to\infty}\tilde M_n(T)=M(T)$ with $M(T)=\sqrt T M_\infty(T)$, 
and since $\tilde V_n(T)\to0$ as $n\to\infty$ relation (\ref{3.13}) 
implies that $\lim\limits_{n\to\infty}(\bar M_n(T)-\tilde M_n(T))=0$. Formula 
(\ref{1.36}) follows from these relations with the above defined number 
$M(T)$. Relations (\ref{3.13}) and (\ref{3.14}) together imply that
$\lim\limits_{n\to\infty}\frac{V_n(T)}{\tilde V_n(T)}=b$. On the other hand,
$\lim\limits_{n\to\infty} l_n \tilde V_n(T)
\lim\limits_{n\to\infty} \frac{c^{(n)}}3 \frac T{c^{(n)}\sqrt TM_n(T)}
=\frac T{3M(T)}$. These relations imply (\ref{1.38}).

Finally to prove relation (\ref{1.39}) let us observe how the functions
$\pi_n(t,T)$ and $\tilde \pi_n(t,T)$ can be expressed with the help
of the function $\bar p_n(t,T)$. Besides, both are probability density
functions, and the integrals 
$\int_{-\infty}^\infty t\tilde\pi_n(t,T)\,dt$ and
$\int_{-\infty}^\infty t^2\tilde\pi_n(t,T)\,dt$ tend to zero and 1 as
$n\to\infty$, while the correponding integrals for $\pi_n(t,T)$ are
equal exactly to these limit values for all parameters $n$. This 
implies that the identity  $\pi_n(t,T)=(1+\varepsilon_n)
\tilde\pi_n((1+\varepsilon_n)t+\delta_n,T)$
with such numbers $\varepsilon_n=\varepsilon_n(T)$ and
$\delta_n=\delta_n(T)$ for which $\lim\limits_{n\to\infty}\varepsilon_n=0$,
and $\lim\limits_{n\to\infty}\delta_n=0$. It can be proved with the 
help of this observation that relation (\ref{3.12c}) remains valid 
if we replace the functions $\tilde \pi_n(t,T)$ by $\pi_n(t,T)$ in it, 
and this means that formula (\ref{1.39}) is valid.

\medskip
Now we formulate a theorem about the {\it high temperature
region}. Put
\begin{equation}
\tilde h_n(\bx,T)=2^{-rn/2}q_n\left(2^{-n/2}\bx,T\right)
=\left(\frac{c^{(n)}}{2^n}\right)^{r/2}
h_n\left(\sqrt{\frac{c^{(n)}}{2^n}}\bx,T\right), \label{3.24}
\end{equation}
and define the probability measures
\begin{equation}
\tilde H_{n,T}(\mathbf A)=\int_{\mathbf A} \tilde h_n(\bx,T)\,d\bx,
\quad {\mathbf A}\subset{\mathbb R}^r \label{3.25}
\end{equation}
on ${\mathbb R}^r$.

\medskip\noindent
{\bf Theorem 3.3.} {\it There exists a number $\kappa_0=\kappa_0(N)$ 
such that for all $0<\kappa<\kappa_0$ and $0<\bar\eta<\eta$ there 
exists a number $L=L(\bar\eta,\kappa)$ such that the following is 
true. Assume that Conditions 1 and 5 (with $\bar\eta$ and this 
$L=L(\bar\eta.\kappa)$) hold, and $T$ is in the high temperature 
region. Then the measures $\tilde H_{n,T}$ defined in (\ref{3.25}) 
converge weakly to the normal distribution on ${\mathbb R}^r$ with 
expectation zero and covariance matrix $\sigma^2(T){\mathbf I}$ 
with some $\sigma^2(T)>0$, where ${\mathbf I}$ denotes the 
identity matrix.

If $T$ belongs to the high temperature region, but the pair
$n=(0,T)$ does not belong to it, (i.e.\ the temperature $T$ is not too
high), then the inequality
\begin{equation}
C_1\frac{2^{\bar n(T)}}{c^{(\bar n(T))}}\le
\sigma^2(T)\le C_2\frac{2^{\bar n(T)}}{c^{(\bar n(T))}} \label{3.26}
\end{equation}
also holds with some $C_2>C_1>0$, where $\bar n(T)$ is defined in
Theorem~3.1.}

\medskip\noindent 
{\it Remark.} Not only the convergence of the measures $\tilde H_{n,T}$
but also the convergence of their density functions $\tilde h_n(\bx,T)$
could be proved. But the proof of the convergence of the distribution is
simpler, and it is also sufficient for our purposes.

\medskip\noindent 
{\bf Corollary.} {\it Let $\bar H_{n,T}$ denote the probability
measure on ${\mathbb R}^r$ with the density function
$$
2^{-rn/2}T^rp_n(2^{-n/2}\sqrt{T}\bx,T).
$$
Under the conditions of Theorem
3.3 the measures $\bar H_{n,T}$ have the same Gaussian limit as the
measures $\tilde H_{n,T}$ defined in Theorem~3.3 as $n\to\infty$.}

\medskip 
Our last theorem concerns the {\it  critical point}. We want to show 
that there is a critical temperature $T_c$ such that
above it all temperatures belong to the high and below it all
temperatures belong to the low temperature region. We also want to
describe the situation in the neighborhood of the critical 
temperature in more detail. In Theorem 3.4 we state such a result.

\medskip\noindent
{\bf Theorem 3.4.} {\it There exists a number 
$\kappa_0=\kappa_0(N)$ such that for all $0<\kappa<\kappa_0$ 
there exists a number $L=L(\bar\eta,\kappa)$ such
that the following is true.
Assume that Conditions~1--4 are
satisfied. Then for a fixed $n$ the set of temperatures $T$ for which
$(n,T)$ belongs to the low temperature region forms an interval 
$(0,T_n]$, and the sequence $T_n$, $n=1,2,\dots$, is  monotone 
decreasing in $n$. Define the critical temperature $T_c$ 
as the limit, $T_c=\lim\limits_{n\to\infty}T_n$. Then 
$c_0A_0/4>T_c>0$. The function
$M_\infty(T)=\lim\limits_{n\to\infty}M_n(T)$ exists in the interval
$(0,T_c]$, and for a fixed $n$ the function $M_n(T)$ is
strictly decreasing on the interval $(0,T_n]$. The relation 
$M_\infty(T_c)=0$ holds. If 
$T_c+\varepsilon>T>T_c$ 
with some $\varepsilon>0$, then the inequality
\begin{equation}
C_1\sum_{k=\bar n(T)}^\infty \frac1{c^{(k)}}<T-T_c<
C_2\sum_{k=\bar n(T)}^\infty \frac1{c^{(k)}}  \label{3.27}
\end{equation}
holds with some appropriate numbers $C_2>C_1>0$,
where $\bar n(T)$ is defined in Theorem 3.1. If
$T_c-\varepsilon<T<T_c$ with a sufficiently
small $\varepsilon>0$, then
\begin{equation}
C_1(T_c-T)^{1/2}< M_\infty(T)<C_2(T_c-T)^{1/2}.
\label{3.28}
\end{equation}
}

\section{Basic Estimates in the Low Temperature Region.}
 
In this section we give some basic estimates on the function $f_n(x,T)$
and its derivatives (with respect to the variable $x$) if the pair
$(n,T)$ is in the low temperature region. These estimates
state in particular, that in the definition of the functions 
$f_n(x,T)$ the right scaling was chosen. With the scaling in formula 
(\ref{2.16}) the function $f_n(x,T)$ is essentially concentrated in 
a finite interval whose size depends only on $M_n(T)$.
Both the results and proofs are closely related to those of
Sections~3---6 in paper~\cite{BM3}.

First we consider the case of {\it  small} indices
$0\le n\le N$, where the number~$N$ defined in~(\ref{1.22}) 
(cf. Section~4 in~\cite{BM3}), and we begin with $n=0$.
Assume that $T<c_0A_0/2$ and $\kappa>0$ is small (exact conditions on the
smallness of $\kappa$ will be given later). In this case
the function $\bar q_0(x,T)$ has its maximum in the points $\bar M_0(T)$ 
(see (\ref{2.13})), where
\begin{equation}
\bar M_0(T)=\left(\frac{A_0 c_0-T}{\kappa T^2}\right)^{1/2} \label{4.1}
\end{equation}
is a large number. From (\ref{2.13}) we obtain that
\begin{eqnarray}
&&\frac1{c^{(0)}} \bar q_0\left(\bar M_0(T)+\frac x{c^{(0)}},T\right)
\label{4.2} \\
&&\qquad =\frac1{Z_0(T)}   
\exp\left\{-\left(A_0 c_0-T\right)\left(\frac{x}{c^{(0)}}\right)^2
\left(1+\frac x
{2c^{(0)} \bar M_0(T)}\right)^2 \right\}, \nonumber 
\end{eqnarray}
where
\begin{equation}
Z_0(T)=\int_{-\bar M_0(T)}^\infty
\exp\left\{-(A_0 c_0-T)\left(\frac x{c^{(0)}}\right)^2
\left(1+\frac x{2c^{(0)}\hat
M_0(T)}\right)^2 \right\}\,dx. \label{$4.2'$}
\end{equation}
It can be proved by means of the identity
\begin{eqnarray}
&&c^{(0)}(M_0(T)-\bar M_0(T)) \label{4.3} \\
&&\qquad =\frac{\int_{-c^{(0)}\bar M_0(T)}^\infty
x\exp\left\{-(c_0A_0-T)\left(\frac x{c^{(0)}}\right)^2
\left(1+\frac x{2c^{(0)} \bar M_0(T)}\right)^2 \right\}\,dx}
{\int_{-c^{(0)}\bar M_0(T)}^\infty \exp\left\{-(A_0 c_0-T)
\left(\frac x{c^{(0)}}\right)^2
\left(1+\frac x{2c^{(0)}\bar M_0(T)}\right)^2 \right\}\,dx} \nonumber 
\end{eqnarray}
that
\begin{equation}
\left|M_0(T)-\bar M_0(T)\right|\le\frac{{\textrm{ const.}\,}}{M_0(T)}\le
{\textrm{ const.}\,} \sqrt \kappa\,T,  \label{4.4}
\end{equation}
where $M_0(T)$ is defined (\ref{2.15}). This shows that 
$\bar M_0(T)$ is a very good approximation to $M_0(T)$.
Straightforward calculation yields with the help of formulas 
(\ref{4.1}) and (\ref{$4.2'$}) that
\begin{equation}
\left|Z_0(T)-\frac{c^{(0)}\sqrt \pi}{\sqrt{(A_0 c_0-T)}}\right|
\le{\textrm{ const.}\,} \sqrt \kappa\,T,  \label{4.5}
\end{equation}
and from (\ref{4.1})--(\ref{4.5}) we obtain that
\begin{eqnarray}
&&\left|\frac{\partial^j}{\partial x^j}\left(f_0(x,T)-\frac{\sqrt{A_0
c_0-T}}{c^{(0)}\sqrt\pi}\exp\left\{-(A_0 c_0-T)\left(\frac
x{c^0}\right)^2\right\}\right)\right| \label{4.6} \\ 
&& \qquad\le{\textrm{ const.}\,} 
\kappa^{1/4}e^{-2|x|/c^{(0)}}  
\quad \textrm{if }|x|<\log \kappa^{-1},\quad j=0,1,2, \nonumber 
\end{eqnarray}
and
\begin{eqnarray}
&&\left|\frac{\partial^j}{\partial x^j} f_0(x,T)\right|\le 
C\exp\left\{-\frac{(A_0 c_0-T)}{4c^{(0)}}\left|2x
+\frac{x^2}{c^{(0)}M_0^2(T)}\right|\right\} \nonumber \\ 
&& \qquad \textrm{for }x\ge -c^{(0)} M_0(T),\quad j=0,1,2. \label{4.7}
\end{eqnarray}
A relatively small error is committed if $M_n$ is very large and the
arguments $\ell_{n,M_n}^\pm(x,u,v)$ (defined
in formula (\ref{2.25})) of the function $f_n$ in the operator 
$\bar{\mathbf Q}_{n,M}^{\mathbf c}f_n$ are replaced by $x\pm u$. 
Exploiting this fact one can prove, using a natural adaptation 
of the proof of Proposition~1 of paper~\cite{BM3}, the following

\medskip\noindent
{\bf Proposition 4.1.} {\it There exists a number $\kappa_0=\kappa_0(N)$
such that if
\begin{enumerate} 
\item [(i)] $0<\kappa<\kappa_0$, and 
\item [(ii)] $0<T\le c_0A_0/2$, 
\end{enumerate}
then the relations 
\begin{eqnarray*}
&& \left|\frac{\partial^j}{\partial x^j}
\left(f_n(x,T)-\frac{\sqrt{A_0 c_0-T}}{\sqrt\pi}
\frac{2^n}{c^{(n)}}
\exp\left\{-2^n(A_0 c_0-T)
\left(\frac x{c^{(n)}}\right)^2\right\}
\right)\right|  \\
&& \qquad\le B(n) \kappa^{1/4} e^{-2^{n+1}|x|/c^{(n)}},
\quad \textrm{if }\;|x|<2^{-n}\log \kappa^{-1}, \quad j=0,1,2, \\
&& \left|\frac{\partial^j}{\partial x^j} f_n(x,T)\right|
\le B(n)\exp\left\{-\frac{(A_0 c_0-T)}4\frac{2^n}{c^{(n)}}
\left|2x+\frac{x^2}{c^{(n)}M_n^2(T)}\right|\right\}  \\
&&\qquad\qquad\qquad \textrm{for }x\ge -c^{(n)} M_n(T), \quad j=0,1,2,
\end{eqnarray*}
and
\begin{equation}
|M_n(T)-\bar M_0(T)|\le B(n)\sqrt \kappa\, T \label{4.10}
\end{equation}
hold  for all $0\le n\le N$ with the function $\bar M_0(T)$ defined in
(\ref{4.1}) and a function $B(n)$ which depends 
neither on $T$ nor on $\kappa$.}

\medskip 
We formulate and prove, similarly to paper~\cite{BM3}, certain inductive
hypotheses about the behaviour of the functions $f_n(x,T)$ for $n\ge N$
if the pair $(n,T)$ is in the low temperature region. In the formulation
of these hypotheses we apply the sequence $\beta_n(T)$ defined in
(\ref{3.1}) and the sequence $\alpha_n(T)$ defined as
\begin{equation}
\begin{array}{rl}
\alpha_N(T)&=\frac1{200}\frac{ (c^{(N)})^2}{2^N}, \\
\alpha_{n+1}(T)&=\left(\frac{\bar c_{n+1}^2}2-\sqrt{\frac{\beta_n(T)}
{c^{(n)}}}\right)
\alpha_n(T) +\frac{10^{-12}}{M_n^2(T)}\quad \textrm{ for }n\ge N.
\end{array} \label{4.11}
\end{equation}
To formulate the inductive hypotheses we also introduce a
regularization of the functions $f_n(x,T)$.

\medskip\noindent
{\bf Definition of the regularization of the functions $f_n(x,T)$.}
{\it Let us fix a $C^\infty$-function
$\varphi(x)$, $-\infty<x<\infty$, such that $\varphi(x)=1$ for  $|x|\le
1$, \ $0\le\varphi(x)\le1$ if $1\le x\le 2$ and $\varphi(x)=0$ for
$|x|\ge2$. Then the regularization of the function $f_n(x,T)$ is
\[
\varphi_n(f_n(x,T))=A_n\varphi\left(\frac{x+B_n}
{100\sqrt{c^{(n)}}}\right)f_n(x+B_n,T),
\]
 with norming constants $A_n$
and $B_n$ such that 
\[
\int_{-\infty}^\infty \varphi_n(f_n(x,T))\,dx=1,\qquad 
\int_{-\infty}^{\infty} x\varphi_n (f_n(x,T))\,dx=0.
\]}


\medskip
Now we formulate the inductive hypotheses.

\medskip\noindent
{\bf Hypothesis $I(n)$.} {\it
\begin{eqnarray*}
\left|\frac{\partial^jf_n(x,T)}{\partial x^j}\right|
&\le&\frac C{\beta_n(T)^{(j+1)/2}}\exp\left\{-\frac1{\sqrt{\beta_n(T)}}
\left|2x+\frac{x^2}{c^{(n)}M_n(T)}\right|\right\}\\
&&\qquad\textrm{ for } j=0,1,2,\quad x\ge -c^{(n)}M_n(T),
\end{eqnarray*}
with a universal constant $C>0$. One could choose, e.g.,\ $C=10^{20}$.}

\medskip\noindent
 {\bf Hypothesis $J(n)$.} {\it
$$
|\tilde \varphi_nf_n(t+is,T)|\le
\frac{e^{\beta_n(T)s^2}}{1+\alpha_n(T)t^2}
\quad\textrm{ if }\quad |s|\le\frac2{\sqrt{\beta_{n+1}(T)}}\,,
$$
where 
\[
\tilde \varphi_nf_n(t+is,T)=\int e^{i(t+is)x} \varphi(f_n(x,T))\,dx.
\] }

\medskip\noindent
{\bf Corollary of Proposition 4.1.} {\it Under the conditions of
Proposition~4.1, the inductive hypotheses $I(n)$ and $J(n)$ hold for
$n=N$ with a universal constant $C>0$ in hypothesis $I(n)$. (For
instance, one can choose $C=10^5$.)}

\medskip 
Before formulating the main result of this Section, we introduce the
operators ${\mathbf T}_n$. They are appropriate scaling of 
the operators $\bar{\mathbf T}_{n,M_n(T)}^{\mathbf c}$ defined in 
formula~(\ref{2.27}), but
these operators will be applied only for the regularization of the
functions $f_n(x,T)$ and not for the functions $f_n(x,T)$ themselves.
Put 
\begin{equation} \label{4.12}
\begin{aligned}
& {\mathbf T}_{n}\varphi_n(f_n(x,T))=
\frac{2}{\bar c_{n+1}\pi^{\frac{r-1}{2}}}
\int_{u\in {\mathbb R}^1,\, \bv\in{\mathbb R}^{r-1}} e^{-\bv^2}  \\ 
&\times \varphi_n\left(f_n\left(\frac{x}{\bar c_{n+1}}
-\frac{r-1}{4M_n(T)}+u+\frac{\bv^2}{2M_n(T)},T\right)\right)   \\
&\times\varphi_n\left(f_n\left(\frac x{\bar c_{n+1}}
-\frac{r-1}{4M_n(T)}-u+\frac{\bv^2}{2M_n(T)},T\right)\right) \,du\,d\bv ,
\end{aligned}
\end{equation}
with the constants $\bar c_n$ defined in (\ref{2.24a}) and the number 
$V(S^{r-2})$ introduced in Proposition~1.4. The main result of this 
section is the following

\medskip\noindent
{\bf Proposition 4.2.} {\it There exists $\kappa_0=\kappa_0(N)>0$ such
that if
\begin{enumerate}
\item  [(i)] the inductive hypotheses $I(n)$ and $J(n)$ hold for
the function  $f_n(x,T)$, 
\item  [(ii)] $0<\kappa<\kappa_0$, ($\kappa$ appears in formula 
(\ref{1.4})), and
\item [(iii)] the pairs $(m,T)$ belong to the low
temperature region for all \hfill\break
$0\le m\le n$, 
\end{enumerate}
then the inductive hypotheses $I(n+1)$ and $J(n+1)$ hold for the 
function $f_{n+1}(x,T)$.
Also there exist universal constants $C_1$, $K_1$, $K_2$ and $K_3$ such that
the following estimates hold:
\begin{enumerate}
\item [(a)]
\begin{equation}\label{4.13} 
\begin{aligned}
&M_{n+1}(T)=M_n(T)-\frac{r-1}{4c^{(n)}M_n(T)}+\frac{\gamma_n(T)}{c^{(n)}},\\
& \textrm{where }\; |\gamma_n(T)|
\le C_1\frac{\beta_{n+1}(T)}{c^{(n+1)}} \sqrt{\beta_{n+1}(T)}\,, 
\end{aligned}
\end{equation}
\item[(b)]
\begin{equation}
1\le\frac{\beta_{n+1}(T)}{\alpha_{n+1}(T)}\le K_1, \label{4.14}
\end{equation}
\item[(c)] For $x>-c^{(n+1)}M_{n+1}(T)$ and $j=0,1,2$,
\begin{equation}\label{4.15} 
\begin{aligned}
&\left|\frac{\partial^j}{\partial x^j}\left[f_{n+1}(x,T)-{\mathbf T}_n
\varphi_n(f_n(x,T))\right]\right|
\le
\frac{K_2C^4}{\beta_{n+1}^{(j+1)/2}(T)} \frac{\beta_n(T)}{c^{(n)}} \\ 
&\qquad \times \Bigg[\exp\left\{-\frac1{\sqrt{\beta_{n+1}(T)}}
\left|2x+\frac{x^2} {c^{(n+1)}M_{n+1}(T)}\right| \right\} \\
&\qquad +\exp
\left\{-\frac{2|x|}{\sqrt{\beta_{n+1}(T)}}\right\}\Bigg]  . 
\end{aligned}
\end{equation}
\item[(d)] For  $x\in {\mathbb R}^1$  and $j=0,1,2,3,4$,
\begin{equation}  \label{4.16}
\left|\frac{\partial^j}{\partial x^j}{\mathbf T}_n
\varphi_n(f_n(x,T))\right|\le
\frac{K_3C^2}{\beta_{n+1}^{(j+1)/2}(T)}
\exp\left\{-\frac{2|x|}{\sqrt{\beta_{n+1}(T)}}\right\}.
\end{equation}
\end{enumerate}
}

\medskip
The proof of Proposition~4.2 is based on the observation that 
the operator ${\mathbf T}_n$ approximates the operator 
${\mathbf Q}_{n,M_n(T)}^{\mathbf c}$ very well, and  it has a 
relatively simple structure. Namely, 
it can be written by
writing the vectors $\bv\in{\mathbb R}^{r-1}$ in formula (\ref{4.12})
in spherical coordinates in the form
\begin{eqnarray*}
&&{\mathbf T}_n\varphi_{n}(f_n(x,T)) 
=\frac4{\bar c_{n+1}\Gamma(\frac{r-1}2)}\int_0^\infty  w^{r-2}e^{-w^2} \\
&&\qquad \qquad \varphi_{n}(f_n)*\varphi_{n}(f_n)\left(\frac{2x}{\bar c_{n+1}}
+\frac{w^2}{M_n(T)}-\frac{r-1}{2M_n(T)},T\right) dw,
\end{eqnarray*}
where $w=|\bv|$. Then we get with the substitution 
$\frac{w^2}{M_n(T)}=u$ that
\begin{eqnarray}
&&{\mathbf T}_n\varphi_{n}(f_n(x,T)) 
=\frac2{\bar c_{n+1}\Gamma(\frac{r-1}2)}
\int_0^\infty M_n(T) 
(M_n(T)u)^{(r-3)/2}e^{-M_n(T)u}  \nonumber \\
&& \qquad\qquad \varphi_{n}(f_n)*\varphi_{n}(f_n)\left(\frac{2x}{\bar c_{n+1}}
+u-\frac{r-1}{2M_n(T)},T\right) du  \nonumber \\
&& \qquad =\frac2{\bar c_{n+1}}
\varphi_n(f_n)*\varphi_n(f_n)*k_{M_n(T)}^-
\left(\frac{2x}{\bar c_{n+1}}-\frac{r-1}{2M_n(T)}\right), \label{4.16a}
\end{eqnarray}
where $*$ denotes convolution, $k^-_{M_n(T)}(x)=k_{M_n(T)}(-x)$, and 
$k_{M_n(T)}(x)=M_n(T)k\left(M_n(T)x\right)$ with 
$k(x)=\frac1{\sqrt{\pi x}}e^{-x}$ 
$k(x)=\frac{x^{(r-3)/2}e^{-x}}{\Gamma(\frac{r-1}2)}$ for $x>0$, 
and $k(x)=0$ for $x\le0$.

The operator ${\mathbf T}_n$ has a certain contraction property which can
be expressed in the Fourier space. 
the Fourier transform 
of $\tilde{\mathbf T}_{n} \tilde\varphi_n( f_n(\xi,T))$ can be 
expressed with the help of formula (\ref{4.16a}). One gets that
\begin{equation}\label{4.17} 
\begin{aligned}
&\tilde{\mathbf T}_{n} \tilde\varphi_n( f_n(\xi,T))\\
&=\exp\left\{ i\frac{(r-1)\bar c_{n+1}}{4M_n(T)}\xi\right\} 
\tilde k\left(-\frac{\bar c_{n+1}\xi}{2M_n(T)}\right) 
\left[\tilde\varphi_n\left(f_n
\left(\frac {\bar c_{n+1}}{2}\xi,T\right)\right)\right]^2,  \\
&= \frac{\exp\left\{i\frac{(r-1)\bar c_{n+1}}{4M_n(T)}\xi\right\}}
{\left(1+i\frac{\bar c_{n+1}\xi}{2M_n(T)}\right)^{(r-1)/2}}
\left[\tilde\varphi_n\left(f_n
\left(\frac {\bar c_{n+1}}{2}\xi,T\right)\right)\right]^2.
\end{aligned}
\end{equation}
In this calculation we
have exploited that $k(x)$ is the density function of the gamma 
distribution with parameter $\frac{r-1}2$, whose characteristic 
function equals $(1-i\xi)^{-(r-1)/2}$. It can be seen from 
formula~(\ref{4.17}) that ${\mathbf T}_n\varphi_{n}(f_n(x,T))$
is the density function of a random variable with expectation zero. 

The proof of Proposition~4.2 is a natural adaptation of the proof of 
the corresponding result (of Proposition~3) in paper~\cite{BM3}. Hence 
we only explain the main points and the necessary modifications.
 
Because of the inductive property $I(n)$ $f_n(x,T)$ is essentially 
concentrated in a neighbourhood of the origin of size 
$\sqrt{\beta_n(T)}$, and if $(n,T)$ is in the low temperature domain 
and $\eta>0$ is chosen sufficiently small, then 
$\frac{|x|}{100\sqrt{c^{(n)}}}\le\frac{\eta}{10}$ for 
$|x|\le \sqrt{\beta_n(T)}$, and the function $f_n(x,T)$ (disregarding
the scaling with the numbers $A_n$ and $B_n$) is not changing in the
typical region by the regularization of the function $f_n(x,T)$. 
This is the reason why such a regularization works well.

The proof of Proposition~4.2 contains several estimates. First we list
those results whose proof apply the bound on $f_n(x,T)$ formulated in
the Inductive hypothesis~$I(n)$. One can bound the differences
$$
\frac{\partial^j}{\partial x^j}
(\bar{\mathbf Q}_{n,M_n(T)}^{\mathbf c}f_n(x,T)-
\bar{\mathbf Q}_{n,M_n(T)}^{\mathbf c}\varphi_n(f_n(x,T)))
\quad \textrm{(Lemma~4 in \cite{BM3}),}
$$
$$
\frac{\partial^j}{\partial x^j}
(\bar{\mathbf Q}_{n,M_n(T)}^{\mathbf c}\varphi_n(f_n(x,T))-
\bar{\mathbf T}_{n,M_n(T)}^{\mathbf c}\varphi_n(f_n(x,T)))
\quad \textrm{(Lemma~5 in \cite{BM3}),}
$$
with the help of Property~$I(n)$ similarly to paper~\cite{BM3}. The absolute
value of these expressions can be bounded for all $\varepsilon>0$ by
$$
\frac{\beta_{n}}{c^{(n)}} \frac{C_1(\varepsilon)C^2}{\beta_n^{(j+1)/2}(T)}
\exp\left\{-\frac{2(1-\varepsilon)}{\bar c_{n+1}\sqrt{\beta_{n}(T)}}
\left|2x+\frac{x^2} {c^{(n+1)}M_{n}(T)}\right|\right\}
$$
with some appropriate constant $C_1(\varepsilon)>0$ if $f_n(x,T)$ satisfies
Condition $I(n)$.

The main difference between these estimates and the analogous results in
paper~\cite{BM3} is that the upper bounds given for the above expressions
contain a small multiplying factor $\frac{\beta_n(T)}{c^{(n)}}$.
In paper~\cite{BM3} the multiplying factors $2^{-n}$ and $1/c^{(n)}$ appear
instead of this term. In the proof of this paper we had to make some
modifications, because while in paper~\cite{BM3} only very low temperatures
were considered when $M_n(T)$ is strongly separated from zero, now we
want to give an upper bound under the weaker condition formulated in the
definition of the low temperature region. The proofs are very similar.
The only essential difference is that in the present case the typical
region, where a good asymptotic approximation must be given is chosen
as the interval $|x|<10\sqrt{c^{(n)}}$, i.e.\ it does not depend on the
value of $M_n(T)$.

Also the expression $\bar{\mathbf Q}_{n,M_n(T)}^{\mathbf c}f_n(x,T)$ can be
bounded together with their first two derivatives with the help of
Property~$I(n)$ in the same way as in Lemma~3 of paper~\cite{BM3}. But this
estimate is useful only for large $x$. It can be proved, similarly to
the proof of the corresponding result in paper~\cite{BM3} (lemma~7) that the
scaling constants which appear in the formulas expressing
${\mathbf Q}_{n,M_n(T)}^{\mathbf c}$ through $\bar{\mathbf Q}_{n,M_n(T)}^{\mathbf
c}$ and ${\mathbf T}_n$ through $\bar{\mathbf T}_{n,M_n(T)}^{\mathbf c}$ are
very close to each other. Here again the multiplying factor
$\frac{\beta_n(T)}{c^{(n)}}$ appears in the error term instead of the
multiplying factor $1/c^{(n)}$ in paper~\cite{BM3}. This Lemma~7 in~\cite{BM3}
is a technical result which expresses the difference of the functions
$\bar{\mathbf T}_{n,M_n(T)}^{\mathbf c}F_1(x)$ and  $\bar{\mathbf
T}_{n,M_n(T)}^{\mathbf c}F_2(x)$ together with its derivatives if we have
a control on the difference of the original functions $F_1(x)$ and
$F_2(x)$. We gain such kind of information from the inductive
hypothesis~$I(n)$. They give a good control on the difference
$f_{n+1}(x,T)-{\mathbf T}_n\varphi_n(f_n(x,t))$. The consequences 
of these results are formulated in Proposition~2 in paper~\cite{BM3}. 
These results also imply an estimate on the Fourier transforms
$\tilde\varphi_{n+1}(f_{n+1}(\xi,T))-\tilde{\mathbf
T}_n\tilde\varphi_n(f_n(\xi,T)))$ and $\tilde{\mathbf
T}_n\tilde\varphi_n(f_n(\xi,T))$ and also on their analytic
continuation. This is done in lemma~8 in paper~[\cite{BM3}. Now again
the analogous result holds under the conditions of the present paper
with the difference that the term $c^{-n}$ must be replaced
$\frac{\beta_n(T)}{c^{(n)}}$. The estimate obtained for
$\tilde{\mathbf T}_n\tilde\varphi_n(f_n(\xi,T))$ in such a way is
relatively weak, it is useful only for large $\xi$.

The above results are not sufficient to prove Proposition~4.2. In
particular, they do not explain why the right scaling was chosen in the
definition of the function $f_n(x,T)$. Their role is to bound the error
which is committed when ${\mathbf Q}_{n,M_n(T)}^{\mathbf c}f_n(x,T)$ is
replaced by ${\mathbf T}_n\varphi(f_n(x,T))$. The function ${\mathbf
T}_n\varphi_n(f_n(x,T))$ together with its derivatives and Fourier
transform can be well bounded by means of formula (\ref{4.17}) and the
inverse Fourier transform. In the estimations leading to such bounds
the inductive hypothesis $J(n)$ plays a crucial role. The proof of
Lemma~9 in paper~\cite{BM3} can be adapted to the present case without any
essential difficulty. But, the parameters $\alpha_n$, $\beta_n$
and~$c$ must be replaced by $\alpha_n(T)$, $\beta_n(T)$ and $\bar c_{n+1}$
in the present case.
 
Proposition 4.2 can be proved similarly to its analog, Proposition~3 in
paper~\cite{BM3}. The notation must be adapted to the notation of
the present paper. Besides, the small coefficient $c^{-n/2}$
appearing in the proof of Proposition~3 in~\cite{BM3} must be replaced by
$\sqrt{\frac{\beta_n(T)}{c^{(n)}}}$. There is one point where a
really new argument is needed in the proof. This argument requires a
more detailed discussion. It is the proof of relation (\ref{4.14}), i.e.\ of
the fact that $\alpha_n(T)$ and $\beta_n(T)$ have the same order of
magnitude. Their ratio must be bounded by a number independent of
$\eta$. The proof of the analogous result in paper~\cite{BM3} exploited the
fact that in the model of that paper the sequence~$c^{(n)}$ tended to
infinity exponentially fast. In the present case this property does not
hold any longer, hence a different argument is needed. The validity
of relation (\ref{4.14}) has a different cause for relatively small and large
indices~$n$.

For large $n$ it can be shown that both $\beta_n(T)$ and $\alpha_n(T)$
have the same order of magnitude as $M_n^{-2}(T)$, and for large $n$
these relations imply~(\ref{4.14}). If $n$ is relatively small and 
$M_0(T)$ is large, then $M_n^{-2}(T)$ is much less than $\alpha_n(T)$ and
$\beta_n(T)$. In this case the above indicated argument does not work,
but it can be proved that for such indices~$n$ the numbers $\beta_n(T)$
are decreasing exponentially fast, and the proof of 
relation~(\ref{4.14}) for such~$n$ is based on this fact.

To distinguish between  small and large indices $n$ define the number
\begin{eqnarray}
N_1(T)&=&\left\{\min n\colon  n\ge N,\textrm{ and } \beta_{n+1}(T)\le
\frac{100}{M_n^2(T)}\right\},  \nonumber \\
&&\qquad \qquad(N_1(T)=\infty \textrm{ if there is no such }n).
\label{4.18} 
\end{eqnarray} 
where the number $N$ was defined in formula~(\ref{1.22}). We shall 
later see that $N_1(T)<\infty$ for all $0<T\le c_0A_0/2$.

First we prove relation~(\ref{4.14}) under the additional condition 
$n\le N_1(T)$. In this case
$\beta_{m+1}(T)\le\frac{c_{m+1}^2}2\beta_m(T)+\frac{\beta_m(T)}{10}$
for $m\le n$, and because of Condition~1
\begin{equation}
\beta_{m+1}(T)\le\frac23\beta_m(T) \quad  \textrm{if }m\le N_1(T)
\label{4.19}
\end{equation}
for all $N\le m\le n$. 

Hence
$\sqrt{\frac{\beta_{m+1}(T)}{c^{(m+1)}}} \le\frac56
\sqrt{\frac{\beta_m(T)}{c^{(m)}}}$, \ 
$\sqrt{\frac{\beta_m(T)}{c^{(m)}}}\le \left(\frac56\right)^{m-N}\!\!\!
\sqrt{\frac{\beta_N(T)}{c^{(N)}}}$,
\begin{eqnarray*}
1\le\frac{\beta_{m+1}(T)}{\alpha_{m+1}(T)}&\le \max\left(\frac
{\frac{c_{m+1}^2}2+\sqrt{\frac{\beta_m(T)}{c^{(m)}}}}
{\frac{c_{m+1}^2}2-\sqrt{\frac{\beta_m(T)}{c^{(m)}}}} \cdot
\frac{\beta_m(T)}{\alpha_m(T)},\,10^{13}\right) \\
&\le\max \left(\exp\left\{5\sqrt{\frac{\beta_m(T)}{c^{(m)}}}\right\}\cdot
\frac{\beta_m(T)}{\alpha_m(T)},\,10^{13}\right)
\end{eqnarray*}
for $N\le m\le n$, and
\begin{eqnarray*}
\frac{\beta_{n+1}(T)}{\alpha_{n+1}(T)}\le\max\left(\frac{\beta_N(T)}
{\alpha_N(T)},\,10^{13}\right)
\exp\left\{5\sum_{m=N}^n\sqrt{\frac{\beta_m(T)}{c^m(T)}}\right\}\le K.
\end{eqnarray*}
The above argument together with the observation that $\beta_N(T)\gg
M_N^{-2}(T)$ if the parameter $t>0$ in~(\ref{1.4}) is sufficiently small and
$T\le c_0A_0/2$ imply that $N<N_1(T)$, and  the pair $(n,T)$ is in the
low temperature region for all $n\le N_1(T)$. The latter property
follows from the fact that by formula~(\ref{4.19}) the sequence
$\frac{\beta_n(T)}{c^{(n)}}$ is monotone decreasing for $N\le n\le N_1(T)$.

In the case $n>N_1(T)$ we can prove by induction with respect to $n$
together with the inductive proof of Proposition~4.2 that
\begin{eqnarray}
\beta_{n+1}(T)&\le&\frac{100}{M_n^2(T)}\quad
\textrm{if }n\ge N_1(T). \nonumber \\
&& \quad \textrm{and $(n,T)$ is in the low temperature region.}
\label{4.20} 
\end{eqnarray} 

By applying formula (\ref{4.20}) for $n-1$ and the fact that
$(n,T)$ is in the low temperature region we get that the term
$\gamma_{n-1}(T)$ in formula (\ref{4.13}) can be bounded as
\begin{equation}
|\gamma_{n-1}(T)|\le\frac{\beta_{n}(T)}{c^{(n)}}\sqrt{\beta_{n}(T)}\le\eta
\frac{10}{M_{n-1}(T)}\le\frac1{8C_1M_{n-1}(T)} \label{4.21}
\end{equation}
with the same number $C_1$ which appears in (\ref{4.13}) if the number
$\eta>0$ was chosen sufficiently small. Then formula (\ref{4.13}) 
implies that
$M_n(T)\le M_{n+1}(T)$. Hence we get by applying again formula~(\ref{4.20})
with $n-1$ that $M_n(T)<M_{n-1}(T)$, and
$$
\beta_{n+1}(T)\le \frac23 \beta_n(T)+\frac{10}{M_n^2(T)}\le
\frac{200}{3M_{n-1}^2(T)}+\frac{10}{M_n^2(T)}\le\frac{100}{M_n^2(T)}.
$$
This means that formula (\ref{4.20}) also holds for $n$. 
Relation (\ref{4.20}) together with the
definition of the sequence $\alpha_n(T)$ 
imply that for $n\ge N_1(T)$
$$
\alpha_{n+1}(T)\ge \frac{10^{-12}}{M_n^2(T)}\ge 10^{-14}\beta_n(T),
$$
i.e.\ formula (\ref{4.14}) is also valid for $n>N_1(T)$ if $(n,T)$ is in the
low temperature domain. With the help of this argument Proposition~4.2
can
be proved by an adaptation of the proof of the corresponding result
in~\cite{BM3}.
 
We formulate and prove a lemma which describes some properties of the
numbers $\beta_n(T)$ in the cases when $n\le N_1(T)$ or $n\ge N_1(T)$.
Several parts of it were already proved in the previous arguments.

\medskip\noindent
{\bf Lemma 4.3.} {\it Let $0<T\le  c_0A_0/2$. If the parameter
$\kappa>0$ in formula~(\ref{1.4}) is sufficiently small, then 
the following statements are valid:
\begin{enumerate}
\item
The
 number $N_1(T)$ defined in (\ref{4.18}) is finite, and $N_1(T)>N$. 
\item
The pair
$(N_1(T),T)$ is in the low temperature region. 
\item
The relations (\ref{4.19}),
(\ref{4.20}) hold.
\item If $n\ge N_1(T)$  and $(n,T)$ is in the low
temperature region then  
\begin{equation}\label{4.22}
   M_{n}(T)-\frac3{8c^{(n)}M_{n}(T)}\le M_{n+1}(T)\le
M_{n}(T)-\frac1{8c^{(n)}M_{n}(T)}.
\end{equation} 
\item If  $N\le n\le N_1(T)$ then
\begin{equation} \label{4.23} 
\begin{aligned}
&M_{n}(T)-\frac1{4c^{(n)}M_{n}(T)}-\eta\left(\frac23\right)^{(n-N)/2}
\le M_{n+1}(T) \\
& \le M_{n}(T)-\frac1{4c^{(n)}M_{n}(T)}+\eta\left(\frac23\right)^{(n-N)/2}.
\end{aligned}  
\end{equation}
\item We have that
\begin{equation}
N_1(T)-N\le 10 \log (1/\kappa T^2).  \label{4.24}
\end{equation} 
\item
If $M_n(T)<10$ then $n\ge N_1(T)$.
\end{enumerate}
}

\medskip\noindent
{\bf Proof of Lemma 4.3.}\/ Formulas (\ref{4.19}) and (\ref{4.20}) were 
already proved in the previous argument, and since $(N,T)$ is in 
the low temperature region, i.e. $\beta_N(T)\ge\eta c^N$, relation 
(\ref{4.19}) implies that $(n,T)$ is in the low temperature region 
for all $N\le n\le N_1(T)$. Formula~(\ref{4.22}) follows from 
formula~(\ref{4.21}) with the replacement of
$n-1$ by $n$ and formula~(\ref{4.13}). By relation (\ref{4.19})
$\beta_n(T)\le\left(\frac23\right)^{n-N}$ if $N\le n\le N_1(T)$. Hence it
follows from~(\ref{4.13}) that
\begin{equation}
M_{n+1}(T)\le M_n(T)+\frac{\beta_{n+1}(T)}
{c^{(n)}}\frac{\sqrt{\beta_{n+1}(T)}}{c^{(n)}}
\le M_n(T)+\eta\left(\frac23\right)^{(n-N)/2}, \label{4.25}
\end{equation}
and even relation (\ref{4.23}) holds in this case.
 
Relation~(\ref{4.25}) and the estimate obtained for $\beta_n(T)$
imply that $M_{n}^2(T)\le (M_N(T)+1)^2 \le 2M_N^2(T)$
and $\beta_{n+1}(T)M^2_n(T)\le 2M_N^2(T)\left(\frac{2}{3}\right)^{n-N}$
if $n\le N_1(T)$. This relation together with the definition of the
index $N_1(T)$ defined in (\ref{4.18}) imply that
$2M_N^2(T)\left(\frac{2}{3}\right)^{n-N}\ge 100$ if $n<N_1(T)$. 
Applying the last formula for $n=N_1(T)-1$ we get that
$(N_1(T)-1-N))\log\frac{3}{2}\le\log\frac {M_N^2(T)}{50}$. Since
$M_N^2(T)\sim {\textrm{ const.}\,}\frac1{\kappa T^2}$ this relation 
implies that $N_1(T)$ is finite, and moreover it satisfies~(\ref{4.24}). 
Finally, if the inequalities
$M_n(T)\le 10$ and $n<N_1(T)$ held simultaneously, then the inequality
$M^2_n(T)\beta_{n+1}(T)\le 100\left(\frac23\right)^{n-N}\le100$ would also
hold. This relation contradicts to the assumption $n<N_1(T)$. Hence
also the last statement of Lemma~4.3 holds. \qed

\medskip 
The previous results enable us to describe the different behaviour of
the model in the cases when the Dyson condition (\ref{1.3}) is 
satisfied and when it is not.
This will be done in Lemma~4.4. It shows that if (\ref{1.3})
{\it is not satisfied} then for all $T$ there is a pair $(n,T)$
which does not belong to the low temperature region, while if
(\ref{1.3}) {\it is satisfied}, then all sufficiently low temperatures $T$
belong to the low temperature region. In the latter case the asymptotic
behaviour of the spontaneous magnetization $M_n(T)$ can be described for
large~$n$. The description of the behaviour of the function $q_n(x,T)$
in the case when $T$ does not belong to the low temperature region needs
further investigation, and this will be done in Sections~5 and~6. A
more detailed investigation of the case when $T$ belongs to the low
temperature region will be done in Section~7. We finish this section
with the proof of a result about the behaviour of the magnetization
$M_n(T)$ at low temperatures~$T>0$ which will be useful in the
subsequent part of the paper.

\medskip\noindent
{\bf Lemma 4.4.} {\it Let $0<T\le c_0A_0/2$,
and let the parameter $\kappa>0$  be 
sufficiently small. If the Dyson condition (\ref{1.3})
 is not satisfied, then for all $T>0$ there is some $n=n(T)$
for which $(n,T)$ does not belong to the low temperature region. If,
on the other hand,
condition (\ref{1.3}) is satisfied, then $T$ belongs to the low 
temperature region for sufficiently small~$T>0$. In this case 
relation (\ref{3.8}) and, under the additional Condition~2, also  
relation (\ref{3.9}) 
hold.} 

\medskip\noindent
{\bf Proof of Lemma 4.4.}\/
It follows from formulas (\ref{4.22}) and (\ref{4.23}) that
\begin{equation}
-\frac1{c^{(n)}}\le M_{n+1}^2(T)-M_n^2(T)\le -\frac1{8c^{(n)}} \label{4.26}
\end{equation}
if $n\ge N_1(T)$ and the pair $(n,T)$ is in the low temperature region,
and
\begin{eqnarray} \label{$4.26'$} 
\begin{aligned}
&-\frac1{2 c^{(n)}}-10\left(\frac23\right)^{n-N}(M_N(T)+1)\le
M_{n+1}^2(T)-M_n^2(T) \\
&\le -\frac1{2 c^{(n)}} +10\left(\frac23\right)^{n-N}(M_N(T)+1)
\end{aligned}
\end{eqnarray}
if $N\le n\le N_1(T)$. Formula (\ref{4.26}) can be obtained by taking 
square in formula (\ref{4.22}) and observing that 
$c^{(n)}M_n(T)^2>10\eta^{-1}$. Formula (\ref{$4.26'$}) can be deduced 
similarly from (\ref{4.23}) by observing first that the right-hand 
side of (\ref{4.23}) implies that $M_n(T)\le M_N(T)+1$ for
$N\le n\le N_1(T)$.

Formulas (\ref{4.26})  and (\ref{$4.26'$}) imply that if a 
temperature $T>0$ is in the low temperature region, then
$$
\sum_{k=N}^n\frac 1{c^{(n)}}\le 8(M_N^2(T)-M_n^2(T))+30(M_N(T)+1)\le
8M_N^2(T)+30(M_N(T)+1)
$$
for all~$n\ge N$, where the number $N$ is defined in (\ref{1.22}). 
Since the right-hand side of the last formula does not depend on $n$, 
this implies that (\ref{1.3}) holds.

In the other direction, if (\ref{1.3}) holds, then since by 
Proposition~4.1
$$
\lim_{T\to\infty}M_0(T)=\lim_{T\to\infty}M_N(T)=\infty,
$$
there is some number $\bar T\le c_0A_0/2$ such that for all
temperatures $0<T\le \bar T$ \ $M_N^2(T)>8\sum\limits_{n=N}^\infty
\frac1{c^{(n)}}+30M_n(T)+31$. If $T>0$ satisfies the above
inequality, then the left-hand side of the inequalities (\ref{4.26}) and
(\ref{$4.26'$}) imply that if the pair $(n,T)$ is in the low temperature
domain and $n\ge N_1(T)$, then
$$
M_n^2(T)>M_N^2(T)-8\sum\limits_{n=N}^n\frac1{c^{(n)}}30(M_n(T)+1))\ge1.
$$
Hence $M_n^2(T)>1$ for all $n$, and $T$ is in the low temperature
region.

Let $T>0$ be in the low temperature region. If $n>m>N_1(T)$, then by
(\ref{4.26})
$$
\left| M_n^2(T)-M_m^2(T)\right|\le \sum_{k=m}^n\frac1{c^{(k)}}.
$$
Since in this case Condition~1 holds, the last relation implies that
$M_n^2(T)$, $n=1,2,\dots$, is a Cauchy sequence, and relation (\ref{3.8})
holds. We claim that if Condition~2 also holds, then for any 
$\varepsilon>0$
\begin{equation}
-\frac{r-1+\varepsilon}{2 c^{(n)}}\le M_{n+1}^2(T)-M_n^2(T)\le 
-\frac{r-1-\varepsilon}{2c^{(n)}}
\label{4.27}
\end{equation}
if $n\ge n(\varepsilon)$. Relation (\ref{3.9}) is a consequence of 
(\ref{4.27}).  Relation (\ref{4.27}) can be deduced from (\ref{4.13}) 
and (\ref{4.20}) if we show 
that for any temperature $T>0$ in the low temperature region
\begin{equation}
\lim\limits_{n\to\infty}\frac{\beta_n(T)}{c^{(n)}}=0. \label{4.28}
\end{equation}
Relation (\ref{4.28}) holds under Condition~2, since by 
(\ref{4.26}) in this case for all $n>N_1(T)$,
$$
M_n^2(T)\ge\lim_{k\to\infty}\left(M_n^2(T)-M_k^2(T)\right)
\ge\frac18\sum_{k=n}^\infty\frac1{c^{(k)}},
$$
and 
\[
\frac{\beta_n(T)}{c^{(n)}}\le\frac{100}{M_{n-1}^2(T)c^{(n)}}\le800
\left(c^{(n)}\sum\limits_{k=n-1}^\infty\frac1{c^{(k)}}\right)^{-1}.
\]
 Under
Condition~2 the last expression tends to zero as $n\to\infty$. This
implies formula~(\ref{4.27}). Lemma~4.4 is proved. \qed

\section{Estimates in the Intermediate Region. The proof of Theorem~3.1.}

In this section we give some estimates on $q_n(\bx,T)$ when the pair
$(n,T)$ belongs neither to the low nor to the high temperature region
and prove Theorem~3.1 with their help.

Let us consider the number $\bar n=\bar n(T)$ introduced in the
formulation of Theorem~3.1, namely
$$
\bar n(T)=\min\{n\colon\; D^2_n(T)<e^{-1/\eta^2}\}.
$$
In Lemmas~5.1 and~5.2 we shall prove some estimates about a scaled 
version of the function $q_{\bar n(T)}(\bx,T)$, where $q_n(\bx,T)$ was 
defined in~(\ref{2.10}). In Lemma~5.1 the case $T\le  c_0A_0$, and 
in Lemma~5.2 the case $T\ge c_0A_0$ will be considered. Lemmas 5.1 
and 5.2 yield some estimates on the tail-behaviour of a scaled 
version of the function $q_{\bar n(T)}(\bx,T)$. This will be needed to 
start an inductive procedure for all  $n\ge \bar n(T)$ which state 
that the functions $q_n(\bx,T)$ become more and more strongly concentrated 
around zero as the index~$n$ is increasing. This procedure is based 
on Lemmas~5.3 and~5.4. The role of Lemma~5.3 is to give an appropriate 
lower bound for the norming constant $Z_n(T)$ in the definition of
the function $q_n(\bx,T)$. Then in Lemma~5.4 we prove some contraction 
property of the operator which maps an appropriate scaled
version of the distribution function  with density function 
${\textrm{ const.}\,}\bar q_{n-1}(|\bx|,T)$ to an appropriate scaled 
version of the distribution function with density 
${\textrm{ const.}\,}\bar q_n(|\bx|,T)$, $\bx\in{\mathbb R}^r$. The 
proof of Lemma~5.4 will exploit the rotation symmetry of the model. 
Theorem~3.1 will be proved by means of these lemmas.

To formulate these results we introduce some notations.
Let us introduce the functions
\begin{equation}
\hat h_n(\bx,T)= \left(c^{(\bar n(T))}\right)^{-r/2}
q_n\left(\frac{\bx}{\sqrt{c^{\bar
n(T)}}},T\right), \quad \bx\in {\mathbb R}^r, \label{5.1}
\end{equation}
and measures
\begin{equation}
\hat H_{n,T}({\mathbf A})=\int_{\mathbf A}\hat h_n(\bx,T)\,d\bx,\quad 
{\mathbf A}\subset{\mathbb R}^r, \label{5.2}
\end{equation}
in the space ${\mathbb R}^r$. Define also the function
\begin{equation}
\hat H_{n,T}(R)=\hat H_{n,T}(\{\bx\colon |\bx|\ge R\}) \quad\textrm{for
}R\ge0. \label{5.3}
\end{equation}
The functions $\hat h_{n,T}$ and measures
$\hat H_{n,T}$ are similar to the functions $h_{n,T}$ and measures
$H_{n,T}$ defined in (\ref{3.5}) and (\ref{3.6}). The only difference 
is that the scaling of $q_n(\bx,T)$ in (\ref{5.2}) and (\ref{5.3}) is made 
by means of $c^{(\bar n(T))}$ instead of $c^{(n)}$. If Condition~5 is 
satisfied with a sufficiently small $\bar\eta$ and sufficiently large 
$L(\bar\eta,T)$,
and $n-\bar n(T)$ is not too large, then 
the approximation of $c^{(n)}$ by $c^{(\bar n(T))}$ is sufficiently
good for our purposes. Hence it will be enough to have a good control
on the measure $\hat H_{n,T}$. In Lemma~5.3 we give a bound on it for
large $|\bx|$ and in Lemma~5.4 we prove an estimate which enables to
bound $\hat H_{n,T}(R)$ for small $R$ too.

With the help of these results we can prove that starting from
$\bar n=\bar n(T)$ after finitely many steps $k$ the pair $(\bar
n+k,T)$ is in the high temperature region. Moreover, this number $k$
can be bounded from above independently of the temperature~$T$. First
we formulate Lemma~5.1.

\medskip\noindent
{\bf Lemma 5.1.} {\it Under the conditions of Proposition~4.2,
the function \hfill\break 
$h_{\bar n(T)}(\bx,T)$ defined in~(\ref{3.5}) satisfies the
inequality
\begin{equation}
h_{\bar n(T)}(\bx,T)\le \exp\left\{\frac
K{\eta}-\frac{|\bx|^2}{10}\right\} \quad \textrm{if }
\quad  T\le  c_0A_0/2 \label{5.4}
\end{equation}
with an appropriate $K>0$. For $ T\le  c_0A_0/2$ the pair $(\bar
n(T),T)$ does not belong to the high temperature region, and there
exists some $\tilde \eta=\tilde \eta(\eta)$ such that the function
$\hat H_{n,T}(\cdot)$ defined in (\ref{5.3}) satisfies the inequality
\begin{equation}
\hat H_{\bar n(T),T}\left(\tilde\eta^{-1}\right)\le 1/2,\quad
\textrm{if }\quad T\le  c_0A_0/2, \label{5.5}
\end{equation}
i.e. for $T\le  c_0A_0/2$ there is a ball with its center in the
origin whose radius depends only on $\eta$, and whose $\hat H_{\bar
n(T),T}$ measure is greater than $1/2$.}

\medskip\noindent
{\bf Proof of Lemma 5.1.}\/  Let us introduce the function
$$
\bar h_n(x,T)=\frac1{\sqrt{c^{(n)}}}\bar q_n\left(\frac
x{\sqrt{c^{(n)}}},T\right),\quad x\ge0  
$$
with the function $\bar q_n$ introduced in (\ref{2.14}).
Observe that
$$
\int_0^\infty\bar h_n(x,T)\,dx=1.
$$
Let us apply 
Proposition~4.2 with the choice $n=\bar n(T)-1$.
Since hypothesis $I(n)$ holds for $n=\bar n$, we obtain that
\begin{eqnarray*}
f_{\bar n(T)}(x,T)&\le& \frac K{\beta_{\bar
n(T)-1}^{1/2}(T)}\exp\left\{-\frac1{\sqrt{\beta_{\bar n(T)}(T)}}
\left|2x+\frac{x^2} {c^{(\bar n(T))}M_{\bar n(T)}(T)}\right|\right\} \\
&&\qquad\qquad \textrm{if } x>-c^{(\bar n(T))}M_{\bar n(T)}(T)
\end{eqnarray*}
with some universal constant $K>0$. It follows from this relation that
the function $\bar h_{\bar n(T)}(x,T)=\sqrt{c^{(\bar n(T))}}f_{\bar
n(T)}\left(\sqrt{c^{(\bar n(T))}}x- c^{(\bar n(T))}M_n(T)),T\right)$ satisfies
the inequality
\begin{eqnarray*}
\bar h_{\bar n(T)}(x,T)&\le& K\left(
\frac {c^{(\bar n(T))}}{\beta_{\bar n(T)-1}(T)}\right)^{1/2} \\ 
&& \qquad \exp\left\{\frac1{\sqrt{\beta_{\bar n(T)}(T)}}
\left(c^{(\bar n(T))}M_{\bar n(T)}(T)-\frac{x^2}
{M_{\bar n(T)}(T)}\right)\right\}.
\end{eqnarray*}
The inequalities $\beta_{\bar n(T)}(T)>\eta c^{(\bar n(T))}$ and
$\beta_{\bar n(T)-1}(T)\le\eta c^{(\bar n(T)-1)}$ hold. 

Lemma~4.3 implies that the fractions $
\frac{\beta_{\bar n(T)}(T)}{\beta_{\bar n(T)-1}(T)}$, 
$\frac{M_{\bar n(T)}(T)}{M_{\bar n(T)-1}(T)}$ and \hfill \break
$\beta_{\bar n(T)}(T)M_{\bar n(T)}(T)^2$ are separated both from
zero and infinity, hence 
$$
\frac {c^{(\bar n(T))}}{\beta_{\bar
n(T)-1}(T)}\le \frac{{\textrm{ const.}\,}}\eta,
$$ 
$$
\frac{c^{(\bar n(T))}M_{\bar n(T)}(T)}
{\sqrt{\beta_{\bar n(T)}(T)}}\le \frac{{\textrm{ const.}\,}}\eta
$$
and
$$
\frac1 {M_{\bar n(T)}(T) \sqrt{\beta_{\bar n(T)}(T)}}\ge \frac1{20}.
$$
These inequalities together with the last relation imply that
\begin{equation}
\bar h_{\bar n(T)}(x,T)\le e^{\bar K/\eta}e^{-x^2/20}\,\quad x\ge 0, 
\label{5.6}
\end{equation}
with an appropriate $\bar K>0$. Since
\begin{equation}
h_{\bar n(T)}(\bx,T)= C(T)\bar h_{\bar n(T)}(|\bx|,T),\quad \bx\in {\mathbb R}^r, 
\label{5.7}
\end{equation}
with an appropriate number $C(T)>0$, estimate (\ref{5.4}) can be 
deduced from (\ref{5.6}) if we give a good upper bound for the 
constant $C(T)$ in~(\ref{5.7}). Observe that because of (\ref{5.7})
\begin{equation}
C(T)^{-1}=\int_{{\mathbb R}^r} \bar h_{\bar n(T)}(|\bx|,T)\,d\bx=\textrm{Vol}(S^{r-1}) 
\int_0^\infty x^{r-1}
\bar h_{\bar n(T)}(x,T)\,dx, \label{5.7a}
\end{equation}
hence
\begin{eqnarray*}
C(T)^{-1}&=&\textrm{Vol}(S^{r-1}) \int_0^\infty x^{r-1}
\bar h_{\bar n(T)}(x,T)\,dx\\
&\ge& \textrm{Vol}(S^{r-1}) R^{r-1}\left(1-\int_0^R \bar h_{\bar
n(T)}(x,T)\,dx\right)
\end{eqnarray*}
for any $R>0$. On the other hand, by formula (\ref{5.6})
$$
\int_0^R \bar h_{\bar n(T)}(x,T)\,dx\le \frac12
$$
if $0<T\le c_0A_0/2$ and 
$R\le e^{-K/\eta}$ with a sufficiently large $K>0$. Hence
$C(T)^{-1}\ge\frac12\textrm{Vol}(S^{r-1})e^{-K(r-1)}$. This means that 
$C(T)\le e^{Kr/\eta}$ 
in (\ref{5.7}), and inequality (\ref{5.4}) follows from (\ref{5.6}). 

We shall prove that $\bar n(T)$ does not belong to the high 
temperature region with the help of the following estimate on
$D^2_{\bar n(T)}(T)$. In its proof we shall apply formula~(\ref{5.7a}).
\begin{eqnarray*}
D^2_{\bar n(T)}(T)&=&\int_{{\mathbb R}^r}|\bx|^2 h_{\bar n(T)}(\bx,T)\,dx=
\textrm{Vol}(S^{r-1})
\int_0^\infty x^{r+1}  h_{\bar n(T)}(x,T)\,dx\\
&=& C(T) \textrm{Vol}(S^{r-1})
\int_0^\infty  x^{r+1} \bar h_{\bar n(T)}(x,T)\,dx\\
&\ge& C(T) \textrm{Vol}(S^{r-1})
\left(\int_0^\infty  x^{r-1} \bar h_{\bar n(T)}(x,T)\,dx\right)^{(r+1/(r-1)}\\
&=& \left(\int_0^\infty  x^{r-1} \bar h_{\bar n(T)}(x,T)\,dx\right)^{2/(r-1)}\\
&\ge& \left(\int_0^\infty  x \bar h_{\bar n(T)}(x,T)\,dx\right)^2
=\left( M_{\bar n(T)}\sqrt{ c^{(\bar n(T))}}\right)^2.
\end{eqnarray*}

Since $M^2_n\ge\frac{10}{\beta_{n+1}}$ if $n\ge N+1$, 
$N+1$ is in the low temperature region if $T\le c_0A_0/2$, (see 
(\ref{3.4}) and the subsequent sentence in our discussion), and 
$\frac{M_{\bar n(T)}}{M_{\bar n(T)-1}}\le\textrm{const.}$, hence 
$$
D^2_{\bar n(T)}(T)\ge M_{\bar n(T)}^2 c^{(\bar n(T))}
\ge\textrm{const.} \frac{c^{(\bar n(T))}}{\beta_{\bar n(T)}}
\ge \frac{\textrm{const.}}\eta.
$$
This implies that $\bar n(t)$ is not in the high temperature region.  \qed

\medskip
In the next Lemma 5.2 we shall formulate some properties of the 
function $h_n(\bx,T)$ defined in (\ref{3.5}) in the case $n=0$.
For the sake of a better discussion we define the function 
$\bar h_n(x,T)$, $x\ge 0$, by the formula $\bar h_n(x,T)=h_n(|\bx|,T)$,
and from now on $\bar h_n(x,T)$ means this function.  
(It differs slightly from the function $\bar h_n(x,T)$ applied in the
proof of Lemma~5.1 where a different norming constant was applied.)

If $T\ge c_0A_0/2$, then $\bar n(T)=0$, and 
$$
\bar h_{\bar n(T)}(|\bx|,T)=
\textrm{const.}\,q_0\left( (c^{(0})^{-1/2}\bx,T\right),
$$
where $q_0(\bx,T)$ is defined in (\ref{2.13}). 
Hence
\begin{equation}
\bar h_{\bar n(T)}(x,T)=\frac1{Z_0(T)}
\exp\left\{\left(\frac{A_0c_0-T}
{c^{(0)}}\right)\frac {x^2}2-\kappa T^2\frac{x^4}{4 (c^{(0)})^2} \right\}
\quad \textrm{if } T\ge c_0A_0/2 \label{5.8}
\end{equation}
with the norming constant (for the function $h_{\bar n(T)}(x,T)$)
\begin{equation}
Z_0(T)=\textrm{Vol}(S^{r-1}) \int_0^\infty x^{r-1} \exp
\left\{\left(\frac{A_0 c_0- T}{c^{(0)}}\right)
\frac {x^2}2-\kappa T^2\frac{x^4}{4 (c^{(0)})^2} \right\}\,dx.
\label{$5.8'$}
\end{equation}
Using formulas (\ref{5.8}) and (\ref{$5.8'$}), we will prove the
following

\medskip\noindent
{\bf Lemma 5.2.} {\it There is a constant $\kappa_0=\kappa_0(N)>0$ such 
that if $0<\kappa<\kappa_0$, Condition~1 is satisfied, and $T\ge c_0A_0/2$, 
then $\bar n(T)=0$, and
\begin{eqnarray}
 \bar h_{\bar n(T)}(x,T)&\le& \textrm{const.}\, T^{r/2}
\exp\left\{\frac1\kappa-\frac T{2c^{(0)}}x^2\right\} \;\;
\textrm{if } T\ge  \frac{c_0A_0}2.   \label{5.9} \\
\bar h_{\bar n(T)}(x,T)&\le& \textrm{const.}\, T^{r/2} 
\exp\left\{\frac {Tx^2}{2c^{(0)}}\right\} \;\;
\textrm{if } T\ge  \frac{c_0A_0}2.   \label{$5.9'$} \\
\bar h_{\bar n(T)}(x,T)&\le&\textrm{const.}\, e^{-Tx^2/4}
\;\; \textrm{if } T\ge 10A_0, \textrm{ and }x\ge T^{-1/3}. \label{$5.9''$}
\end{eqnarray}
The $\textrm{const.}$ in formulas (\ref{5.9})---(\ref{$5.9''$})
depend only on the dimension~$r$ of the model.

The pair $(\bar n(T),T)$ belongs to the high temperature region if $T$
is very large, e.g.\ if $T\ge e^{-1/\eta^9}$, and it does not belong to
it if $T>0$ is relatively small, e.g.\ if $T\le \eta^{-100}$. If $(\bar
n(T),T)$ does not belong to the high temperature region, then the
function $h_{\bar n(T)}(\bx,T)$ defined in formula~(\ref{3.5}) satisfies the
inequality
\begin{equation}
h_{\bar n(T)}(\bx,T)\le \exp\{K(\eta,\kappa)-\alpha|\bx|^2\} \label{5.10}
\end{equation}
with a constant $\alpha=\alpha(\eta)>0$ and an appropriate number
$K(\eta,\kappa)$ depending only on $\kappa$ and $\eta$. In this 
case there is a
constant $B=B(\eta,\kappa)>0$ in such a way that the quantity $\hat
H_{\bar n(T),T}(\cdot)$ defined in (\ref{5.3}) satisfies the inequality
\begin{equation}
\hat H_{\bar n(T),T}(B)\le \frac12. \label{5.11}
\end{equation}
This means that if the pair $(\bar n(T),T)$ is not in the high
temperature region (and $T\ge  c_0A_0/2$), then there is a radius
$B=B(\eta,\kappa)$ such that the $\hat H_{\bar n(T),T}$ measure of the ball
$\{x\colon |x|\le B(\eta,\kappa)\}$ is bigger than $1/2$.

If $(\bar n(T),T)=(0,T)$ is in the high temperature region, then
\begin{equation}
\hat H_{\bar n(T),T}(x)\le K_1e^{-K_2\eta^2 x^2}\quad\textrm{for all } x>0
\label{5.12}
\end{equation}
with some universal constants $K_1>0$ and $K_2>0$.}

\medskip\noindent
{\bf  Proof of Lemma 5.2.}\/ First we estimate the norming factor 
$Z_0(T)$ from below. Let us observe that 
$$
\frac{A_0c_0-T}{c^{(0)}}\frac {x^2}2-\kappa T^2\frac{x^4}{4 (c^{(0)})^2}
\ge -Tx^2\left(\frac1{2c^{(0)}}+\frac1{4(c^{(0)})^2}\right)\ge-10 Tx^2, 
$$
if $\kappa Tx^2\le1$ and Condition~1 holds. Hence
\begin{eqnarray}
Z_0(T)&\ge& \textrm{Vol}\,(S^{r-1})
\int_0^{1/\sqrt{\kappa T}} x^{r-1}e^{-10 Tx^2}\,dx \label{5.13} \\
&=& \textrm{Vol}\,(S^{r-1})\int_0^{1/\sqrt{\kappa}} 
\frac{x^{r-1}e^{-10x^2}}{ T^{r/2}}\,dx \ge\textrm{const.} T^{-r/2}. \nonumber
\end{eqnarray}
Now, if $T\ge c_0A_0/2$, then 
\begin{eqnarray*}
\frac{A_0c_0-T}{c^{(0)}}\frac {x^2}2-\kappa T^2\frac{x^4}{4 (c^{(0)})^2}
&\le& -\frac12 \frac{Tx^2}{c^{(0)}}
+\max_{x\colon\;x\ge0}\left( \frac{Tx^2}{c^{(0)}}
-\frac \kappa4\left(\frac{Tx^2}{c^{(0)}}\right)^2\right) \\
&=& -\frac T{2c^{(0)}}x^2+\frac1\kappa,
\end{eqnarray*}
and combining this with (\ref{5.13}), we obtain (\ref{5.9}). The
estimate 
$\frac{A_0c_0-T}{c^{(0)}}\frac {x^2}2-\kappa T^2\frac{x^4}{4 (c^{(0)})^2}
\le  \frac{Tx^2}{2c^{(0)}}$ yields (\ref{$5.9'$}).

If $T\ge 10A_0$, then 
$$
\frac{A_0c_0-T}{c^{(0)}}\frac {x^2}2-\kappa T^2\frac{x^4}{4 (c^{(0)})^2}
\le -\frac T3x^2\le-\frac T4x^2-\frac{T^{1/3}}{12},
\quad  \textrm{for $|x|\ge T^{-1/3}$,}
$$
which together with (\ref{5.13}) imply inequality~(\ref{$5.9''$}).

Furthermore, (\ref{$5.9''$}) implies that if $T>e^{-1/\eta^9}$, then 
the pair $(0,T)$ belongs to the high temperature region. Indeed,
if we estimate the integral expressing $D_0^2(T)$, then by this relation 
the contribution of the domain $\{\bx\colon\; |\bx|\ge T^{-1/3}\}$ to this 
integral is very small. On the other hand, the contribution of the 
domain $\{\bx\colon\; |\bx|\le T^{-1/3}\}$ is less than $T^{-2/3}$ which is 
also very small in this case. To see that for $T<\eta^{-100}$ the pair 
$(0,T)$ does not belong to the high temperature domain it is enough 
to observe that in this case by formula (\ref{$5.9'$}) the $H_{0,T}$ 
measure of the ball $\{\bx\colon |\bx|\le \eta^{100}\}$ is less than 
$\textrm{const.}\,T^{r/2}\eta^{100r}\le 1/2$. Hence in this case
$D_0^2(T)\ge\frac12 \eta^{200}$. (We get this estimate by restricting
the integral expressing $D_0^2(T)$ to the domain
$\{\bx\colon |\bx|\ge \eta^{100}\}$). This means that $T$ is not in the high 
temperature region. 

Inequality (\ref{5.9}) together with the fact that if the
pair $(0,T)$ does not belong to the high temperature region then $T\le
e^{-1/\eta^{9}}$ imply relations (\ref{5.10}) and (\ref{5.11}).

Since $T>\eta^{-100}$ if the pair $(0,T)$ is in the high temperature
region, relation (\ref{$5.9''$}) implies relation (\ref{5.12}). 
Lemma~5.2 is proved. \qed

\medskip
To formulate Lemmas 5.3 and 5.4 we rewrite formula (\ref{2.12}) for the
functions $\hat h_n(x,T)$ defined in (\ref{5.1}). It has the form
\begin{equation}
\hat h_{n+1}(\bx,T)=\frac1{Z_n(T)}\int_{{\mathbb R}^r}
\exp\left\{-\frac{c^{(n)}}{c^{(\bar n(T))}} \bu^2\right\}
\hat h_n(\bx-\bu,T)\hat h_n(\bx+\bu,T)\,d\bu \label{5.14}
\end{equation}
with
\begin{equation}
{Z_n(T)}=\int_{{\mathbb R}^r\times{\mathbb R}^r}
\exp\left\{-\frac{c^{(n)}}{c^{(\bar n(T))}} \bu^2\right\}
\hat h_n(\bx-\bu,T)\hat h_n(\bx+\bu,T)\,d\bu\,d\bx \label{$5.14'$}
\end{equation}
for all $n\ge \bar n(T)$.

Let us also introduce the moment generating function of
the measures $\hat H_{n,T}$, defined in (\ref{5.2}):
$$
\varphi_{n,T}(\bu)=\int_{{\mathbb R}^r} e^{\bu\bx}\hat h_{n,T}(\bx)\,d\bx, \quad
\bu\in {\mathbb R}^r,
$$
where $\bu\bx$ denotes scalar product. By studying the properties 
of the moment generating function $\varphi_{n,T}(\bu)$, we get an
upper bound for the function $\hat H_{n,T}(R)$ for large values $R$. 
Namely, we have the following result:

\medskip\noindent
{\bf Lemma 5.3.} {\it There exists some $\kappa_0=\kappa_0(N)$ 
such that for all $0<\kappa<\kappa_0$ and $0<\bar\eta<\eta$ 
(with the numbers $N$ and $\eta$ defined in (\ref{1.22})) the
following relations hold. If we have such a positive integer $L$ 
for which Conditions~1 and~5 (with $\bar\eta$ and this number $L$) 
are satisfied, then for all such temperatures $T>0$ for which the 
number $\bar n(T)$ exists, and the pair$(\bar n(T),T)$ does not
belong to the high temperature region, the inequality
\begin{equation}
\hat H_{\bar n(T)+l,T}(R)\le e^{-2^l\alpha R^2/5r}
\quad\textrm{if }R\ge D\textrm{ and }0\le l\le L \label{5.15}
\end{equation}
holds with appropriate constants $\alpha>0$ and $D>0$, and also
the norming factor $Z_n(T)$ in (\ref{$5.14'$}) can be estimated as
\begin{equation}
Z_{\bar n(T)+l}(T)\ge D_1\quad \textrm{for } 0\le l\le L \label{5.16}
\end{equation}
with some constant $D_1>0$.  These constants can be chosen as some 
functions of $\kappa$ and $\bar\eta$, i.e. $\alpha=\alpha(\kappa,\bar\eta)$,  
$D=D(\kappa,\bar\eta)>0$ and $D_1=D_1(\kappa,\bar\eta)>0$. This means
in particular that they do not depend on the temperature $T$.}

\medskip\noindent 
{\bf Proof of Lemma 5.3.}\/ It follows from formulas (\ref{5.4}) 
and (\ref{5.10}) that
$$
\varphi_{\bar n(T),T}(\bu)\le \exp\left\{ K_0+\frac
{\bu^2}\alpha\right\}\quad\textrm{for all }\bu\in{\mathbb R}^r
$$
with some $K_0=K_0(\eta,\kappa)>100$ and 
$\alpha=\alpha(\eta)>0$. It can be seen by induction with respect to
$l$ that
\begin{equation}
\varphi_{\bar n(T)+l,T}(\bu)\le \exp\left\{2^l K_l+\frac
{\bu^2}{2^l\alpha}\right\}\quad\textrm{for all } 0\le l \le L \textrm{ and }
\bu\in{\mathbb R}^r \label{5.17}
\end{equation}
with some $K_0>0$ and
\begin{equation}
K_l=K_{l-1}-\frac{\log Z_{\bar n(T)+l-1,T}}{2^l}, \quad 1\le l\le L. \label{$5.17'$}
\end{equation}
Indeed, the function $\hat h_{\bar n(T)+l+1,T}(\bx,T)$ is increased if the
kernel term \hfill\break
$\exp\left\{-\frac{c^{(n)}}{c^{(\bar n(T))}} \bu^2\right\}$ is omitted from the
integral in (\ref{5.14}), and the integral turns into the convolution
$2\hat h_{\bar n(T)+l,T}*\hat h_{\bar n(T)+l,T}(2\bx)$ after this change.
By computing this convolution with the help of the inductive hypothesis
and dividing it by $Z_{\bar n(T)+l+1}$ we get an upper bound
for $\varphi_{\bar n(T)+l+1,T}(\bu)$. Formulas (\ref{5.17}) and (\ref{$5.17'$})
follow from these calculations. We will prove formulas 
(\ref{5.15}) and (\ref{5.16})
from these relations by induction for $l$ together with the inductive
hypothesis that
\begin{equation}
K_l\le B\quad\textrm{for all }0\le l\le L \label{5.18}
\end{equation}
with some constants $B>10$ depending only on $\kappa$ and $\bar \eta$.

Observe that formula (\ref{5.17}) with the choice of vectors
of the form $(u,0)$, $u\in {\mathbb R}^1$, $u>0$, $0\in{\mathbb R}^{r-1}$. 
implies that the function $\hat H_{\bar n(T)+l,T}(R)$ defined in formulas 
(\ref{5.2}) and (\ref{5.3}) satisfies the
inequality
\begin{eqnarray*}
\hat H_{\bar n(T)+l,T}(R)&\le& r\hat H_{\bar n(T)+l,T} 
\left(\{\bx=(x_1,\bx_2)\in {\mathbb R}^r,\; x_1>\frac R{\sqrt r}\right)\\
&\le& r \exp\left\{-\frac{uR}{\sqrt r}+2^lK_l+\frac{u^2}{2^l\alpha}\right\}
\end{eqnarray*}
for all real numbers~$u$. In particular, 
\begin{equation}
\hat H_{\bar n(T)+l,T}(R)
\le r\exp\left\{2^l\left(K_l-\frac {R^2\alpha}{4r}\right)\right\}
\label{5.19}
\end{equation}
with the choice $u=\frac{2^lR\alpha}{2\sqrt r}$. Hence 
\begin{equation}
\hat H_{\bar n(T)+l,T}\left(\sqrt{\frac{ 4r(r+1)B}\alpha}\right) \le re ^{-2^lr B}
\le\frac12 \label{5.20}
\end{equation}
with the number $B>0$ appearing in (\ref{5.18}). Formula (\ref{5.20}) 
implies that
$$
\hat H_{\bar n(T)+l,T}\left(\left\{\bx\colon  \bx\in{\mathbb R}^r,\,|\bx|\le
\sqrt{\frac{ 4r(r+1)B}\alpha}\right\}\right)\ge \frac 12.
$$
For $\bz\in {\mathbb R}^r$ and $u>0$ let 
$K(\bz,u)=\{\bx\colon  \bx\in {\mathbb R}^r,
|\bx-\bz|\le u\}$ denote the ball with center $\bz$ and radius~$u$.
Since the ball 
$\left\{\bx\colon\ \bx\in{\mathbb R}^r,\,
|x|\le\sqrt{\frac{ 4r(r+1)B}\alpha}\right\}$ 
can be
covered by $C(r)B(\alpha \bar\eta)^{-1}$ balls of radius
$\sqrt{\bar\eta}$, where $C(r)>0$ depends only on $r$, there is a
ball $K\left(\bz,\sqrt{\bar\eta}\right)$ of radius $\sqrt{\bar\eta}$ 
whose $\hat H_{n,T}$ measure (this measure was defined in (\ref{5.2})) 
is greater than $\frac{\alpha\bar\eta}{C(r) B}$. Hence
$$
\hat H_{\bar n(T)+l,T}\times\hat H_{\bar n(T)+l,T}
\left(K(\bz,\sqrt{\bar\eta})\times
K(\bz,\sqrt{\bar\eta})\right)\ge\frac{\alpha^2\bar\eta^2}{(C(r) B)^2},
$$
and because of Condition 5 the expression $Z_n(T)$ defined 
in (\ref{$5.14'$}) can be estimated for $n=\bar n(T)+l$ as follows:
\[
\begin{aligned}
Z_{\bar n(T)+l}(T)&\ge 2^{-r}\int_{\bx\in K\left(\bz,\sqrt{\bar\eta}\right),\,
\bu\in K\left(\bz,\sqrt{\bar\eta}\right)}
\exp\left\{-
\frac {c^{(\bar n(T)+l)}}{ c^{\bar n(T)}}
\frac{(\bx-\bu)^2}4\right\} \\
&\times \hat h_{\bar n(T)+l,T}(\bx) 
\hat h_{\bar n(T)+l,T}(\bu)\,d\bx\,d\bu  \\ 
&\ge e^{-5}\hat H_{\bar n(T)+l,T}\times\hat H_{\bar n(T)+l,T}
\left(K\left(\bz,\sqrt{\bar\eta}\right)\times 
K\left(\bz,\sqrt{\bar\eta}\right)\right) \\ 
&\ge  e^{-5}\frac{\alpha^2\bar\eta^2}{(C(r) B)^2}.
\end{aligned}
\]
In the above estimation we have exploited that because of Condition~5
and Condition~1 
$\frac {c^{(\bar n(T)+l)}}{ c^{\bar n(T)}}\le\frac5{\bar\eta}$, hence
$\frac {c^{(\bar n(T)+l)}}{ c^{\bar n(T)}}\frac{(\bx-\bu)^2}4\le 5$ if
$\bx\in K\left(\bz,\sqrt{\bar\eta}\right)$ and
$\bu\in K\left(\bz,\sqrt{\bar\eta}\right)$.

The last relation implies (\ref{5.16}) with
$D_1=\frac{e^{-5}\alpha^2\bar\eta^2}{(C(r) B)^2}$, although we still
have to show that the number $D_1$ (depending on $B$) can be bounded
by a number which does not depend on the parameter~$T$. We show with
the help of  (\ref{5.16}), (\ref{$5.17'$}) and the inductive hypothesis 
(\ref{5.18}) that $K_l\le (1-2^{-(l+1)})B$ if the number $B$ is chosen as 
$B=\max(2K_0, K^*)$, where $K^*$ is the larger solution of the equation 
$x=2\log \frac{x^2}{\bar D_1}$ with 
$\bar D_1=\frac {e^{-5}\alpha^2\bar\eta^2}{C(r)^2}$. This means that $B$ in 
(\ref{5.18}) can be chosen as a number not depending on $T$. 

Indeed, this relation holds for $l=0$, and  if it holds for $l-1$, then 
$$
K_l\le(1-2^{-(l+1)})B-2^{-(l+1}\left(B-2\log \frac {B^2}{\bar D_1}\right)
\le(1-2^{-(l+1)})B
$$ 
if $B\ge K^*$.

This implies (\ref{5.16}) (with the constant $D_1$ not depending on~$T$
and the validity of the inductive hypothesis (\ref{5.18}) for 
$0\le l\le L$. Finally, relation (\ref{5.15}) follows from (\ref{5.18}) 
and (\ref{5.19}). Lemma~5.3 is proved. \qed

\medskip
Formulas (\ref{5.14}) and (\ref{$5.14'$}) can be rewritten for 
the function $\hat H_{n,T}(R)$ defined in (\ref{5.3}) as
\begin{equation} \label{5.21}
\begin{aligned}
\hat H_{n+1,T}(R)&=\frac{2^r}{Z_n(T)}\int_{|\bx|\ge R}
\int_{\bu\in{\mathbb R}^r} \exp\left\{-\frac
{c^{(n)}}{c^{(\bar n(T))}}\bu^2 \right\}  \\
& \times \hat h_n(\bx-\bu,T)\hat h_n(\bx+\bu,T)\,d\bu\,d\bx    \\
&=\frac1{Z_n(T)} \int_{\left|\frac{\bx+\bu}2\right|\ge R}
\int_{\bu\in{\mathbb R}^r} \exp\left\{-\frac { c^{(n)}}{c^{(\bar n(T))}} 
\frac{(\bx-\bu)^2}4\right\}  \\
&\times \hat H_{n,T}(\,d\bx) \hat H_{n,T}(\,d\bu)
\end{aligned}
\end{equation}
with
$$
Z_{n}(T)=\int_{\bx\in{\mathbb R}^r}\int_{\bu\in{\mathbb R}^r}
\exp\left\{-\frac
{c^{(n)}}{ c^{(\bar n(T))}}\frac{(\bx-\bu)^2}4\right\}
\hat H_{n,T}(\,d\bx) \hat H_{n,T}(\,d\bu).
$$
for all $R\ge0$. We apply these formulas in the proof of the following
Lemma~5.4. The proof of Lemma~5.4 also exploits the rotational
invariance of the measure $\hat H_{n,T}$.

\medskip\noindent
{\bf Lemma 5.4.} {\it Let the conditions of Lemma~5.3 hold. Then
there exist some numbers $\delta=\delta\left(\bar\eta,D_1\right)>0$
and $M=M\left(\bar\eta,D_1\right)>0$ depending only on the numbers $D_1$ 
in formula (\ref{5.16}) and $\bar\eta$ in Condition~5 in such a way that
\begin{eqnarray}
\hat H_{\bar n(T)+l+1,T}((1-\delta) R)&\le&\frac12\hat
H_{\bar n(T)+l,T}((1-\delta)R)+M\hat H_{\bar n(T)+l,T}(R)) \nonumber \\
&&\qquad \textrm{for all } R>0 \textrm { and } 0\le l\le L.
\label{5.22}
\end{eqnarray}
} 

\medskip\noindent
{\bf Proof of Lemma 5.4.}\/ Observe that
\begin{eqnarray*}
&&\left\{\left|\frac{\bx+\bu}2\right|\ge (1-\delta)R\right\}
\subset\{|\bx|\ge R\} \cup\{|\bu|\ge R\} \\
&&\quad \cup\{|\bx|\ge (1-\delta)R,\;\textrm{arg} 
(\bx,\bu)\le \alpha\} 
\cup\{|\bu|\le (1-\delta)R,\;\textrm{arg} (\bx,\bu)\ge \alpha\}
\end{eqnarray*}
for all $R>0$ and $0<\delta<1$ with $\alpha=2\arccos(1-\delta)$. Indeed,
if $\left|\frac{\bx+\bu}2\right|\ge (1-\delta)R$, then either $|\bx|>R$ or
$|\bu|>R$ or both $|\bx|$ and $|\bu|$ is less than $R$, but in this case
either $|\bx|>(1-\delta)R$ or $|\bu|>(1-\delta)R$, and the angle between
the vectors $\bx$ and $\bu$ must be small. On the other hand, because of
the rotational invariance of the measure $\hat H_{n,T}$
\begin{eqnarray*}
&&\hat H_{\bar n(T)+l,T}\times \hat H_{\bar n(T)+l,T}
\left(\{(\bx,\by)\colon |\bx|\ge(1-\delta)R,   
\textrm{ arg}(\bx,\bu)\le\alpha\}\right)\\
&&\qquad\le \frac\alpha{\pi}
\hat H_{\bar n(T)+l,T}  (\{\bx\colon |\bx|\ge(1-\delta)R\}) 
=\frac\alpha{\pi}\hat H_{\bar n(T)+l,T}((1-\delta)R).
\end{eqnarray*}
The last two relations together with (\ref{5.21}) and the inequality
$\frac\alpha\pi\le\sqrt \delta$ imply that
\begin{eqnarray}
&&\hat H_{\bar n(T)+l+1,T}((1-\delta)R) \label{5.23} \\
&&\qquad \le \frac1{Z_n(T)}\left(2\sqrt \delta
\hat H_{\bar n(T)+l,T}((1-\delta)R)+2\hat H_{\bar n(T)+l,T}(R)\right). \nonumber
\end{eqnarray}
Relation (\ref{5.22}) follows from (\ref{5.23}) and (\ref{5.16}) if we 
choose $\delta>0$ so small that the inequality 
$\frac{2\sqrt \delta} {D_1}\le \frac12$ holds. Lemma~5.4 is proved.\qed

\medskip
Next we prove Theorem 3.1 with the help of the previous results.

\medskip\noindent
{\bf Proof of Theorem 3.1.}\/ First we give a good estimate on
$H_{\bar n(T)+l}(R)$ if the conditions of Lemma~5.3 hold with a sufficiently
large $L=L(\kappa,\bar\eta)$ and $l\le L$ is sufficiently large.
For this goal we introduce the following quantities.

Put $P(j,l)=P(j,l,T)=\hat H_{\bar n(T)+l}((1-\delta)^j D)$,
$j=0,1,\dots$, $0\le l\le L$ with the number $D$ appearing in (\ref{5.15})
and $\delta$ in Lemma 5.4. Clearly, $P(j,l)\le 1$ for all $j$ and $l$.
By Lemma~5.4
\begin{equation}
P(j,l+1)\le\frac12 P(j,l)+MP(j-1,l),\quad j\ge1, \label{5.24}
\end{equation}
and by relation (\ref{5.15}) $P(0,l)\le e^{-\alpha 2^l  D^2/5r}$ if 
$l\le L$. Hence there is a constant $k_0>0$ in such a way that 
$P(0,k_0+l) \le \left(\frac23\right)^{l}$ if $k_0+l\le L$. 
Because of this relation, the inequality $P(j,l)\le1$ and formula
(\ref{5.24}) there is a constant $k_1\ge k_0$ in such a way that
$P(1,k_1+l) \le \frac1{3M}\left(\frac23\right)^{l}$ and
$P(1, k_1+l)\le \left(\frac23\right)^{l}$ if $k_1+l \le L$. 
Similarly, there is a constant
$k_2$ such that $P(2,k_2+l) \le \frac1{3M}\left(\frac23\right)^{l}$,
and $P(2, k_2+l)\le \left(\frac23\right)^{l}$ if $k_2+l \le L$. This
procedure can be continued, and we get a sequence $k_0\le k_1\le
k_2\le\cdots$ in such a way that the inequality $P(p,k_p+l)\le
 \left(\frac23\right)^{l}$ holds if
$k_p+l \le L$. The numbers $k_p$ depend only on the parameter $\kappa$ in
(\ref{1.4}) and the number $\bar\eta$ in 
Condition~5. The above procedure can
be continued till $k_p\le L$.  
In such a way we have proved that for all fixed $j\ge0$
$$
\hat H_{\bar n(T)+l}\left((1-\delta)^pD\right)\le C(p)\left(\frac23\right)^l,
$$
if $0\le l \le L$.
The above relation together with formula
(\ref{5.15}) imply that if 
Condition~5 holds with a sufficiently large
constant $L=L(\bar\eta,t)$, then an integer $k>0$ can be chosen
independently of the parameter $T$ in such a way that 
\begin{equation}
\hat H_{\bar n(T)+l,T}(R)\le
2\exp\left\{-\frac{e^{1/\eta^3}R^2}{\bar\eta/5}\right\}
\quad\textrm{for all }R>0\textrm{ and }k\le l\le L(\bar\eta,t). 
\label{5.25}
\end{equation}
Indeed, by (\ref{5.15}) relation (\ref{5.25}) holds for $R\ge D$
if $l\ge k'_0$ with some $k_0'>0$ and by the previous inequality for
all $j=1,2,\dots$ it also holds for 
$(1-\delta)^j D\le R< (1-j)^{j-1}D$ if $l\ge k'j$ with a 
sufficiently large $k'j\ge$. On the other hand, it is enough
to demand this inequality for finitely many indices $j$, since
relation~(\ref{5.25}) automatically holds if
$\frac{e^{1/\eta^3}R^2}{\bar\eta/5}\le\log2$.

Since the measure $H_{n,T}$ defined in (\ref{3.6}) satisfies the relation
$$
H_{\bar n(T)+l,T}\{\bx\colon  |\bx|>R\}=
\hat H_{\bar n(T)+l,T}\left(\sqrt{\frac{c^{(\bar n(T))}} {c^{(\bar n(T)+l)}}}R\right)
\le\hat H_{\bar n(T)+l,T}\left(\sqrt{\frac{\bar\eta}5} R\right)
$$
relation (\ref{5.25}) implies that 
\begin{equation}
H_{\bar n(T)+l,T}(R)\le
2\exp\left\{-e^{1/\eta^3}R^2\right\}\quad\textrm{for all }R>0,
\textrm{ and }l^*\le l \le L \label{5.26}
\end{equation}
with some appropriate $l^*\ge0$. Relation (\ref{5.26}) implies in particular 
that $D^2_{\bar n(T)+l}(T)<e^{-1/\eta^2}$, i.e.\ $(\bar n(T))+l,T)$ is in
the high temperature region if $l^*\le l\le L$. The inequality
$D^2_{\bar n(T)}(T)<K$ follows from (\ref{5.15}) with $l=0$.

\medskip
To complete the proof of Theorem 3.1 we have to give a lower bound 
for $D^2_{\bar n(T)+k}(T)$. Let us introduce the following notation: 
Given two positive numbers $R_2>R_1>0$ let ${\mathbf K}(R_1,R_2)
=\{\bx\colon\;\bx\in {\mathbb R}^r,\, R_1\le |x|\le R_2\}$ denote the
annulus between the concentrical balls with center in the origin and
radii $R_1$ and $R_2$. We claim that for any $0\le l\le L$ there exist
some positive numbers $R_1(l)=R_1(l,\bar\eta,\kappa)$, 
$R_2(l)=R_2(l,\bar\eta,\kappa)$ and  $A(l)=A(l,\bar\eta,\kappa)>0$ 
such that the measure of the annulus determined by these numbers 
satisfies the inequality
\begin{equation}
\hat H_{\bar n(T)+l,T}({\mathbf K}(R_1(l),R_2(l))\ge A(l),\quad 0\le
l\le L \label{5.27}
\end{equation}
if the pair $(\bar n(T),T)$ does not belong to the high temperature 
region. Observe that the relation between the measures 
$\hat H_{\bar n(T)+l,T}$ and $H_{\bar n(T)+l,T}$ implies that relation 
(\ref{5.27}) also holds for $H_{\bar n(T)+l,T}({\mathbf K}(R_1(k),R_2(k))$
(i.e. the function $\hat H(\cdot)$ can be replaced by $H(\cdot)$ in
formula (\ref{5.27})) if the radii $R_2(k)$ and $R_1(k)>0$ are 
multiplied with an appropriate number. This implies that the variance 
$D^2_{\bar n(T)+k,T}$ can be bounded from below by a positive number 
which depends only on $k$ and~$\bar\eta$. Hence we can choose
e.g. $k=L=L(\kappa,\bar\eta)$ as the number $k$ satisfying the properties
demanded in Theorem~3.1.

We shall prove a slightly stronger statement than relation (\ref{5.27}) 
which will be useful in later applications. We shall prove that
\begin{equation}
\hat H_{\bar n(T)+l,T}\left({\mathbf K}\left(\frac1{2^l}R_1,\left(\frac{\sqrt
3}2\right)^lR_2\right)\right)\ge A(l),\quad 0\le l\le L. \label{$5.27'$}
\end{equation}
with some numbers $R_2>R_1>0$ and $A(l)>0$ if the pair $(0,T)$ does not
belong to the high temperature region. The numbers $R_j$ can be chosen
in such a way that $R_j=R_j(\eta,\kappa)$, $j=1,2$.

Let us first observe that relation (\ref{$5.27'$}) holds for $l=0$ 
if $\bar n(T)$ is not in the high temperature region. This follows from
relations (\ref{5.4}) and (\ref{5.5}) in the case $T\le  c_0
A_0/2$ and from (\ref{5.9}) and (\ref{5.11}) if $T\ge  c_0 A_0/2$, but 
$(\bar n(T),T)$ does not belong to the high temperature region. Indeed,
formulas (\ref{5.5}) and (\ref{5.11}) make possible to choose 
the number $R_2$ in such a way that the $H_{\bar n(T),T}$ measure of 
the ball with center in the origin and radius $R_2=R_2(\eta)$ is 
greater than $1/2$. By formulas (\ref{5.4}) and (\ref{5.9}) we can 
choose the number $R_1=R_1(\eta)$ in such a way that by cutting out 
from this ball the ball with radius $R_2$ and center in the origin 
the remaining annulus ${\mathbf K}(R_1,R_2)$ has a measure greater 
than $1/4$. In the case $T\ge c_0A/2$ we also exploited in the above
argument that $T$ cannot be very large if $(\bar n(T),T)$ is not in 
the high temperature region. By Lemma~5.2 $T\le e^{-1/\eta^9}$ in this 
case. 

We claim that
\begin{equation}
\hat H_{\bar n(T)+l+1,T}\left({\mathbf K}\left(\frac{\bar R_1}2,\frac{\sqrt 3}2
\bar R_2\right)\right)\ge
B(\bar R_1,\bar R_2,\bar\eta)\hat H_{\bar n(T)+l,T}({\mathbf
K}(\bar R_1,\bar R_2))^2 \label{5.28}
\end{equation}
for all $0\le l\le L$ and $\bar R_2>\bar R_1>0$ and an appropriate
constant 
$$
B(\bar R_1,\bar R_2,\bar\eta)>0.
$$ 
Relation (\ref{$5.27'$}) 
follows from (\ref{5.28}) and the fact that it holds for $l=0$.

In the proof of relation (\ref{5.28}) we exploit the relation
\begin{eqnarray*}
&& \!\!\!\!\!\!\!\!
\left\{(\bu,\bx)\colon  \bu\in {\mathbb R}^r, \bx\in {\mathbb R}^r, 
\frac{\bar R_1}2\le
\left|\frac{\bx+\bu}2\right|\le\frac{\sqrt 3}2\bar R_2,\; \frac \pi3\le
\textrm{arg}\,(\bx,\bu)\le \frac \pi2\right\}\\
&&\; \supset \left\{(\bu,\bx)\colon  \bu
\in {\mathbb R}^r, \bx\in {\mathbb R}^r,\;
\bar R_1\le |\bx|,|\bu|\le \bar R_2,\; \frac \pi3\le
\textrm{arg}\,(\bx,\bu)\le \frac \pi2\right\}.
\end{eqnarray*}
It follows from relation (\ref{$5.14'$}) that $Z_{\bar n(T)+l+1}(T)\le1$,
since we get an upper bound for it by omitting the kernel term
$\exp\left\{-\frac{c^{(n)}}{c^{(\bar n(T))}} \bu^2\right\}$ from the integral
in (\ref{$5.14'$}). Hence the previous relation together 
with (\ref{5.21}) and the
rotational invariance of the measure $\hat H_{\bar n(T)+l,T}$ yield that
\begin{eqnarray*} 
&& \!\!\!\!\!\!\!\!\!\!\!\!
\hat H_{\bar n(T)+l+1,T}\left({\mathbf K}\left(\frac{\bar R_1}2,\frac{\sqrt 3}2
\bar R_2\right)\right) =\frac1{Z_{\bar n(T)+l+1}(T)} \int\int
_{\frac{\sqrt 3}2\bar R_2\ge\left|\frac{\bx+\bu}2\right|
\ge \frac{\bar R_1}2,\,\bx,\bu\in{\mathbb R}^r} \\
&&\qquad\qquad\exp\left\{-\frac {c^{(\bar n(T)+l)}}{\bar
c^{(\bar n(T))}}\frac{(\bx-\bu)^2}4\right\}
\hat H_{\bar n(T)+l,T}(\,d\bx)\hat H_{\bar n(T)+l,T}(\,d\bu)\\
&&\quad\ge e^{-\bar R_2^2/\bar \eta} \int\int
_{\substack{\frac{\sqrt3}2\bar R_2
\ge\left|\frac{\bx+\bu}2\right|\ge\frac{\bar R_1}2,\;\bx,\bu\in{\mathbb R}^2,\\
\frac{\pi}3\le\arg(\bx,\bu)\le\frac \pi 2}}
\hat H_{\bar n(T)+l,T}(\,dx) \hat H_{\bar n(T)+l,T}(\,d\bu)\\
&&\quad\ge e^{-\bar R_2^2/\bar \eta}
\int\int_{\bar R_2\ge|\bx|,|\bu|\ge
\bar R_1,\frac{\pi}3\le\arg(\bx,\bu)\le\frac\pi2}
\hat H_{\bar n(T)+l,T}(\,d\bx) \hat H_{\bar n(T)+l,T}(\,d\bu)\\
&&\quad=C(r) e^{-\bar R_2^2/\bar \eta}
\hat H_{\bar n(T)+l,T}({\mathbf K}(\bar R_1,\bar R_2))^2
\end{eqnarray*}
with an appropriate constant $C(r)>0$.
The last estimate implies relation (\ref{5.28}) with
$B(\bar R_1,\bar R_2,\bar\eta)=C(r)
e^{-\bar R_2^2/\bar \eta}$. Theorem 3.1 is proved. \qed

\section{Estimates in the High Temperature Region. The proof of
Theorem 3.3.}

To study the behaviour of the function $f_n(x,T)$ in the high
temperature region we need a starting index $n={\tilde n(T)}$ for 
which a good estimate is known about the tail behaviour of the 
measure $H_{{\tilde n(T)},T}$. We also need a lower bound for the 
variance $D_n^2(T)$ defined in (\ref{3.5}) for $n\ge{\tilde n(T)}$. 
This requirement will be also taken into consideration in
the definition of ${\tilde n(T)}$. Let us first define the number
\begin{equation}
l_0=l_0(T)=\min \left\{l\colon  \left(\frac{\sqrt 3}2\right)^l R_2\le
\frac{\bar\eta}{10}\right\} \label{6.1}
\end{equation}
if the pair $(0,T)$ is not in the high temperature region, where
$\bar\eta$ appeared in 
Condition~5, and the number $R_2$ was introduced
in formula (\ref{$5.27'$}). Now define
\begin{equation}
{\tilde n(T)}=
\left\{ \begin{array}{l}
0\quad\textrm{if $(0,T)$ is in the high temperature region}\\
\bar n(T)+l \textrm{ with the smallest $l$ satisfying both 
(\ref{5.26}) and the }\\
\qquad\qquad\,\textrm{inequality $l\ge l_0$ with $l_0$ defined in
(\ref{6.1})}\\
\qquad\qquad\,\textrm{if $(0,T)$ is not in the high temperature
region.} 
\end{array} 
\right.  
\label{6.2}
\end{equation}

It follows from the results of the previous section that for a
temperature~$T$ which is not in the low temperature region the
inequality $0\le{\tilde n(T)}-\bar n(T)\le L(\bar\eta,t)$ 
holds if the number $L$ in Condition~5 is chosen sufficiently 
large. The measure $H_{{\tilde n(T)},T}$ introduced in formula 
(\ref{3.6}) is strongly concentrated around the origin. Indeed, 
formulas (\ref{5.12}) and (\ref{5.26}) give a good estimate 
for the $H_{{\tilde n(T)},T}$ measure of the sets 
$\{\bx\colon\;|\bx|\le R\}$ for all $R\ge 0$.
 
Let us introduce the moments of the functions $h_{{\tilde n(T)}+l}(\bx,T)$ 
defined in (\ref{3.5}).
$$
M_k(l,T)=\int_{{\mathbb R}^r}|\bx|^k h_{{\tilde n(T)}+l}(\bx,T)\,d\bx
\quad l\ge0,\;k\ge1.
$$
We shall estimate the moments $M_2(l,T)$ and $M_4(l,T)$. It follows from
relations (\ref{5.12}) and (\ref{5.26}) that
\begin{equation}
M_2(0,T)\le\eta^*\quad\textrm{and}\quad M_4(0,T)\le\eta^*\quad\textrm{with }
\eta^*=e^{-1/\eta^2} \label{6.3}
\end{equation}
for all $T>0$ which is not in the low temperature region. To get lower
bounds for the second moments $M_2(l,T)$ let us introduce the truncated
second moments
$$
M_{2,\textrm{tr.}}(l,T)=M_{2,\textrm{tr.}}\left(\frac1{10},l,T\right)
=\int_{|\bx|\le \frac1{10}}|\bx|^2
h_{{\tilde n(T)}+l}(\bx,T)\,d\bx.
$$
It follows from (\ref{$5.9''$}) if $(0,T)$ is in the high 
temperature region and from (\ref{$5.27'$}) and the definition of 
${\tilde n(T)}$ if $(0,T)$ is not in the high temperature region that
\begin{eqnarray}
M_{2,\textrm{tr.}}(0,T)>0,  &&\quad\textrm{for all }
T\ge c_0A_0/2 \nonumber \\
M_{2,\textrm{tr.}}(0,T)>\tilde\eta, &&\quad\textrm{if }
T\ge c_0A_0/2  \textrm{ and } (0,T) \textrm{ is not in the}
\nonumber \\ 
&& \quad  \textrm{high temperature region}
\label{6.5}
\end{eqnarray}
with some $\tilde\eta=\tilde\eta(\bar\eta,\kappa)>0$. First we shall bound
$M_2(l,T)$ and $M_4(l,T)$ from above in Lemma~(\ref{6.1}) for all $l\ge0$.
Then the second moment $M_2(l,T)$ will be bounded from below in
Lemma~6.2. These estimates enable us to prove the central limit
theorem for $g_{{\tilde n(T)}+l}(x,T)$ by means of the characteristic function
technique.
 
Simple calculation yields that
\begin{eqnarray}
M_k(l+1,T)&=&\frac{2^r}{Z_l(T)}\int e^{-\bu^2} |\bx|^k h_{{\tilde n(T)}+l}
\left(\frac {\bx}{\sqrt{\bar c_{{\tilde n(T)}+l+1}}}-\bu,T\right) \nonumber \\
&&\qquad\qquad h_{{\tilde n(T)}+l}
\left(\frac {\bx}{\sqrt{\bar c_{{\tilde n(T)}+l+1}}}+\bu,T\right)\,d\bx\,d\bu
\label{6.6}
\end{eqnarray}
for all $l\ge0$ and $k\ge1$ with
\begin{eqnarray}
Z_l(T)&=&{2^r}\int e^{-\bu^2} h_{{\tilde n(T)}+l}
\left(\frac {\bx}{\sqrt{\bar c_{{\tilde n(T)}+l+1}}}-\bu,T\right) \nonumber \\
&& \qquad\quad h_{{\tilde n(T)}+l}
\left(\frac {\bx}{\sqrt{\bar c_{{\tilde n(T)}+l+1}}}+\bu,T\right)\,d\bx\,d\bu 
\label{$6.6'$} 
\end{eqnarray}
with the constants $\bar c_n$, $n=1,2,\dots$ defined in~(\ref{2.24a}).
These formulas will be used in the proof of the following

\medskip\noindent
{\bf Lemma 6.1.} {\it Under the conditions of Theorem~3.3 the
inequalities
\begin{eqnarray}
M_2(l,T)&\le&\eta^*\left(\frac23\right)^l, \label{6.7} \\
Z_l(T)&\ge& \left(\bar c_{\tilde n(T)+l}\right)^{r/2}
\left(1-8\sqrt{\eta^*}\left(\frac56\right)^l\right), \label{$6.7'$} \\
M_2(l+1,T)&\le& \frac{\bar c_{{\tilde n(T)}+l+1}}2
\left(1+10\sqrt{\eta^*}\left(\frac5{6}\right)^l\right)M_2(l,T), 
\label{$6.7''$}
\end{eqnarray}
\begin{equation}
M_2(l,T)\le 2\cdot 2^{-l}\frac{c^{{(\tilde n(T)}+l)}}
{c^{{(\tilde n(T))}}}\eta^* \quad \textrm{and}\quad
M_4(l,T)\le 5\cdot 4^{-l}\left(\frac{c^{{(\tilde n(T)}+l)}}
{c^{{(\tilde n(T))}}}\right)^2\eta^* \label{6.8}
\end{equation}
hold for all $l\ge0$ with the same number $\eta^*$ which appears in
(\ref{6.3}). }

\medskip\noindent
{\bf Proof of Lemma 6.1.}\/ Relation (\ref{6.7}) holds for $l=0$ by 
relation (\ref{6.3}). We shall prove that if relation (\ref{6.7}) 
holds for an integer $l$, then relations (\ref{$6.7'$}) and 
(\ref{$6.7''$}) also hold for this~$l$. Then we prove that if
relations (\ref{6.7}) and (\ref{$6.7''$}) hold for some $l$, then 
relation (\ref{6.7}) holds also for $l+1$. These statements imply 
relations (\ref{6.7}), (\ref{$6.7'$}) and~(\ref{$6.7''$}). We 
prove them with the help of the following calculations.

It follows from formulas (\ref{6.6}) and (\ref{$6.6'$}) that
\begin{eqnarray}
&&M_k(l+1,T)=\frac{\left(\bar c_{\tilde n(T)+l+1}\right)^{r/2}}{Z_l(T)}
\int\exp\left\{-\frac{(\bx-\bu)^2}4\right\}\left|\frac{\bx+\bu}2\right|^k 
\label{6.9}  \\ 
&&\qquad\qquad\qquad\qquad \left(\bar c_{{\tilde n(T)}+l+1}\right)^{k/2} 
h_{{\tilde n(T)}+l}(\bx,T)h_{{\tilde n(T)}+l}\left(u,T\right)\,d\bx\,d\bu \nonumber \\
&&\qquad \le \frac{\left(\bar c_{{\tilde n(T)}+l+1}\right)^{(k+r)/2}}{2^k Z_l(T)}
\int |\bx+\bu|^k h_{{\tilde n(T)}+l}(\bx,T)h_{{\tilde n(T)}+l}
\left(\bu,T\right)\,d\bx\,d\bu
\nonumber
\end{eqnarray} 
for all $l\ge0$ and $k\ge1$, and
\begin{eqnarray*}
Z_l(T)&=&(\bar c_{{\tilde n(T)}+l+1})^{r/2}
\int\exp\left\{-\frac{(\bx-\bu)^2}2\right\} \\
&& \qquad\qquad\qquad h_{{\tilde n(T)}+l}(\bx,T)h_{{\tilde n(T)}+l}
\left(\bu,T\right)\,d\bx\,d\bu\\
&\ge& (\bar c_{{\tilde n(T)}+l+1})^{r/2} e^{-4\sqrt{M_2(l,T)}} \int_{|\bx|\le
M_2(l,T)^{1/4},\,|\bu|\le M_2(l,T)^{1/4}}\\
&&\qquad\qquad\qquad h_{{\tilde n(T)}+l}(\bx,T)
h_{{\tilde n(T)}+l}\left(\bu,T\right)\,d\bx\,d\bu \\
&\ge& (\bar c_{{\tilde n(T)}+l+1})^{r/2} e^{-4\sqrt{M_2(l,T)}}
\biggl(1-2\int_{|\bx|\ge M_2(l,T)^{1/4}}  
h_{{\tilde n(T)}+l}(\bx,T)d\bx\biggr)  \\ 
&\ge& (\bar c_{{\tilde n(T)}+l+1})^{r/2} e^{-4\sqrt{M_2(l,T)}} \\
&& \qquad\qquad \left(1-\frac2{\sqrt{M_2(l,T}}\int_{|\bx|\ge M_2(l,T)^{1/4}}  
x^2 h_{{\tilde n(T)}+l}(\bx,T)\,d\bx\right). 
\end{eqnarray*}
Hence
$$
Z_l(T)\ge (\bar c_{{\tilde n(T)}+l+1})^{r/2}
e^{-4\sqrt{M_2(l,T)}}\left(1-2\sqrt{M_2(l,T)}\right).
$$
The last relation and formula (\ref{6.7}) for $l$ together imply that
\begin{eqnarray*}
Z_l(T)&\ge&(\bar c_{{\tilde n(T)}+l+1})^{r/2}
\left(1-5\sqrt{M_2(l,T)}\right) \left(1-2\sqrt{M_2(l,T)}\right)\\
&\ge& (\bar c_{{\tilde n(T)}+l+1})^{r/2} \left(1-8\sqrt{M_2(l,T)}\right)\\
&\ge& (\bar c_{{\tilde n(T)}+l+1})^{r/2}
\left(1-8\sqrt{\eta^*}\left(\frac56\right)^l\right),
\end{eqnarray*}
and this is relation (\ref{$6.7'$}) for the number $l$. Relation 
(\ref{6.9}) for
$k=2$ and formula (\ref{$6.7'$}) for $l$ together yield that
\begin{eqnarray*}
M_2(l+1,T)&\le& \frac{(\bar c_{{\tilde n(T)}+l+1})^{(2+r)/2}}
{2Z_l(T)}M_2(l,T) \\
&\le& \frac{\bar c_{\tilde n(T)+l+1}}2\left(1+10\sqrt{\eta^*}
\left(\frac5{6}\right)^l\right) M_2(l,T),
\end{eqnarray*}
and this is formula (\ref{$6.7''$}) for $l$. Finally, if $\eta$ is 
chosen sufficiently small, then formulas (\ref{6.7}) and 
(\ref{$6.7''$}) for $l$ imply (\ref{6.7}) for $l+1$. Thus formulas 
(\ref{6.7})~---~(\ref{$6.7''$}) are proved.

The first relation in (\ref{6.8}) follows from the first relation in 
(\ref{6.3}) and (\ref{$6.7''$}). Formula (\ref{6.9}) with the choice $k=4$, 
(\ref{$6.7'$}) and the first formula in (\ref{6.8}) imply that
\begin{eqnarray*}
M_4(l+1,T)&\le& \frac {(\bar c_{{\tilde n(T)}+l+1})^{(r+4)/2}}{8Z_l(T)}
\left(3M_2(l,T)^2+M_4(l,T)\right)\\
&\le&\frac18 \bar c_{{\tilde n(T)}+l+1}^2
\left(1+10\sqrt{\eta^*}\left(\frac5{6}\right)^l\right)
\left(3M_2(l,T)^2+M_4(l,T)\right)\\
&\le& 2\cdot 4^{-l}\left(\frac{\bar c^{({\tilde n(T)}+l+1)}}
{c^{({\tilde n(T)})}}\right)^2{\eta^*}^2+\frac{M_4(l,T)}6.
\end{eqnarray*}
The second relation in (\ref{6.8}) follows by induction from the last
inequality and the second inequality in (\ref{6.3}). Lemma~6.1 is proved.
\qed

\medskip\noindent 
{\it Remark.}\/ The Corollary formulated after Theorem~3.1 follows from
Theorem~3.1, formula (\ref{6.8}) and Lemma 4.4. Indeed, if $T$ is not in the
low temperature, then by Theorem~3.1 the pair $({\tilde n(T)},T)$ with the
definition of ${\tilde n(T)}$ given in (\ref{6.1}) is in the high 
temperature domain. By
formula (\ref{6.8}) all pairs $(n,T)$, $n\ge{\tilde n(T)}$, are in the 
high temperature
region, i.e.\ if $T>0$ is not in the low temperature region, then it is
in the high temperature region. The remaining statements of the
Corollary are contained in Lemma~4.4. 

\medskip
In the next lemma we prove an estimate from below for $M_2(l,T)$.

\medskip\noindent
{\bf Lemma 6.2.} {\it Put
$$
\sigma^2(l,T)=2^l\frac{c^{({\tilde n(T)})}}{c^{({\tilde n(T)}+l)}} M_2(l,T), 
\quad l\ge0.
$$
Under the conditions of Theorem~3.3 the limit
$$
\bar\sigma^2(T)=\lim_{l\to\infty} \sigma^2(l,T)>0
$$
exists, and it is positive for all $T>0$. If ${\tilde n(T)}\neq 0$, i.e.
\ if $(0,T)$ is not in the high temperature region, then there exist two
constants $C_2>C_1>0$ depending only on the parameter $\tilde\eta$ in
formula (\ref{6.5}) 
in such a way that the inequalities
\begin{equation}
C_1\le \bar\sigma^2(T)\le C_2 \label{$6.11'$}
\end{equation}
hold. The upper bound in (\ref{$6.11'$}) holds for all $T>0$ 
which is not in the low temperature region.} 

\medskip\noindent
 {\bf Proof of Lemma 6.2.}\/ The hard part of the proof is to show 
that $\sigma^2(l,T)$ has a non-negative $\liminf$. It follows simply 
from formula (\ref{$6.6'$}) that $Z_l(T)\le (\bar c_{{\tilde n(T)}+l+1})^{r/2}$. 
A natural lower bound  for $M_2(l,T)$ can be obtained in the following 
way. By formula (\ref{6.6}) and the upper bound for $Z_l(T)$
\begin{eqnarray}
M_2(l+1,T)&\ge&\bar  c_{{\tilde n(T)}+l+1}\int e^{-{(\bx-\bu)^2}/4}
\left|\frac{\bx+\bu}2\right|^2 \nonumber \\ 
&&\qquad\qquad
h_{{\tilde n(T)}+l}(\bx,T)h_{{\tilde n(T)}+l}\left(\bu,T\right)\,d\bx\,d\bu 
\nonumber \\
&=& \frac {\bar c_{{\tilde n(T)}+l+1}}4\biggl(2M_2(l,T)-\int
|\bx+\bu|^2\left(1-e^{-(\bx-\by)^2/4}\right) \nonumber \\
&&\qquad\qquad\qquad\qquad
h_{{\tilde n(T)}+l}(\bx,T)h_{{\tilde n(T)}+l}\left(\bu,T\right)\,d\bx\,d\bu\biggr) 
\nonumber \\
&\ge& \frac {\bar c_{{\tilde n(T)}+l+1}}2\biggl(M_2(l,T)
-\int\frac14 |\bx+\bu|^2|\bx-\bu|^2 \nonumber \\
&&\qquad\qquad\qquad\qquad
h_{{\tilde n(T)}+l}(\bx,T)h_{{\tilde n(T)}+l}\left(\bu,T\right)\,d\bx\,d\bu\biggr)
\nonumber \\
&\ge& \frac {\bar c_{{\tilde n(T)}+l+1}}2\biggl(M_2(l,T)
-\int\frac12(|\bx|^4+|\bu|^4) \nonumber \\
&&\qquad\qquad\qquad\qquad
h_{{\tilde n(T)}+l}(\bx,T)h_{{\tilde n(T)}+l}\left(\bu,T\right)\,d\bx\,d\bu\biggr)
\nonumber \\
&=&\frac {\bar c_{{\tilde n(T)}+l+1}}2\left(M_2(l,T)-{M_4(l,T)}\right). 
\label{6.12}
\end{eqnarray}
However, this estimate is useful only if we know that the right-hand
side in it is non-negative. We do not know such an estimate for small
$l$, hence in this case we apply a different argument. Clearly
$$
M_2(l,T)\ge M_{2,\textrm{tr.}}(l,T),
$$
where $M_{2,\textrm{tr.}}(l,T)$ is the truncated moment.
On the other hand, we get by using an argument similar to the previous
calculation and making the observation
\begin{eqnarray*}
&&\left\{(\bx,\bu)\colon  \bx\in {\mathbb R}^r, 
\bu\in {\mathbb R}^r, \bar c_{{\tilde n(T)}+l+1} 
\left|\frac{\bx+\bu}2\right|\le \frac1{10} \right\}\\
&&\qquad \supset\left\{(\bx,\bu)\colon  \bx\in {\mathbb R}^r, \bu\in {\mathbb
R}^2, |\bx|\le\frac1{10}, |\bu|\le\frac1{10},\textrm{arg}\,(\bx,\bu)\subset I
\right\}
\end{eqnarray*}
with $I=\left(\frac\pi{50},\frac{49\pi}{50}\right)\cup
\left(\frac{51\pi}{50},\frac{99\pi}{50}\right)$ that
%
%
\begin{eqnarray*}
M_{2,\textrm{tr.}}(l+1,T)
&\ge& \bar c_{{\tilde n(T)}+l+1} \int_{\bar c_{{\tilde n(T)}+l+1}
\left|\frac{\bx+\bu}2\right|\le
\frac{1}{10}} e^{-{(\bx-\bu)^2}/4}
\left|\frac{\bx+\bu}2\right|^2\\
&&\qquad\qquad\qquad h_{{\tilde n(T)}+l}(\bx,T)h_{{\tilde n(T)}+l}
\left(\bu,T\right)\,d\bx\,d\bu\\
&\ge& \bar c_{{\tilde n(T)}+l+1}e^{-1/100} \int_{ |\bx|\le\frac1{10},
|\bu|\le\frac1{10},\textrm{arg}\,(\bx,\bu)\subset I}
\left|\frac{\bx+\bu}2\right|^2\\
&&\qquad\qquad\qquad h_{{\tilde n(T)}+l}(\bx,T)h_{{\tilde n(T)}+l}
\left(\bu,T\right)\,d\bx\,d\bu\\
&=&\frac{\bar c_{{\tilde n(T)}+l+1}}4e^{-1/100}\int_{ |\bx|\le\frac1{10},
|\bu|\le\frac1{10},\textrm{arg}\,(\bx,\bu)\subset I} (\bx^2+\bu^2)\\
&&\qquad\qquad\qquad h_{{\tilde n(T)}+l}(\bx,T)h_{{\tilde n(T)}+l}
\left(\bu,T\right)\,d\bx\,d\bu\\
&=&\bar c_{{\tilde n(T)}+l+1}e^{-1/100}\frac{12}{25}M_{2,\textrm{tr.}}(l,T)
\ge\frac13 \bar c_{{\tilde n(T)}+l+1} M_{2,\textrm{tr.}}(l,T).
\end{eqnarray*}
The last estimate implies that
\begin{equation}
\sigma^2(l,T)=2^l\frac {c^{({\tilde n(T)})}}{c^{({\tilde n(T)}+l)}}M_2(l,T)\ge
2^l\frac {c^{({\tilde n(T)})}}{c^{({\tilde n(T)}+l)}}
M_{2,\textrm{tr.}}(l,T)\ge\left( \frac23\right)^l M_{2,\textrm{tr.}}(0,T). 
\label{6.13}
\end{equation}
On the other hand, it follows from (\ref{6.12}) and the second 
inequality in (\ref{6.8}) that
\begin{eqnarray}
\sigma^2(l+1,T)&\ge&\sigma_2(l,T)-2^{l+1} \frac
{c^{({\tilde n(T)})}}{c^{{\tilde n(T)}+l+1}}M_4(l,T) \label{6.14}  \\
&\ge&\sigma^2(l,T)-\frac{5\eta^*}{2^lc_{{\tilde n(T)}+l+1}}
\frac{c^{({\tilde n(T)}+l)}} {c^{({\tilde n(T)})}}
\ge\sigma^2(l,T)-50\eta^*\left(\frac34\right)^l. \nonumber 
\end{eqnarray}
Because of (\ref{6.13}) and (\ref{6.5}) an index $\bar l\ge0$ 
can be chosen in such a way that
$$
\sigma^2(\bar l,T)\ge 1000\eta^*\left(\frac34\right)^{\bar l},
$$
and if the pair $(0,T)$ is not in the high temperature region,
then we may choose $\bar l$ so that $\bar l\le K(\bar \eta,\kappa)$ 
with some appropriate $K(\bar\eta,\kappa)$. Hence relation 
(\ref{6.14}) implies that
$$
\frac{\sigma^2(\bar l+l+1,T)}{\sigma^2(\bar l,T)}
\ge \frac{\sigma^2(\bar l+l,T)}{\sigma^2(\bar l,T)}
-\frac 1{20}\left(\frac34\right)^l .
$$
This relation and the bound on $\sigma^2(\bar l,T)$ imply that
$\liminf\limits_{l\to\infty}\sigma^2(l,T)>0$, and this $\liminf$ can be
bounded by a positive number which depends only on $\bar\eta$ and 
$\kappa$ if $(0,T)$ is not in the high temperature region. The 
analogous result for $\limsup$ follows from (\ref{$6.7''$}). To 
complete the proof it is enough to show that the $\liminf$ is 
actually $\lim$. To prove this let us observe that for any 
$\varepsilon>0$ and $N>0$ there is some $m>N$ such that
$\sigma^2(m,T)<\liminf\limits_{n\to\infty} \sigma^2(n,T)+\varepsilon$. 
Then by formula (\ref{$6.7''$})
\begin{eqnarray*}
\sigma^2(n,T)&\le&\sigma^2(m,T)\prod_{l=m}^n
\left(1+10\sqrt{\eta^*}\left(\frac5{6}\right)^l\right) \\
&\le&\liminf_{n\to\infty} \sigma^2(n,T)+2\varepsilon,\quad n>m
\end{eqnarray*}
for any $\varepsilon>0$ if $N=N(\varepsilon)$ is chosen 
sufficiently large. Lemma 6.2 is proven. \qed

\medskip
To prove Theorem 3.3 let us introduce the characteristic functions
$$
\varphi_n(s,T)=\int_{{\mathbb R}^r} e^{i s\bx}\tilde h_n(\bx,T)\,d\bx,\quad s\in
{\mathbb R}^r 
$$
and moments
$$
\tilde M_k(n,T)=\int_{{\mathbb R}^r} |\bx|^k\tilde h_n(\bx,T)\,d\bx,
$$
where the function $\tilde h(\bx,T)$ was defined in (\ref{3.24}). Clearly,
$$
\tilde M_k(n,T)=\left(\frac{2^n}{c^{(n)}}\right)^{k/2}M_k(n-{\tilde n(T)},T)
\quad \textrm{if }n\ge {\tilde n(T)}.
$$
In particular, $\tilde M_2(n,T)=\frac
{2^{{\tilde n(T)}}}{c^{({\tilde n(T)})}}\sigma^2(n-{\tilde n(T)},T)$.  
We shall prove Theorem~3.3 by means
of the usual characteristic function technique. The following 
lemma plays a crucial role in the proof.

\medskip\noindent
{\bf Lemma 6.3.} {\it  Under the conditions of Theorem 3.3 the relation
\begin{equation}
\lim_{n\to\infty}\frac{c^{({\tilde n(T)})}}{2^{{\tilde n(T)}}}\tilde M_2(n,T)
=\bar\sigma^2(T) \label{6.16}
\end{equation}
holds with the constant $\bar\sigma^2(T)$ appearing in Lemma 6.2, and
\begin{equation}
\lim_{n\to\infty}\sup_{|\bs|\le A}\left|\log\varphi_n\left(\bs,T\right)+
\frac{2^{{\tilde n(T)}}}{c^{({\tilde n(T)})}}\bar\sigma^2(T)
\frac{\bs^2}2\right|\to0 \label{6.17}
\end{equation}
for all $A>0$.}

\medskip\noindent
{\bf Proof of Lemma 6.3.}\/ Relation (\ref{6.16}) follows from 
Lemma~6.2,  and it follows from the second relation in (\ref{6.8}) 
that $\tilde M_4(n,T)\le5\left(\frac{2^{{\tilde n(T)}}}
{c^{({\tilde n(T)})}}\right)^2\eta^*$. Hence the characteristic function
$\varphi$ can be estimated as
\begin{eqnarray}
&&\left|\varphi_n(\bs,T)-\left(1-
\tilde M_2(n,T)\frac{\bs^2}2\right)\right|\le 
\left(\frac{2^{{\tilde n(T)}}}{c^{({\tilde n(T)})}}\right)^2
\eta^*|\bs|^4 \label{6.18} \\ 
&&\qquad\qquad\qquad\qquad\textrm{for }n\ge{\tilde n(T)}\textrm{ and }
\bs \in {\mathbb R}^r. \nonumber
\end{eqnarray}

In the proof of formula (\ref{6.18}) we exploit that 
$\int (\bs,\bx)\tilde h_n(\bx,T)\,d\bx=0$ and 
$\int (\bs,\bx)^3\tilde h_n(\bx,T)\,d\bx=0$
because of the rotational symmetry of $\tilde h_n(\bx,T)$.
The coefficient of $|\bs|^4$ in (\ref{6.18}) is bounded by a constant 
(depending on $T$), and the coefficient at $|\bs|^2$ converges to the 
positive constant $\frac{2^{{\tilde n(T)}}}{c^{({\tilde n(T)})}}\bar\sigma^2(T)$. 
Hence formula (\ref{6.18}) implies that for any $\varepsilon>0$,
\begin{equation}
\left|\log\varphi_n\left(\bs,T\right)
+\frac{2^{{\tilde n(T)}}}{c^{({\tilde n(T)})}}\bar\sigma^2(T)
\frac{\bs^2}2\right|\le\varepsilon \quad\textrm{if } n>n_1 
\textrm{ and }|\bs|\le\delta
\label{6.19}
\end{equation}
with some $n_1=n_1(\varepsilon,T)$ and $\delta=\delta(\varepsilon,T)$. 
By a rescaled version of the recursive formula (\ref{2.12}) we can write
\begin{eqnarray*}
&&\varphi_{n+1}(\sqrt 2\bs,T)=\frac1{Z_n(T)}\int \exp\left\{i\bs(\bx+\bu)
-\frac{c^{(n)}(\bx-\bu)^2}{4\cdot2^n}\right\}  \\
&& \qquad\qquad\qquad\qquad\qquad \tilde h_n(\bx,T)\tilde
h_n(\bu,T)\,d\bx\,d\bu\\
&&\qquad\quad =\frac1{Z_n(T)}\biggl[\varphi_{n}\left(\bs,T\right)^2-\int
e^{i\bs(\bx+\bu)}\left(1-\exp\left\{
-\frac{c^{(n)}(\bx-\bu)^2}{4\cdot2^n}\right\}\right)\\
&&\qquad\qquad\qquad\qquad\qquad
\tilde h_n(\bx,T)\tilde h_n(\bu,T)\,d\bx\,d\bu\biggr]
\end{eqnarray*}
with
$$
Z_n(T)=\int \exp\left\{-\frac{c^{(n)}(\bx-\bu)^2}{4\cdot2^n}\right\}
\tilde h_n(\bx,T)\tilde h_n(\bu,T)\,d\bx\,d\bu.
$$
The estimates
\begin{eqnarray*}
&&\left|\int e^{i\bs(\bx+\bu)}\left(1-\exp\left\{
-\frac{c^{(n)}(\bx-\bu)^2}{4\cdot2^n}\right\}\right)
\tilde h_n(\bx,T)\tilde h_n(\bu,T)\,d\bx\,d\bu\right|\\
&&\qquad\le \int\frac{c^{(n)}(\bx-\bu)^2}{4\cdot2^n}
\tilde h_n(\bx,T)\tilde h_n(\bu,T)\,d\bx\,d\bu=
\frac{c^{(n)}}{2\cdot 2^n}\tilde M_2(n,T)
\end{eqnarray*}
and similarly
$$
1\ge Z_n(T)\ge1-\frac{c^{(n)}}{2\cdot 2^{n}}\tilde M_2(n,T)
$$
hold. Hence
$$
\varphi_{n}^2\left(\bs,T\right)-\frac{c^{(n)}}{2\cdot 2^n}\tilde M_2(n,T)
\le\varphi_{n+1}(\sqrt2\bs,T)\le\frac{\varphi_{n}^2\left(\bs,T\right)
+\frac{c^{(n)}}{2\cdot 2^n}\tilde M_2(n,T)}
{1-\frac{c^{(n)}}{2\cdot 2^n}\tilde M_2(n,T)}.
$$
The term $\frac{c^{(n)}}{2^n}\tilde M_2(n,T)$ is much less than
$\left(\frac23\right)^n$ for large $n$. If we have a 
positive lower bound on $\varphi_n(\bs)$ then we get by fixing some
$K>0$ and taking logarithm in the last relation that 
\begin{equation}
\left|\log\varphi_{n+1}\left(\sqrt2\bs,T\right)-2\log\varphi_n(\bs,T)\right|
\le\left(\frac23\right)^n\;\textrm{ if }n>n_2\textrm{ and }
\varphi_n(\bs,T)\ge\frac1K \label{6.20}
\end{equation}
with some $n_2=n_2(K,T)$. Formula (\ref{6.17}) can be deduced from 
(\ref{6.19}) and (\ref{6.20}). Indeed, define an index $k$ by the relation
$A\le\delta2^{k/2}<\sqrt2A$ with the numbers $A$ and $\delta$ in 
(\ref{6.17}) and (\ref{6.18}). Put 
$K=2e^{-2^{{\tilde n(T)}}\bar\sigma^2(T)A^2/c^{({\tilde n(T)})}}$ and let
$\varepsilon\le\frac1{8K}$. Choose a number $n_3$ such that
$\left(\frac23\right)^{n_3}\le\varepsilon$, and let us consider 
such indices $n$ for
which $n\ge \max(n_1(\varepsilon,T),n_2(K,T), n_3)$. Then simple 
induction yields that 
\begin{eqnarray*}
&&\left|\varphi_{n+j}(\bs,T)+
\frac{2^{{\tilde n(T)}}}{c^{({\tilde n(T)})}}\bar\sigma^2(T)
\frac{\bs^2}2\right|\le \varepsilon+
3\left(\frac23\right)^n\left(1-\left(\frac23\right)^{j+1}\right)
\le 4\varepsilon \\ 
&&\qquad\qquad \textrm{ and}\left|\varphi_{n+j}(\bs,T)\right|\ge\frac1K
\end{eqnarray*}
for $j\le k$ and $|\bs|\le\delta 2^{j/2}$. Since $\varepsilon$ can be 
chosen arbitrary small in the last relation, it implies (with $j=k$) 
relation (\ref{6.17}). Lemma 6.3 is proved. \qed

\medskip
Theorem 3.3 follows from Lemmas 6.2 and 6.3. Indeed, Lemma~6.3
implies that the measures $\tilde H_{n,T}$ converge in distribution to
the normal law with expectation zero and covariance
$\frac{2^{{\tilde n(T)}}}{c^{({\tilde n(T)})}}\bar\sigma^2(T){\mathbf I}$. 
The bounds obtained
for the variance follow from Lemma 6.2 and the observation that the
difference ${\tilde n(T)}-\bar n(T)$ can be bounded by a number 
depending only on $\bar\eta$ and $\kappa$.

%
Let us finally show that Corollary to Theorem~3.3 follows from 
Theorem ~3.3. By formulas (\ref{2.10}), and 
(\ref{3.24}) we can write
\begin{equation}
2^{-n}p_n(2^{-n/2}\sqrt T\bx,T)=C(n)\exp
\left\{-\frac{l_n A_n}{2^{n+1}}\bx^2\right\}\tilde h_n(\bx,T) \label{6.21}
\end{equation}
with an appropriate norming constant $C(n)$. Observe that the
expressions at both sides of this identity  are density functions, the
measures with density function $\tilde h_n(\bx,T)$ have a limit as
$n\to\infty$, the term 
%
$\left\{- \frac{l_nA_n}{2^{n+1}}\bx^2\right\}$ 
is bounded, and it tends to 1 uniformly in
any compact set as $n\to\infty$. These facts imply that $C(n)\to1$ in
(\ref{6.21}), and the measures with density functions
$2^{-n}p_n(2^{-n/2}\sqrt T\bx,T)$ have the same limit as the measures
with density functions $\tilde h_n(\bx,T)$. Hence the Corollary
of Theorem~3.3 holds. \qed

\section{Estimates in the Low Temperature Region. The proof of
Theorem 3.2.}

The proof of Theorem~3.2 heavily exploits the results of Section~4. 
These results show that the replacement of the operator ${\mathbf Q}_n$ 
whose application makes possible to compute the function $f_{n+1}(x,T)$ 
by its linearization ${\mathbf T}_n$ causes only a negligible error. 
Formula (\ref{4.17}) enables one to investigate the operator 
${\mathbf T}_n$ in the Fourier space. In such a way good estimates 
can be obtained for the Fourier transform of a regularized version 
of the function $f_{n+1}(x,T)$. The results of Theorem~3.2 can be 
proved by means of these estimates with the help of inverse 
Fourier transformation.

It is simpler to work with an appropriately scaled version of the
functions $f_n(x,T)$. Put
$$
\bar f_n(x,T)=\frac1{M_n(T)}f_n\left(\frac x{M_n(T)},T\right)
$$
and
$$
\bar\varphi_n(f_n(x,T))
=\frac1{M_n(T)}\varphi_n\left(f_n\left(\frac
x{M_n(T)},T\right)\right).
$$
We defined the function $\bar\varphi_n(f_n(x,T))$ 
by means of the definition of the regularization of 
the function $f_n(x,T)$ introduced in Section~4. 

Let us also introduce the functions
$$
\psi_{n+1}(f_n(x,T))=\frac1{M_n(T)}{\mathbf T}_n \varphi_n
\left(f_n\left(\frac x{M_n(T)},T\right)\right).
$$
The estimates of Proposition 4.2 and relation (\ref{4.17}) can be rewritten
for these new functions. We shall rewrite formulas (\ref{4.15}) and
(\ref{4.16}) only in the case when $n>N_1(T)$ with the number $N_1(T)$ defined
in formula (\ref{4.18}), i.e.\ in the case when $\beta_n(T)$ and $M_n^{-1}(T)$
have the same order of magnitude. In this case
$M_n(T)\sqrt{\beta_{n+1}(T)}\le10$,
\begin{equation} \label{7.3} 
\begin{aligned}
&\left|\frac{\partial^j}{\partial x^j}\left(\bar f_{n+1}(x,T)-\psi_{n+1}
(f_n(x,T))\right)\right|  \\
&\le K_1 \frac{\beta_n(T)}{c^{(n)}}
\left[\exp\left\{-\frac1{10} \left|2x+\frac{x^2}
{c^{(n+1)}}\right| \right\}+\exp \left\{-\frac{|x|}{5}\right\}\right] \\
&\le K_2 \frac{\beta_{n}(T)}{c^{(n)}} e^{-|x|/10}, 
\qquad\ x>-c^{(n+1)}M_{n+1}^2(T),\quad j=0,1,2, 
\end{aligned}
\end{equation}
and
\begin{equation} \label{7.4}
\left|\frac{\partial^j}{\partial x^j}\psi_{n+1}
(f_n(x,T))\right|\le K_3e^{-|x|/5},
\quad x\in {\mathbb R}^1,\quad j=0,1,2,3,4,
\end{equation}
with some universal constants $K_1$,~$K_2$ and~$K_3$. Formula 
(\ref{4.17}) can be rewritten as
\begin{equation} \label{7.5} 
\begin{aligned}
\tilde\psi_{n+1}(f_n(\xi,T))&=\tilde{\mathbf T}_{n} \tilde\varphi_n(
f_n(M_n(T)\xi,T)) \\ 
&= \frac{\exp\left\{i\frac {(r-1)\bar c_{n+1}}{4}\xi
\right\}}{\left(1+i\frac {\bar c_{n+1}}{2}\xi\right)^{(r-1)/2}} \,\,
\tilde{\bar\varphi}_n^2
\left(f_n\left(\frac {\bar c_{n+1}}{2}\xi,T\right)\right). 
\end{aligned}
\end{equation}
We claim that under the conditions of Theorem~3.2,
\begin{equation}\label{7.6}
\begin{aligned}
\lim_{n\to\infty}\sup_{x\ge-c^{(n)}M_n^2(T)}&\left|\frac{\partial^j}{\partial
x^j}\left(\bar
f_n(x,T)-\bar\varphi(f_n(x,T))\right)\right|e^{|x|/20} =0,\\
& j=0,1,2.
\end{aligned}
\end{equation}
Indeed, by relations (\ref{7.3}) and (\ref{7.4})
\begin{equation}
\left|\frac{\partial^j}{\partial x^j}\bar f_n(x,T)\right|\le
e^{-|x|/10},\quad j=0,1,2,\quad\textrm{if } \; x\ge -c^{(n)}M_n^2(T) ,
\label{7.7}
\end{equation}
and $\bar\varphi_n(f_n(x,T))$ is the appropriate scaling of the function
$$
\varphi\left(\frac x{\sqrt{c^{(n)}}M_n(T)}\right)f_n\left(\frac
x{\sqrt{c^{(n)}}M_n(T)}\right).
$$
Under the conditions of Theorem~3.2, formula (\ref{4.28})
holds, which implies that
$$
\lim_{n\to\infty}\sqrt{c^{(n)}}M_n(T)=\infty .
$$
This fact together
with (\ref{7.7}) allow us to give a good bound on the difference between
the functions $\bar\varphi_n(f_n(x,T))$ and $\varphi\left(\frac
x{\sqrt{c^{(n)}}M_n(T)}\right)f_n\left(\frac x{\sqrt{c^{(n)}}M_n(T)}\right)$. 
\hfill\break
Relation (\ref{7.6})
can be deduced from this bound and formula~(\ref{7.7}).

It follows from Lemma~4.4 that
$\lim\limits_{n\to\infty}\frac{M_{n+1}(T)}{M_n(T)}=1$.
Relations (\ref{7.3}) and (\ref{7.6}) together with this fact imply that
\begin{equation}\label{7.8}
\begin{aligned}
\lim_{n\to\infty}\sup_{|x|<\infty}&\left|\frac{\partial^j}{\partial
x^j}(\psi_{n}(f_{n-1}(x,T))-\bar\varphi_n(f_n(x,T))) \right|e^{|x|/20}
=0, \\
& j=0,1,2. 
\end{aligned}
\end{equation}

Now we prove, using an adaptation of the proof of Lemmas~14  and~15 in
\cite{BM3}, that the Fourier transforms of the functions
$\psi_{n+1}(f_n(x,T))$ converge to the Fourier transform of the 
function $g(x)$, and this convergence is uniform in all compact
domains. First we prove a modified version of this statement, where
$\psi_n$ is replaced with $\bar\varphi_n$ in a small neighbourhood
of the origin. We want to work with the functions $\log\tilde
{\bar\varphi}_{n}(f_n(\xi,T)))$. To do this, observe first that for
$n>N_1(T)$  there is some constant  $A>0$ such that all functions
$\tilde{\bar\varphi}_{n+1}(f_n(\xi,T)))$ are separated from zero in the
interval $|\xi|\le A$. Indeed,
\begin{eqnarray*}
|1-\tilde{\bar \varphi}_{n}(f_n(\xi,T)))|&\le&
\int |e^{ix\xi}-1|\bar\varphi_{n}(f_n(x,T)))\,dx\\
&\le& \int|\xi| |x|\bar\varphi_{n}(f_n(x,T)))\,dx
\le {\textrm{ const.}\,} |\xi|.
\end{eqnarray*}
Similarly,
$$
\left|\frac{\partial^j}{\partial \xi^j}
\tilde{\bar \varphi}_{n}(f_n(\xi,T))\right|\le C(j) \quad\textrm{for
all }j\ge0 \textrm{ and }n\ge N_1(T).
$$
Hence a constant $A>0$ can be chosen in such a way that
$$
\sup_{|\xi|\le 2A}\max\left( |1-\tilde g(\xi)|, \sup_{n\ge
N_1(T)}|1-\tilde{\bar\varphi}_{n}(f_n(\xi,T))| \right)\le \frac12.
$$
These estimates imply that
\begin{equation}
\sup\sup_{|\xi|\le A}\left|\frac{\partial^2}{\partial \xi^2}
\log\tilde{\bar \varphi}_{n}(f_n(\xi,T)))\right|\le C(T), \label{7.10}
\end{equation}
with a constant $C(T)<\infty$ independent of $n$. We claim that
\begin{equation}
\sup_{|\xi|\le A}\left|\frac{\partial^2}{\partial \xi^2}
\log\tilde{\bar \varphi}_{n}(f_n(\xi,T)))-\frac{d^2}{d^2\xi}\log
\tilde g(\xi)\right|\to0 \quad \textrm{as }n\to\infty. \label{7.11}
\end{equation}
To prove (\ref{7.11}) let us first observe that 
$\lim\limits_{n\to\infty}\bar c_n=1$ by Condition~1. By (\ref{7.8}),
$$
\lim_{n\to\infty}\sup_{|\xi|\le A}\left|\frac{\partial^2}{\partial
\xi^2}\left(\log\tilde
\psi_{n+1}(f_{n}(\xi,T))-\log\tilde{\bar\varphi}_{n+1}
(f_{n+1}(\xi,T))\right)\right|=0,
$$
and because of the estimates obtained for the derivatives of
$\tilde{\bar\varphi}_n(\xi,T)$
\begin{eqnarray*}
&&\left|\frac{\partial^2}{\partial \xi^2}\log\tilde{\bar
\varphi}_{n}(f_n(\xi_1,T)))-\frac{\partial^2}{\partial \xi^2}
\log\tilde{\bar \varphi}_{n}(f_n(\xi_2,T)))\right|\le
{\textrm{ const.}\,}|\xi_1-\xi_2|\\
&&\qquad\qquad\qquad\qquad\qquad
\textrm{if } |\xi_1|\le A\textrm{ and }|\xi_2|\le A
\end{eqnarray*}
for all large indices $n$ with a constant independent of $n$.
Taking logarithm and then differentiating twice in formulas (\ref{7.5}) 
and (\ref{1.30aa}) we get with the help of the above observations that
\begin{eqnarray*}
&&\sup_{|\xi|\le A}\left|\frac{\partial^2}{\partial \xi^2}
\log\tilde{\bar \varphi}_{n+1}(f_{n+1}(\xi,T)))-\frac{d^2}{d^2\xi}\log
\tilde g(\xi)\right|\\
&&\qquad\le\frac12 \sup_{|\xi|\le A}\left|\frac{\partial^2}{\partial
\xi^2} \log\tilde{\bar \varphi}_{n}(f_n(\xi,T)))-\frac{d^2}{d^2\xi}\log
\tilde g(\xi)\right|+\delta_n(T)
\end{eqnarray*}
with some sequence $\lim\limits_{n\to\infty}\delta_n(T)=0$. This
relation together with (\ref{7.10}) imply (\ref{7.11}). Since
\begin{eqnarray*}
&&\left.\frac{\partial}{\partial \xi}
\log\tilde{\bar
\varphi}_{n}(f_{n}(\xi,T)))\right|_{\xi=0}=\left.
\frac{d}{d\xi}\log
\tilde g(\xi)\right|_{\xi=0}=0 \\  
&&\qquad\qquad \textrm{ and }
\log\tilde{\bar \varphi}_{n}(f_{n}(0,T)))=\log\tilde g(0)=0,
\end{eqnarray*}
relation (\ref{7.11}) also implies that
\begin{equation}
\lim_{n\to\infty}\sup_{|\xi|\le A}\left|\tilde{\bar
\varphi}_{n}(f_{n}(\xi,T)))-\tilde g(\xi)\right|=0. \label{7.12}
\end{equation}
Moreover, relation (\ref{7.12}) holds for all $A>0$. This can be proved
similarly to the argument of Lemma~15 in~\cite{BM3}. One has to observe that
because of the structure of formulas (\ref{7.5}) and (\ref{1.30aa}), 
the relation $\bar c_n\to1$ as $n\to\infty$, the continuity of the 
function $\tilde g(\xi)$ and the relation
$$
\lim_{n\to\infty}\sup_{|\xi|<\infty}\left|\tilde
\psi_{n+1}(f_{n}(\xi,T))-\tilde{\bar\varphi}_{n+1}
(f_{n+1}(\xi,T)))\right|=0,
$$
the validity of relation 
(\ref{7.12}) in an interval $|\xi|\le A$ also implies its validity 
in the interval $|\xi|\le(2-\varepsilon)A$ for any $\varepsilon>0$. 
In relation (\ref{7.12}) the function
$\tilde{\bar\varphi}_{n}(f_{n}(\xi,T)))$ can be replaced by
$\tilde \psi_{n+1}(f_{n}(\xi,T))$, i.e.\ the relation
\begin{equation}
\lim_{n\to\infty}\sup_{|\xi|\le A}\left|\tilde\psi_{n+1}
(f_{n}(\xi,T))) -\tilde g(\xi)\right|=0 \label{$7.12'$}
\end{equation}
holds for all $A>0$. It can be proved from (\ref{$7.12'$}), 
by means of inverse Fourier transformation, that
\begin{equation}
\lim_{n\to\infty}\sup_{|x|<\infty}\left|\frac{\partial^j}{\partial x^j}
\psi_{n+1} (f_{n}(x,T))) -\frac{d^j}{dx^j} g(x)\right|=0\quad j=0,1,2.
\label{7.13}
\end{equation}
To prove (\ref{7.13}) we need, besides the estimate (\ref{$7.12'$}), 
some bound about the
decrease of the functions $\tilde g(\xi)$ and $\tilde\psi_{n+1}
(f_{n}(\xi,T)))$ as $\xi\to\pm\infty$. The estimate (\ref{1.30a}) gives 
a good bound for the Fourier transform of the function $g(x)$. We can 
get a good estimate for the Fourier transform of
the function $\psi_{n+1}(f_n(x,T))$ 
with the help of the inductive hypothesis $J(n)$ in Section~4 and 
relation (\ref{7.5}). Rewriting the inductive hypothesis $J(n)$ for 
the function $\bar \varphi_n(f_n(x,T))$ we get with the help of some 
standard calculation that the Fourier transform 
$\tilde\psi_{n+1}(f_n(\xi,T))$ decreases at infinity faster than 
$|\xi|^{-4}$.  These estimates are sufficient for the proof of 
(\ref{7.13}). Relations (\ref{7.13}) and (\ref{7.3}) give an 
estimate on the function $\bar f_n(x,T)$ and its derivatives, 
which is equivalent to (\ref{3.10}). Theorem~3.2 is proved. \qed

\section{Estimates Near the Critical Point.
The proof of Theorems 3.4, 1.3, and 1.5.}

Our previous results suggest that $M_{n+1}^2(T)\sim
M_n^2(T)-\frac{r-1}{2c^{(n)}}$, hence the derivative $\frac {dM_n^2(T)}{dT}$,
as a function of $n$, changes very little if the pair $(n,T)$ is in the
low domain region (observe that $c^{(n)}$ does not depend on
$T$). Therefore, it is natural to expect that
$\frac{dM^2_\infty(T)}{dT}$ is of constant order below the critical
value $T_c$, and $M_\infty^2(T)-M_\infty^2(T_c)\sim
{\textrm{ const.}\,}(T_c-T)$ for $T<T_c$. If $T_n$ 
denotes the smallest $T$ for which the pair $(n,T)$ leaves the 
low temperature region at the $n$-th step, then the following 
heuristic argument may suggest the magnitude of $T_n-T_{n+1}$ for 
large~$n$. Since both $c^{(n)}M_n^2(T_n)\sim\eta^{-1}$ and
$c^{(n+1)}M_{n+1}^2(T_{n+1})\sim\eta^{-1}$, besides this
$M_{n}^2(T_{n+1})-M_{n+1}^2(T_{n+1})\sim\frac{r-1}{2c^{(n)}}$,
$M_{n}^2(T_{n+1})-M_n^2(T_n)\sim\frac{r-1}{2c^{(n)}}$. On the other hand,
$M_{n}^2(T_{n+1})-M_n^2(T_n)\sim T_{n+1}-T_n$. This argument suggests
that  $T_{n+1}-T_n\sim\frac{r-1}{2c^{(n)}}$ and $T_n-T_c\sim
\sum\limits_{k=n}^\infty \frac{r-1}{2c^{(k)}}$. In this section we 
justify these heuristic arguments. The proofs are based on 
the following result:

\medskip\noindent
{\bf Theorem 8.1.} {\it  There exists $\kappa_0=\kappa_0(N)$ such that if
(i) $0<\kappa<\kappa_0$ in formula (\ref{1.17}), (ii) Conditions~1---4 
are satisfied, (iii)  $0<\bar T<c_0A_0/2$, and 
(iv) the integer $n\ge1$ has the property that the pair $(n,\bar T)$ 
belongs to the low temperature region, then for all $0<T<\bar T$ the 
pair $(n,T)$ also belongs to the low temperature region, and  the 
following inequalities hold for $T\le\bar T$:

\medskip\noindent
a.) If $0\le n\le N$, then
$$
\frac{C_1}{\sqrt \kappa T^2}<-\frac {dM_{n+1}(T)}{dT}<\frac{C_2}{\sqrt \kappa
T^2} \quad \textrm{with some }\infty>C_2>C_1>0.
$$

\noindent
b.) If $n\ge N$, then
\begin{eqnarray*}
&&\frac{dM_{n+1}(T)}{dT}=
\frac{dM_n(T)}{dT}\left(1+\frac{r-1}{4c^{(n)}M_n^2(T)}
+\frac{\delta_n(T)}{c^{(n)}}\right),\\
&&\qquad \textrm{where }|\delta_n(T)|\le
C\frac{\beta_{n+1}(T)}{c^{(n+1)}}\beta_n(T)
\textrm{ with some appropriate } C>0.
\end{eqnarray*}
}

\medskip
We will prove Theorem 8.1 in Appendix A below with the help of
Proposition A which is proved also there. This result can be 
interpreted in an informal way as the ``differentiation" of the 
asymptotic identity~(\ref{4.13}). 
The main difficulty in the proof of Proposition~A is to bound the 
error caused by the linear approximation of the operator 
${\mathbf Q}_n$ by ${\mathbf T}_n$ when differentiating with 
respect to $T$. To overcome this difficulty we need a good 
control not only on the functions $f_n(x,T)$ but also on their
derivatives $\frac{\partial}{\partial T}f_n(x,T)$. Hence we have to
work out the estimation of these derivatives. In particular, we 
have to find the inductive hypotheses describing their behaviour. 
These are the analogs of the inductive hypotheses $I(n)$ and $J(n)$ 
formulated in Section~4. It demands fairly much work to work out 
the details, but after the formulation and proof of these 
inductive hypotheses the proof of Proposition~A is simple.

\medskip\noindent
{\bf Proof of Theorem 3.4.}\/ We prove with the help of Proposition~A 
that if the conditions of Theorem~3.4 hold, $0<\bar T<c_0A_0/2$ and 
the pair $(n-1,\bar T)$ belongs to the low temperature region, then 
there exist some constants $0<C_1<C_2$ independent of $T$ such that
\begin{equation}
\frac{C_1}{\kappa T^3}<-\frac {dM^2_n(T)}{dT}<\frac{C_2}{ \kappa T^3}. 
\label{8.1}
\end{equation}
for all $0<T\le \bar T$.

For $0\le n\le N$ $(n,T)$ is in the low temperature region for 
$0<T<c_0A_0/2$. In this case Properties $K_1(n)$---$K_4(n)$ hold
by Proposition~A, and the validity of (\ref{8.1}) for $n=0$
follows from relations (\ref{4.1}), (\ref{4.4}) and (\ref{A1}). 
Its validity for $0\le n\le N$ can be proved by induction with
the help of Properties  $K_1(n)$ and $K_3(n)$, $1\le n\le N$. 

To prove formula (\ref{8.1}) for $n>N$ first we show that
\begin{eqnarray}
-\frac{dM_n^2(T)}{dT}
\exp\left\{-K\left(\frac{\beta_{n+1}(T)}{c^{(n)}}\right)^2\right\}
&\le&-\frac{dM_{n+1}^2(T)}{dT}  \label{8.2} \\
&\le& -\frac{dM_n^2(T)}{dT}\exp
\left\{K\left(\frac{\beta_{n+1}(T)}{c^{(n)}}\right)^2\right\} \nonumber
\end{eqnarray}
for all $T<\bar T$ and $n\ge N$ with an appropriate $K>0$. Relation
(\ref{8.2}) is a consequence of Part b) of Proposition~A, formula 
(\ref{4.13}), the inequality $\beta_{n+1}(T)M_n^2(T)\ge10$ and the relation
$\frac{\beta_{n+1}(T)}{c^{(n)}}\le \eta$ if $(n,T)$ is in the low
temperature domain. Indeed, since 
$$
-\frac{dM_{n+1}^2(T)}{dT}=-
\left(1-\frac{m_n(T)}{c^{(n+1)}M_n(T)}\right) 
\left(1-\frac{\frac{dm_n(T)}{dT}}{c^{(n+1)}\frac{dM_n(T)}{dT}}\right) 
\frac{dM_n^2(T)}{dT} 
$$
these relations imply that
\begin{eqnarray*}
&&
-\left(1-\frac{r-1}{4c^{(n+1)}M^2_n(T)}
-C_1\frac{\beta_{n+1}^{3/2}(T)}{{c^{(n+1)}}^2M_n(T)}\right) \\
&& \qquad\qquad \left(1+\frac{r-1}{4c^{(n+1)}M_n^2(T)}
-C\frac{\beta^2_{n+1}(T)}{(c^{(n+1)})^2}\right)\frac{dM^2_n(T)}{dT}\\
&&\qquad\le-\frac{dM^2_{n+1}(T)}{dT}\\
&&\qquad\le -\left(1-\frac{r-1}{4c^{(n+1)}M^2_n(T)}
+C_1\frac{\beta_{n+1}^{3/2}(T)}{{c^{(n+1)}}^2M_n(T)}\right)  \\
&&\qquad\qquad \left(1+\frac{r-1}{4c^{(n+1)}M_n^2(T)}
+C\frac{\beta^2_{n+1}(T)}{{c^{(n+1)}}^2}\right)\frac{dM^2_n(T)}{dT}.
\end{eqnarray*}
In this calculation we have exploited that 
$c^{(n+1)}M_n^2(T)\ge\frac{\beta_{n+1}(T)}\eta M_n^2(T)\gg1$.

The left and right-hand side of this inequality can be bounded by
$$
-\left(1\pm K\frac{\beta_{n+1}^2(T)}{{c^{(n)}}^2}\right)\frac{dM^2_n(T)}{dT},
$$
and formula (\ref{8.2}) can be deduced from these relations. 

For $N\le n\le N_1(T)$ with the number $N_1(T)$ defined in relation 
(\ref{4.18}) relation (\ref{8.1}) follows from (\ref{8.2}) and 
(\ref{4.19}). Since by (\ref{4.20}) $\beta_{n+1}M_n^2(T)\le100$ if 
$n\ge N_1(T)$ and the pair $(n,T)$ is in the low temperature domain, 
to prove formula (\ref{8.1}) with the help of (\ref{8.2}) for 
$n>N_1(T)$ it is enough to show that
\begin{eqnarray*}
&&\sum_{k=N_1(T)}^n\frac1{\left(c^{(k)}M_k^2(T)\right)^2}\le L
\textrm{ if }n\ge N_1(T) \\ 
&&\qquad \textrm{ and }(n,T) \textrm{ is in the low temperature domain}
\end{eqnarray*}
with a constant $L>0$ independent of $T$ and $n$. Since
$M_n^2(T)\ge\frac1{10\beta_{n+1}(T)}\ge\frac1{10\eta c^{(n)}}$ and
$M_k^2(T)=M_n^2(T)+(M_k^2(T)-M_n^2(T))\ge\frac1{10\eta c^{(n)}}
+\sum\limits_{j=k}^{n-1}\frac1{8c^{(j)}}$,
$$
\sum_{k=N_1(T)}^n\frac1{\left(c^{(k)}M_k^2(T)\right)^2}\le {\textrm{ const.}\,}
\sum_{k=N_1(T)}^n\frac1{\left(c^{(k)}\sum\limits_{j=k}^{n}\frac1{c^{(j)}}\right)^2}
\le L
$$
because of Condition 3. Hence formula (\ref{8.1}) holds.

It follows from (\ref{1.3}), Condition 4, and the results of 
Section~4 that all $T>c_0A_0/4$ belong to the high temperature 
region. Indeed, it follows from formulas (\ref{4.26}), 
(\ref{$4.26'$}), (\ref{4.1}), (\ref{4.4}) and (\ref{4.10}),
that if $T>0$ is in the low temperature region, then
$$
0\le M_n^2(T)\le
M_N^2(T)-30(M_N(T)+1)-\sum_{n=1}^\infty\frac1{8c^{(n)}}
\le \frac3{\kappa T^2}-\sum_{n=1}^\infty\frac1{8c^{(n)}}
$$
for all $n\ge N$, and $T\le \left(\sum\limits_{n=1}^\infty\frac
\kappa {24c^{(n)}}\right)^{-1/2}$. Hence Condition~4 implies that $T\le
c_0A_0/4$.

It follows from (\ref{8.2}) that for a fixed $n$ the function 
$M_n^2(T)$ is strictly monotone decreasing. Hence a simple 
induction with respect to $n$ yields that the function 
$\beta_n(T)$ is a monotone increasing, continuous function 
of~$T$ for all $n>N$. Put
\begin{equation}
T_n=\sup\{ T\colon  (T,n) \textrm{ is in the low temperature region}\}.
\label{8.2a}
\end{equation}
The sequence $T_n$ is monotone decreasing, hence the limit
$T_c=\lim\limits_{n\to\infty}T_n$ exists, and by Lemma~4.4
$T_c>0$ under Dyson's condition (\ref{1.3}). We want to
show that
\begin{equation}
{C_1}\sum_{k=n}^\infty\frac1{c^{(k)}}\le T_n-T_c\le
{C_2}\sum_{k=n}^\infty\frac1{c^{(k)}}. \label{8.3}
\end{equation}
Since we can handle the sequence $M_n(T)$ better than the sequence
$\beta_{n}(T)$ we also introduce besides the sequence $T_n$ defined
in (\ref{8.2a}) the sequence $T(n)$
$$
T(n)=\sup \left\{ T\colon  M^2_n(T)\ge \frac{100}{c^{(n)}\eta}\;\right\}.
$$
We will show that
\begin{equation}
T_{n+K}\le T(n)\le T_n \label{8.4}
\end{equation}
for all sufficiently large $n$ with an appropriate $K>0$, and
\begin{equation}
\frac{C_1}{c^{(n)}}\le T(n)-T(n+1)\le \frac{C_2}{c^{(n)}} \label{8.5}
\end{equation}
with some appropriate $C_2>C_1>0$ for all sufficiently large $n$.
Because of Condition~5 relation (\ref{8.3}) follows from (\ref{8.4}) 
and (\ref{8.5})
together with the relation $\lim\limits_{n\to\infty}T_n=T_c$.

If $T\le T(n)$, and $m\le n$ then either $m\le N_1(T)$ with the number
$N_1(T)$ defined in (\ref{4.18}) or $\beta_{m+1}(T)\le \frac{100}{M_m^2(T)}
\le \frac{100}{M_n^2(T)} \le c^{(n)}\eta$. This implies that for $T\le
T(n)$ the pair $(m,T)$ is in the low temperature region for all $m\le
n$, and $T(n)\le T_n$. This is the right-hand side of relation (\ref{8.4}).
 
To prove its left-hand side observe that because of Condition~5 there is
some $K$ such that
$$
\sum_{k=n}^{n+K-1}\frac1{8c^{(n)}}> \frac{100}{c^{(n)}\eta}
$$
for all sufficiently large $n$ with appropriate $K>0$. We claim that
for $T\ge T(n)$ the pair $(n+K,T)$ is not in the low temperature 
region. This relation implies the left-hand side of (\ref{8.4}). If 
$(n+K,T)$ were in the low temperature region, then we would get with 
the help of formula (\ref{4.26}) that
$$
M_{n+K}^2(T)\le M_n^2(T)-\sum_{k=n}^{n+K-1}\frac1{8c^{(n)}}<
\frac{100}{c^{(n)}\eta}-\sum_{k=n}^{n+K}\frac1{8c^{(n)}}<
\frac{100}{c^{(n)}\eta}-\frac{100}{c^{(n)}\eta}=0,
$$
and this is a contradiction.

To prove formula (\ref{8.5}) let us first observe that because of the
continuity and strict monotonicity of the function $M_n^2(T)$, \
$M_n^2(T(n))=\frac{100}{c^{(n)}\eta}$. It follows from the
last statement of 
Lemma~4.3 and formula (\ref{8.1}) that $N_1(T)\le n$
for all $T(n)-\varepsilon<T<T(n)$ with an appropriately small 
$\varepsilon>0$.  (The number $N_1(T)$ was defined in (\ref{4.18}).). 
Hence we get with the help of formula (\ref{8.1}) that for 
sufficiently large~$n$ and $T(n)-\varepsilon<T<T(n)$
\begin{eqnarray*}
&&\frac{100}{c^{(n)}\eta}-\frac2{c^{(n)}}+\bar C_1(T(n)-T)\le
\frac{100}{c^{(n+1)}\eta}-\frac1{c^{(n)}}+\bar C_1(T(n)-T)\\
&&\qquad\le M_{n+1}^2(T)\\
&&\qquad\le \frac{100}{c^{(n)}\eta}-\frac1{8c^{(n)}}+\bar
C_2(T(n)-T) \\
&& \qquad\le\frac{100}{\eta c^{(n+1)}}-\frac1{9c^{(n)}}+\bar C_2(T(n)-T)
\end{eqnarray*}
with some appropriate constants $\bar C_2>\bar C_1>0$. Hence the
solution of the equation $M_{n+1}^2(T)=\frac{100}{c^{(n+1)}\eta}$
satisfies the inequality $K_1<c^{(n)}(T-T(n))<K_2$ with appropriate
constants $K_2>K_1>0$. Since the solution of this equation is $T(n+1)$,
this fact implies relation~(\ref{8.5}).

It is not difficult to see that $T_c$ is in the low
temperature region. Since the inequality 
$M_n^2(T_c)=M^2_n(T(n))+(M^2_n(T_c)-M^2_n(T(N)))\le
\frac{100}{c^{(n)}\eta}+{\textrm{ const.}\,} (T(n)-T_c)$ 
holds for all large~$n$ because of (\ref{8.1}), 
$\lim\limits_{n\to\infty}M_n(T_c)=0$. 
Then relation (\ref{8.1}) implies that
$$
C_1\left(T_c-T)\right)\le M_n^2(T_c)-M_n^2(T)\le
C_2\left(T_c-T\right)
$$
with some positive constants $C_2>C_1>0$ if
$T_c\ge T\ge T_c-\varepsilon$. Letting $n$ tend to
infinity in the last relation we get formula (\ref{3.28}). 
Since formula (\ref{8.3}) is equivalent to (\ref{3.27}) 
Theorem~3.4 is proved. \qed

\medskip\noindent 
{\bf Proof of Theorem 1.3.}\/ By Corollary of Theorem 3.1, if the Dyson
condition (\ref{1.3}) is violated then all temperatures $T>0$ belong 
to the high temperature region. By Corollary of Theorem 3.3 relation 
(\ref{1.24}) holds, and the measures $\tilde\nu_{n,T}(dx)$ tends to 
the standard normal distribution as $n\to\infty$. Theorem 1.3 is proved.
\qed 

\medskip\noindent
{\bf Proof of Theorem 1.5.}

\medskip\noindent
Part 1). The convergence of
$\tilde\nu_{n,T}(dx)$ to the $r$-dimensional standard Gaussian 
distribution and relation (\ref{1.32}) follow from Corollary of 
Theorem 3.3. The asymptotics (\ref{1.33}) follows from 
(\ref{3.26}) and (\ref{3.27}).

\medskip\noindent
{\it Part 2).}\/  Formula (\ref{1.34}) follows from (\ref{3.9}), and 
the convergence of $\tilde\nu_{n,T_c}(dx)$ to the uniform distribution 
on the sphere follows from Theorems 3.2 and 3.4. Namely, Theorem 3.4 
tells us that the critical temperature $T_c$ belongs to the {\it low 
temperature region}. Then formula (\ref{3.10}) proves that the probability 
distribution $\tilde\nu_{n,T_c}(dx)$ converges to the uniform 
distribution on the sphere. As a matter of fact, (\ref{3.10}) proves much 
more: it proves the convergence at $T=T_c$ of the distribution of 
normalized fluctuations of the mean spin along the radius to 
a limit. Indeed, by (\ref{3.10}),
$$
\lim_{n\to\infty}
\left\|\frac1{M_n(T_c)}
f_n\left(\frac t{M_n(T_c)},T_c\right)-
g\left(t\right)\right\|=0,
$$
where
$$
\| f(t)\|=\sum_{j=0}^2\,\sup_{t\ge -c^{(n)}M^2_n(T)}e^{|t|}
\left| \frac{d^jf(t)}{d\,t^j} \right|
$$
and the probability density $g(t)$ is defined as a solution of the
fixed point equation (\ref{1.30}). By (\ref{2.16}),
$$
f_n(t,T_c)=\frac1{c^{(n)}}\bar q_n\left(M_n(T_c)
+\frac t{c^{(n)}},T_c\right),
$$
hence
$$
\lim_{n\to\infty}\left\|\frac
{\bar q_n\left(M_n(T_c)+\frac t{c^{(n)}M_n(T_c)},T_c\right)}
{c^{(n)}M_n(T_c)}
-g\left(t\right)\right\|=0.
$$
To obtain a scaling limit of $q_n$ near $M_n(T_c)$, let us rewrite 
the latter formula as
\begin{equation}\label{8.6}
\begin{aligned}
\lim_{n\to\infty}\left\|\frac
{\bar q_n\left(M_n(T_c)\left(1
+\frac t{c^{(n)}M^2_n(T_c)}\right),T_c\right)}
{c^{(n)}M_n(T_c)}
-g\left(t\right)\right\|=0.
\end{aligned}
\end{equation}
Let us evaluate the asymptotics of $c^{(n)}M^2_n(T_c)$ as $n\to\infty$.
By (\ref{3.9}),
\begin{equation}
\lim_{n\to\infty}\frac{M_n^2(T_c)}{\di\sum_{k=n}^\infty 
\frac{1}{c^{(k)}}}=\frac{r-1}{2},
\label{8.7}
\end{equation}
since $M_\infty(T_c)=0$. Define
\begin{equation}
\lambda_n=l_n \sum_{k=n}^\infty \frac{1}{l_k}.
\label{8.8}
\end{equation}
By Condition 2 on the sequence $\{l_n\}$, 
\begin{equation}
\lim_{n\to\infty} \lambda_n=\infty. \label{8.8a}
\end{equation}
By (\ref{2.31})
$$
\lim_{n\to\infty} \frac{c^{(n)}}{l_n}=3,
$$
hence
$$
\lim_{n\to\infty} \frac{\sum_{k=n}^\infty \frac{1}{c^{(k)}}}
{\sum_{k=n}^\infty \frac{1}{l_k}}=\frac{1}{3}, 
$$
and by (\ref{8.7}), (\ref{8.8}),
\begin{equation}
\lim_{n\to\infty}\frac{M_n^2(T_c)}{\sum_{k=n}^\infty \frac{1}{l_k}}
=\lim_{n\to\infty}\frac{M_n^2(T_c)l_n}{\lambda_n}=\frac{r-1}{6},
\label{8.12}
\end{equation}
which relation is equivalent to (\ref{1.34}). Therefore,
$$
\lim_{n\to\infty} \frac{c^{(n)}M^2_n(T_c)}{\lambda_n}=
\frac{r-1}{2}.
$$
Substituting this limit into (\ref{8.6}), we obtain that 
\begin{equation}
\lim_{n\to\infty}
\left\|\frac{2M_n(T_c)}{(r-1)\la_n}\bar q_n\left(M_n(T_c)\left(1+\frac {t}
{\frac{r-1}{2}\,\lambda_n}\right),T_c\right)-g\left(t\right)\right\|=0.
\label{8.13}
\end{equation}
This implies that the probability density 
$\bar Z_n(T_c)^{-1}\,\bar q_n(M_n(T_c)x,T_c)$ 
 is localized in a 
neighborhood of order  $\lambda_n^{-1}$ of the point 1, and 
after the proper scaling it converges to $g\left(t\right)$
as $n\to\infty$.

Let us consider now the scaling limit of the probability
density  $\bar p_n(x,T_c)$.
By equations (\ref{2.10}) and (\ref{2.14}),
$$
\bar p_n(x,T_c)=Z_n(T_c)^{-1}
\exp\left(-\frac{A_nl_n T_c^{-1}x^2}2\right)
\bar q_n(T_c^{-1/2}x,T_c),\quad x\ge 0,
$$
where  $Z_n(T_c)^{-1}$ is a norming factor, hence
\[
\begin{aligned}
&\bar p_n(T_c^{1/2}M_n(T_c)x,T_c) \\
& =Z_n(T_c)^{-1}
\exp\left(-\frac{A_nl_n M^2_n(T_c)x^2}2\right)\bar q_n(M_n(T_c)x,T_c).
\end{aligned}
\]
Applying the same scaling as in (\ref{8.13}), we obtain that
\begin{equation}\label{8.16} 
\begin{aligned}
&\bar p_n\left(T_c^{1/2}M_n(T_c)
\left(1+\frac {t}{\frac{r-1}{2}\,\lambda_n}\right),T_c\right) \\ 
& =Z_n(T_c)^{-1} 
\exp\left(-\frac{A_nl_n M^2_n(T_c)}{2}
\left(1+\frac{t}{\frac{r-1}{2}\,\lambda_n}\right)^2 \right) \\
&\times  \bar q_n\left(M_n(T_c)
\left(1+\frac{t}{\frac{r-1}{2}\,\lambda_n}\right),T_c\right).
\end{aligned}
\end{equation}
Consider the expression in the exponent,
\begin{equation}\label{8.17} 
\begin{aligned}
\frac{A_nl_n M^2_n(T_c)}{2}
\left(1+\frac{t}{\frac{r-1}{2}\,\lambda_n}\right)^2
&=\frac{A_nl_n M^2_n(T_c)}{2}
+\frac{2A_nl_n M^2_n(T_c)t}{(r-1)\lambda_n}\\
&+\frac{2A_nl_n M^2_n(T_c)t^2}{(r-1)^2\lambda_n^2}.
\end{aligned}
\end{equation}
By (\ref{8.12}), (\ref{2.30}) and (\ref{8.8a}) 
\begin{equation}
\lim_{n\to\infty} \frac{2A_nl_n M^2_n(T_c)}{(r-1)\lambda_n}=\frac{2}{3},\quad
\lim_{n\to\infty} \frac{2A_nl_n M^2_n(T_c)}{(r-1)^2\lambda^2_n}=0.
\label{8.17a}
\end{equation}
The constant term in (\ref{8.17}) is not important, because it changes 
in (\ref{8.16}) the norming constant only. 
Therefore, from (\ref{8.13})-(\ref{8.17a}) we obtain that
\begin{equation}\label{8.18} 
\begin{aligned}
\lim_{n\to\infty}
\biggl\| {Z_n'}^{-1} e^{(2/3)t}\bar p_n\left(T_c^{1/2}M_n(T_c)
\left(1+\frac {t}{\frac{r-1}{2}\,\lambda_n}\right),T_c\right)  
-g\left(t\right)\biggr\|=0
\end{aligned}
\end{equation}
This implies that
\begin{equation}\label{8.19} 
\begin{aligned}
&\lim_{n\to\infty}
\biggl\| 
\frac{2T^{1/2}_{c}M_n(T_c)} {(r-1)\lambda_n}
\,\bar p_n\left(T_c^{1/2}M_n(T_c)\left(1+\frac {t}
{\frac{r-1}{2}\,\lambda_n}\right),T_c\right)   \\
&\qquad\qquad\qquad\quad 
-Z^{-1} e^{-(2/3)t} g\left(t\right)\biggr\|'=0.
\end{aligned}
\end{equation}
where $Z^{-1} e^{-(2/3)t} g\left(t\right)$ 
is a probability density (this determines the constant $Z$), and 
\begin{equation}
\| f(t) \|'=\sum_{j=0}^2\,\sup_{t\ge -\frac{r-1}2\lambda_n} e^{|t|/3}
\left| \frac{d^jf(t)}{d\,t^j} \right|.
\label{8.20}
\end{equation}
Substituting $t$ for $(bt-a)$ in \eqref{8.19}, we obtain that
\begin{equation}\label{8.21} 
\lim_{n\to\infty}
\left\|\frac{A_n}{d_n} \bar p_n\left(A_n\left(1+\frac{t}{d_n}\right)\right)-\pi(t)\right\|'=0. 
\end{equation}
where $A_n>0$ is some constant. Since 
\[
\int_{-\infty}^\infty
t\, \bar p_n\left(\hat M_n(T_c)\left(1+\frac{t}{d_n}\right)\right)=0,
\quad \int_{-\infty}^\infty t\, \pi(t)\,dt=0,
\]
we can replace $A_n$ by $\hat M_n(T_c)$ in \eqref{8.21}.  This proves 
equation \eqref{1.35a}.

\medskip\noindent 
{\it Part 3).}  The results of Part 3) of Theorem 3 were already
proved with the exception of relation (\ref{1.37}) in the 
discussion after the formulation of Theorem 3.2. But 
relation  (\ref{1.37}) is a direct consequence of relation
(\ref{3.28}) proved in Theorem~3.4 and the identity 
$M(T)=\sqrt TM_\infty(T)$ for $T<T_c$ which was also proved
in the above discussion. Theorem 1.5 is proved. \qed

\appendix 
\section{The Proof of Theorem 8.1.}
\label{introA}

To prove Theorem 8.1 we need good  estimates on the partial derivatives 
$$
g_n(x,T)=\frac{\partial}{\partial T}f_n(x,T),
$$ 
of a scaled version of the functions $q_n(x,T)$. This can be done 
similarly to the estimation of the functions $f_n(x,T)$, done in 
Section~4. First we give estimates for the starting function
$g_0(x,T)$, then prove that similar estimates hold for small 
indices $n$, more explicitly for $n\le N$ with the index $N$ defined 
in (\ref{1.22}). Then inductive hypotheses can be formulated and 
proved for the functions $g_n(x,T)$. In Section~4 we have introduced 
certain operators $\bar{\mathbf Q}_n$, their normalization
${\mathbf Q}_n$ and the linearization of these operators denoted by
$\bar{\mathbf T}_n$ and ${\mathbf T}_n$. The inductive hypotheses 
formulated there were closely related to the properties of these 
operators. Now we want to work similarly. To do this we have to 
introduce some new operators. We introduce certain operators 
$\bar {\mathbf R}_n$ and ${\mathbf R}_n$ which are the derivatives 
of the operators $\bar{\mathbf Q}_n$ and ${\mathbf Q}_n$ with respect 
to the variable~$T$. We also need their linear approximation which 
we shall denote by $\bar{\mathbf U}_n$ and ${\mathbf U}_n$. We have 
to study the action of these operators on the functions 
$g_n(x,T)=\frac{\partial}{\partial T}f_n(x,T)$ and their Fourier
transform. 

An appropriate description of the  asymptotic behaviour of the  
starting functions $f_0(x,T)$ and numbers $M_0(T)$ were already 
given in formulas (\ref{4.2})~---~(\ref{4.7}). Some more 
calculation yields, with the help of some formulas in Section~4, 
the following estimates for the derivatives of the magnetization 
$M_0(T)$ and the norming constant $Z_0(T)$ if $T<c_0A_0/2$.
$$
\left|\frac d{dT}\left(M_0(T)-\bar M_0(T)\right)\right|
\le{\textrm{ const.}\,}\sqrt \kappa.
$$
\begin{equation}
\frac{C_1}{\sqrt \kappa\,T^2}<-\frac {dM_0(T)}{dT}<\frac{C_2}{\sqrt
\kappa\,T^2} \quad
\textrm{with some }\infty>C_2>C_1>0, \label{A1}
\end{equation}
and
$$
\left|\frac {dZ_0(T)}{dT}-
\frac{\sqrt \pi}{2(A_0-T)^{3/2}}\right|\le{\textrm{ const.}\,} 
\sqrt \kappa.
$$
The derivatives of the functions $\bar q_0(x,T)$ and $f_0(x,T)$ satisfy
the inequalities
\begin{eqnarray*}
&&\left|\frac{\partial \bar q_0(x+M_0(T),T)}{\partial
T}-\frac{\sqrt{A_0-T}}{\sqrt\pi}\left(x^2-\frac1{2(A_0-T)}\right)e^
{-(A_0-T)x^2}\right|  \\
&&\qquad  \le{\textrm{ const.}\,} \kappa^{1/4}, 
\quad \textrm{if } |x|<\log \kappa^{-1}, 
\end{eqnarray*}
and
\begin{eqnarray}
&&\left|\frac{\partial \bar q_0(x+M_0(T),T)}{\partial T}\right|\le
C\exp\left\{
-\frac{(A_0-T)}4\left|2x+\frac{x^2}{M_0^2(T)}\right|\right\} 
\label{A2} \\
&&\qquad \textrm{for }x\ge - M_0(T). \nonumber  
\end{eqnarray}
We shall apply the notation
\begin{equation}
g_n(x,T)=\frac {\partial f_n(x,T)}{\partial T},\quad  n=0,1,\dots.
\label{A3}
\end{equation}
Since $f_0(x,T)=q_0(x+M_0(T),T)$ the previous estimates together with
the results of Section~4 yield a sufficiently good control on
$g_0(x,T)$. The functions $g_n(x,T)$, $n=1,2,\dots$, can be
estimated inductively with respect to the parameter~$n$.

Put
$$
\bar {\mathbf R}_n f_n(x,T)=\frac{\partial}{\partial T}
\bar {\mathbf Q}_{n,M_n(T)}^{\mathbf c} f_n(x,T)
$$
and
$$
{\mathbf  R}_n f_n(x,T)=\frac{\partial}{\partial T}
{\mathbf  Q}^{\mathbf c}_{n,M_n(T)} f_n(x,T)=g_{n+1}(x,T).
$$
Then
\begin{equation}
\bar{\mathbf R}_{n}f_n(x,T)=\bar  {\mathbf R}_{n}^{(1)}f_n(x,T)+
\bar{\mathbf R}_{n}^{(2)}f_n(x,T) \label{A2a}
\end{equation}
with
\begin{eqnarray*}
\bar  {\mathbf R}_{n}^{(1)}f_n(x,T)
&=&2 \int_{u\in{\mathbb R}^1,\,\bv\in{\mathbb R}^{r-1}}
\exp\left\{-\frac{u^2}{c^{(n)}}-\bv^2\right\}
f_n(\ell_{n,M_n(T)}^{{\mathbf c}, +}(x,u,\bv),T)\\
&&\qquad g_n(\ell_{n,M_n(T)}^{{\mathbf c},-}(x,u,\bv),T)\,du\,d\bv,
\end{eqnarray*}
where the functions $g_n(x,T)$ and $\ell_{n,M_n(T)}^{\mathbf{c},\pm}(x,u,\bv),T)$
were defined in (\ref{A3}) and (\ref{2.25}), and
\begin{eqnarray*}
\bar  {\mathbf R}_{n}^{(2)}f_n(x,T)
\! &=& \!  -2\int_{u\in{\mathbb R}^1,\,\bv\in{\mathbb R}^{r-1}} \!\!
\exp\left\{-\frac{u^2}{c^{(n)}}-\bv^2\right\}
f_n(\ell_{n,M_n(T)}^{{\mathbf c},+}(x,u,\bv),T) \\
&&\qquad
h_n(x,u,\bv,T)\frac\partial{\partial x}f_n(\ell_{n,M_n(T)}^{{\mathbf c},-}
(x,u,\bv),T) 
\,du\,d\bv\end{eqnarray*}
with
\begin{eqnarray*}
&&h_n(x,u,\bv,T)=-\frac{\partial  \ell_{n,M_n(T)}^{{\mathbf c},-}(x,u,\bv)}
{\partial T} \\
&&\qquad=\frac{M_n'(T)\bv^2}{\sqrt{\left(M_n(T)+\frac x{c^{(n+1)}}
-\frac u{ c^{(n)}} \right)^2+\frac{\bv^2}{c^{(n)}}}}\\
&&\qquad\qquad\frac1{\sqrt{\left(M_n(T)+\frac x{c^{(n+1)}}
-\frac u{ c^{(n)}} \right)^2+\frac{\bv^2}{c^{(n)}}}+
\left(M_n(T)+\frac x{c^{(n+1)}}-\frac u{c^{(n)}}\right)}.
\end{eqnarray*}
The function $g_{n+1}(x,T)$ can be expressed as
\begin{eqnarray}
g_{n+1}(x,T)&=&{\mathbf R}_n f_n(x,T) \nonumber \\
&=&\frac{\bar{\mathbf R}_nf_n(x+m_n(T),T)}{Z_n(T)}
+\frac{\frac{\partial}{\partial x}
\bar{\mathbf Q}_n f_n(x+m_n(T),T)}{Z_n(T)}\frac{dm_n(T)}{dT} \nonumber \\
&&\qquad -\frac{\bar{\mathbf Q}_n
f_n(x+m_n(T),T)}{Z_{n}^2(T)}\frac{dZ_n(T)}{dT} \label{A3a}
\end{eqnarray}
with
$$
Z_n(T)=\int_{-c^{(n)}M_n(T)}^\infty \bar{\mathbf Q}_n f_n(x,T)\,dx.
$$

If the parameter $\kappa>0$ in formula (\ref{2.13}) is sufficiently small, 
then $\bar q_0(x,T)$ and the functions $g_n(x,T)$, $n\le N$, can be estimated 
similarly to the proof of Proposition~4.1 or Proposition~1 in \cite{BM3}. 
Relation (\ref{A5}) formulated below can be deduced from formula 
(\ref{A2}) similarly to the proof of Lemma~1 of that paper. Then an 
argument similar to the proof of Lemma~2 in~\cite{BM3} enables one to 
prove formula (\ref{A4}) formulated below. In this argument one can 
observe that a negligible error is committed if in the integrals 
appearing in the definition of $\bar{\mathbf R}_nf_n(x,T)$ the
arguments $\ell^{{\mathbf c},\pm}_{n,M_n(T)}(x,u,\bv)$ defined in 
formula (\ref{2.25}) are replaced by $\frac x{c_{n+1}}\pm u$. Some 
calculation also shows that we commit a negligible error by replacing 
${\mathbf R}_n f_n(x,T)$ with   
$\frac{\bar{\mathbf R}_n^{(1)}f_n(x,T)}{Z_n(T)}$. In such a way we get that
\begin{eqnarray}
&&\biggl|g_n(x,T)-
\frac{\sqrt{A_0-T}}{\sqrt\pi}\frac{2^{n/2}}{c^{(n)}}
\left(x^2-\frac1{2(A_0-T)}\frac{{c^{(n)}}^2}{2^n}\right) \nonumber \\
&& \qquad\qquad \exp\left\{-(A_0-T)
\frac{2^nx^2}{{c^{(n)}}^2}\right\}\biggr| \nonumber \\ 
&&\qquad \le C(n) \kappa^{1/4}
\exp\left\{-\frac{(A_0-T)}4 \frac{2^n}{c^{(n)}}
\left|2x+\frac{x^2}{M_n^2(T)}\right|\right\} \nonumber \\
&&\qquad\qquad \textrm{if }|x|<2^{-n}\log \kappa^{-1}, \label{A4} 
\end{eqnarray}
\begin{equation}
|g_n(x,T)|\le C(n)
\exp\left\{-\frac{(A_0-T)}4 \frac{2^n}{c^{(n)}}
\left|2x+\frac{x^2}{M_n^2(T)}\right|\right\}
\quad \textrm{for }x\ge - M_n(T), \label{A5}
\end{equation}
$$
|M_n(T)- M_0(T)|\le C(n)\kappa^{1/2},\qquad
\left|Z_n(t)-\frac{\sqrt \pi}{\sqrt{A_0-t}}\right|\le C(n)\kappa^{1/2}
$$
with some constant $C(n)$ which may depend on $n$ but not on the
parameter $\kappa$ of the model.

The previous results are sufficient to handle the functions
$g_n(x,T)$ for small indices $n\le N$. To work with
indices $n\ge N$ we have to introduce, similarly to the argument
in Section~4, the regularization of the functions $g_n(x,T)$, the
linearization $\bar{\mathbf U}_n$ and ${\mathbf U}_n$  of the operators
$\bar{\mathbf R}_n$ and ${\mathbf R}_n$ and to describe their action in the
Fourier space.

Define the regularization of the function $g_n(x,T)$ as
\begin{equation}
\varphi_n(g_n(x,T))=\frac{\partial \varphi_n(f_n(x,T))}{\partial T}.
\label{A5a}
\end{equation}
We want to approximate the operator ${\mathbf R}_n$ with a simpler operator
${\mathbf U}_n$ in analogy with the approximation of ${\mathbf Q}_n$ by 
${\mathbf T}_n$. Then we formulate and prove some inductive hypothesis 
about the behaviour of the operators ${\mathbf R}_n$ and ${\mathbf U}_n$.

A natural approximation of the operators $\bar{\mathbf R}_n$ and 
${\mathbf R}_n$ by some operators $\bar{\mathbf U}_n$ and 
${\mathbf U}_n$ can be obtained by differentiating 
$\bar{\mathbf T}_n\varphi(f_n(x,T))$ and
${\mathbf T}_n\varphi_n(f_n(x,T))$ with respect to the variable $T$. 
These considerations suggest the definition of the operators
\begin{eqnarray*}
&& \!\!\!\!\!\!\!
\bar{\mathbf U}_n\varphi_n(f_n(x,T))=
2\int_{u\in{\mathbb R}^1,\bv\in\mathbb R^{r-1}} e^{-\bv^2}
\varphi_n\left(f_n\left(\frac x{\bar c_{n+1}}+u
+\frac{\bv^2}{2M_n(T)},T\right)\right)\\
&&\biggl\{\varphi_n\left(g_n\left(\frac
x{\bar c_{n+1}}-u+\frac{\bv^2}{2M_n(T)},T\right)\right)\\
&&\qquad\quad -\bv^2\frac{M_n'(T)}{2M_n(T)^2}
\frac\partial{\partial x}\varphi_n\left(f_n\left(\frac
x{\bar c_{n+1}}-u+\frac{\bv^2}{2M_n(T)},T\right)\right) \Biggr\} \,du\,d\bv
\end{eqnarray*}
with the functions $g_n(x,T)$ and $\varphi_n(g_n(x,T))$ defined in 
(\ref{A3}) and (\ref{A5a}) and
$$
{\mathbf U}_n\varphi_n(f_n(x,T))={\mathbf U}_n^{(1)}\varphi_n(f_n(x,T))+
{\mathbf U}_n^{(2)}\varphi_n(f_n(x,T))
$$
with
\begin{eqnarray*}
&&{\mathbf U}_n^{(1)}\varphi_n(f_n(x,T)) = 
\frac8{\bar c_{n+1}\Gamma(\frac{r-1}2)V(S^{r-2})} \\
&&\qquad \int_{u\in{\mathbb R}^1,\,\bv\in{\mathbb R}^{r-1}} 
e^{-\bv^2}\varphi_n\left(f_n\left(\frac x{\bar c_{n+1}}+u-\frac{r-1}{4M_n(T)}+
\frac{\bv^2}{2M_n(T)},T\right)\right)\\
&&\qquad\qquad \varphi_n\left(g_n\left(\frac
x{\bar c_{n+1}}-u-\frac{r-1}{4M_n(T)}+\frac{\bv^2}{2M_n(T)},T\right)\right)
\,du\,d\bv,
\end{eqnarray*}
and
\begin{eqnarray*}
&&{\mathbf U}_n^{(2)}\varphi_n(f_n(x,T)) 
=\frac8{\bar c_{n+1}\Gamma(\frac{r-1}2)V(S^{r-2})} \\
&&\qquad \int_{u\in{\mathbb R}^1,\,\bv\in{\mathbb R}^{r-1} } 
e^{-\bv^2}\left(\frac{(r-1)M'_n(T)}{4M_n^2(T)}-\bv^2
\frac{M_n'(T)}{2M_n(T)^2}\right)\\
&&\qquad\qquad \varphi_n\left(f_n\left(\frac
x{\bar c_{n+1}}+u-\frac{r-1}{4M_n(T)}+\frac{\bv^2}{2M_n(T)},T\right)\right)\\
&&\qquad\qquad \frac\partial{\partial x}\varphi_n\left(f_n\left(\frac
x{\bar c_{n+1}}-u-\frac{r-1}{4M_n(T)}+\frac{\bv^2}{2M_n(T)},T\right)\right)
\,du\,d\bv.
\end{eqnarray*}

We can calculate the Fourier transform of the functions 
$\bar{\mathbf U}_n\varphi_n(f_n(x,T))$, 
${\mathbf U}_n^{(1)}\varphi_n(f_n(x,T))$ and
${\mathbf U}_n^{(2)}\varphi_n(f_n(x,T))$ 
by expressing them with the help of convolutions.
This is similar to the proof of formula (\ref{4.17}).
In the calculations we exploit the following identity.
As simple integration by parts shows
$ \widetilde{\frac{\partial}{\partial x}}\varphi_n(f_n(\xi))
=\int e^{i\xi x}\frac{\partial}{\partial x}\varphi_n(f_n(x))\,dx
=-i\xi\tilde\varphi(f_n(\xi))$. 
Hence we get that
\begin{eqnarray*}
&&\tilde{\bar{\mathbf U}}_{n} \tilde\varphi_n (f_n(\xi,T)) \\
&&\qquad =\frac{\bar c_{n+1}\Gamma(\frac{r-1}2)V(S^{r-2})}2
\frac{\tilde\varphi_n\left(f_n\left(\frac {\bar c_{n+1}}{2}\xi,T\right)\right)}
{\left(1+i\frac {\bar c_{n+1}}{2M_n(T)}\xi\right)^{(r-1)/2}}
\tilde\varphi_n \left(g_n\left(\frac {\bar c_{n+1}}{2}\xi,T \right)\right)\\
&&\qquad\qquad 
+i\frac{\bar c^2_{n+1}\Gamma(\frac{r+1}2)V(S^{r-2})}8
\frac{M_n'(T)}{M_n(T)^2} \xi
\frac{\tilde\varphi_n^2\left( f_n\left
(\frac {\bar c_{n+1}}{2}\xi,T \right)\right)}
{\left(1+i\frac {\bar c_{n+1}}{2M_n(T)}\xi\right)^{(r+1)/2}},
\end{eqnarray*}
\begin{eqnarray}
&&\tilde{\mathbf U}_{n}^{(1)} \tilde\varphi_n (f_n(\xi,T)) 
\label{A6} \\ 
&&\qquad=2\frac{\exp\left\{i\frac{(r-1)\bar c_{n+1}\xi}{4M_n(T)}\right\}}
{\left(1+i\frac{ \bar c_{n+1}}{2M_n(T)}\xi\right)^{(r-1)/2}}
\tilde\varphi_n\left(f_n\left(\frac{\bar c_{n+1}}{2}\xi,T \right)\right)
\tilde\varphi_n \left(g_n\left(\frac{\bar c_{n+1}}{2}\xi,T \right)\right) 
\nonumber 
\end{eqnarray}
and
\begin{eqnarray}
&&\tilde{\mathbf U}_{n}^{(2)} \tilde\varphi_n (f_n(\xi,T))
=-i\frac{\bar c_{n+1}(r-1)M_n'(T)}{2M_n(T)^2}
\frac{\exp\left\{i\frac{(r-1)\bar c_{n+1}\xi}{4M_n(T)}\right\}}
{\left(1+i\frac {\bar c_{n+1}}{2M_n(T)}\xi\right)^{(r-1)/2}}\xi \nonumber  \\
&&\qquad\qquad \tilde\varphi_n^2
\left( f_n\left(\frac {\bar c_{n+1}}{2}\xi,T \right)\right)
\left(1-\frac1{1+i\frac{\bar c_{n+1}}{2M_n(T)}\xi}\right). \label{A7}
\end{eqnarray}

The above relation can also be extended to a larger set of the
variables $\xi$ in the complex plane by means of analytic continuation.

Now we formulate the inductive hypotheses we want to prove in the
Appendix.

\medskip\noindent
{\bf Property} $K_1(n)$. 
$$
-\frac{dM_n(T)}{dT}>0.
$$

\medskip\noindent
{\bf Property} $K_2(n)$.
\begin{eqnarray*}
|g_n(x,T)|&=&\left|\frac{\partial}{\partial T} f_n(x,T) \right| \\
&<&K\left|\frac {dM_n(T)}{dT}\right|
\exp \left\{-\frac1{\sqrt{\beta_n(T)}}\left|2x+\frac{
x^2}{c^{(n)}M_n(T)}\right|\right\} \\
&&\qquad \quad \textrm{if } x>-c^{(n)}M_n(T)
\end{eqnarray*}
{\it with a universal constant $K$.}

\medskip\noindent
{\bf Property} $K_3(n)$.
\begin{eqnarray*}
&&|g_n(x,T)-{\mathbf U}_{n-1}\varphi_{n-1}(f_{n-1}(x,T))|\\
&&\qquad<K\left|\frac
{dM_n(T)}{dT}\right|\frac{\beta_n(T)}{c^{(n)}}
\exp \left\{-\frac{1.4}{\sqrt{\beta_n(T)}}\left|2x+\frac{
x^2}{c^{(n)}M_n(T)}\right|\right\} \\ 
&&\qquad\qquad\qquad \textrm{if } x>-c^{(n)}M_n(T)
\end{eqnarray*}
{\it with a universal constant $K$. The inequality remains valid 
if the function $g_n(x,T)$ is replaced by its regularization 
$\varphi_n(g_n(x,T))$.} 

\medskip
The following property $K_4(n)$ which gives a bound on the Fourier
transform of $\varphi_n(g_n(x,T))$ is an analog of Property~$J(n)$.

\medskip\noindent
{\bf Property} $K_4(n)$.
\begin{eqnarray*}
\left|\tilde\varphi_n(g_n(-is,T)\right|&=&\left|\int
e^{sx}\varphi_n(g_n(x,T)\,dx\right|\le
\beta_n^{3/2}(T)s^2\left|\frac{dM_n(T)}{dT}\right|
e^{\beta_n(T)s^2}\\
&&\qquad\qquad \textrm{if } |s|< \frac{2}{\sqrt{\beta_{n+1}(T)}}.
\end{eqnarray*}

\medskip
In Property $K_4(n)$ we formulated a weaker estimate than in $J(n)$. It
is enough to have a good bound on the moment generating function, i.e.\
on the analytic continuation of the Fourier transform to the imaginary
axis together with the trivial estimate $|\tilde\varphi_n
(g_n(-is+t,T)|\le \tilde\varphi_n(g_n(-is,T)$ for all~$t$.
 
The main result of the Appendix is the following Proposition~A.

\medskip\noindent
{\bf Proposition A.} {\it Let the properties $K_1(m)$,
$K_2(m)$, $K_3(m)$ and $K_4(m)$ hold in a neighbourhood of a parameter
$T$  together with the property $\frac{\beta_m(T)}{c^m}\le \eta$ (with
the same small number $\eta>0$ which appeared in the proof of 
Propositions~4.1 and~4.2) for all $N\le m\le n$, and let also the
inductive hypotheses $I(n)$ and $J(n)$ be also satisfied. Then the
properties $K_{1}(n+1)$, $K_{2}(n+1)$, $K_{3}(n+1)$ and $K_{4}(n+1)$
also hold for this parameter $T$. The expression
$$
\delta_n(T)=\frac{d}{dT}\left(m_n(T)-\frac{r-1}{4M_n(T)}\right)
=\frac{dm_n(T)}{dT}+\frac{r-1}{4M^2_n(T)}\frac{dM_n(T)}{dT}
$$
satisfies the inequality
\begin{equation}
|\delta_n(T)|\le C\left|\frac{dM_n(T)}{dT}\right|
\frac{\beta_{n+1}(T)}{c^{(n+1)}}\beta_{n+1}(T) \label{A8}
\end{equation}
with an appropriate $C>0$, where $m_n(T)$ was
defined in (\ref{2.22}).}

\medskip 
If we want to apply Proposition~A, then first we have to show that
properties $K_1(n)$, $K_2(n)$, $K_3(n)$ and $K_4(n)$ hold for $n=N$ if
$T<c_0A_0/2$. This can be done with the help of an argument similar
to the proof in the Corollary of Lemma~1 in~\cite{BM3}. Property $K_1(N)$
holds since $\frac{dM_N(T)}{dT}$ hardly differs from
$\frac{dM_0(T)}{dT}$. Property $K_2(N)$ can be proved by means of
relations (\ref{A4}) and (\ref{A5}). In the proof of 
Property $K_3(N)$ still the
following additional observation is needed. Relation (\ref{A4}) 
remains valid if the function $g_N(x,T)={\mathbf R}_Nf_{N-1}(x,T)$ 
is replaced by ${\mathbf U}_N\varphi_{N-1}(f_{N-1}(x,T))$ in this 
formula. (The term $\frac{dM_n(T)}{dT}$ on the right-hand side of 
the inductive hypotheses do not play an important role for $n=N$. 
It is strongly separated from zero if $T\le c_0A_0/2$.)
 
Relation $K_4(N)$ can be proved again with the help of formulas
(\ref{A4}), (\ref{A5}) and the relations 
$$
\int \varphi_n(g_n(x,T))\,dx=\int x
\varphi_n(g_n(x,T))\,dx=0.
$$ 
These relations imply that the value of the
function $\tilde\varphi_n(g_n(s,T)$ and of its first derivative is
zero in the point $s=0$. Hence it is enough to give a good estimate of
the second derivative of $\tilde\varphi_n(g_n(s,T)$.
 
Let us formulate the following Corollary of Proposition~A.

\medskip\noindent
{\bf Corollary.} {\it Under the Conditions of Theorem~3.4 the set of
the points $T$ for which $(n,T)$ is in the low temperature region is an
interval $(0,T_n)$ for all $n\ge0$. The inductive hypotheses $K_1(n)$,
$K_2(n)$, $K_3(n)$ and $K_4(n)$ hold for all $T\in (0,T_n)$.}

\medskip\noindent
{\bf Proof of the Corollary.}\/ The Corollary simply follows from
Proposition~A by means of induction with respect to $n$. In this induction
we assume the statement of the Corollary for a fixed $n$ together with
the assumption that $\beta_n(T)$ is monotone increasing in the variable
$T$ for $0<T<T_n$. The Corollary and the additional assumption hold for
$n=N$ with $T_N=c_0A_0/2$. If properties $K_1(n)$, $K_2(n)$, $K_3(n)$
and $K_4(n)$ hold for $n$, then because of Property~$K_1(n)$ the
function $M_n(T)$ is monotone decreasing and $\beta_{n+1}(T)$ is
monotone increasing in the variable~$T$. Then $T_{n+1}=\min(T_n,\max(T\colon 
\beta_{n+1}(T)<\eta))$, and by Proposition~A the statements of the
Corollary  hold for $n+1$. \qed 

\medskip
Before turning to the proof of Proposition~A we prove Theorem~8.1 
with its help.

\medskip\noindent
{\bf Proof of Theorem 8.1}.\/ The proof of Part a.) is contained in the
previous estimates of the Appendix. Part b.) can be obtained by
differentiating the second  formula in (\ref{2.24}), and applying 
formula (\ref{A8}). \qed

\medskip\noindent
{\bf Proof of Proposition A.}\/ Some calculation yields that because 
of properties $K_4(n)$, $J(n)$ relations (\ref{A6}) and (\ref{A7}) 
the Fourier transforms 
$$
\tilde{\mathbf U}_{n}^{(1)} \tilde\varphi_n(f_n(\xi,T)), \quad
\tilde{\mathbf U}_{n}^{(2)}\tilde\varphi_n(f_n(\xi,T))
$$
satisfy the inequalities
\begin{eqnarray*}
&&\left|\tilde{\mathbf U}_{n}^{(1)}\tilde\varphi_n(f_n(t+is,T))\right|\\
&&\qquad\le
2\left|\frac{dM_n(T)}{dT}\right|
\left(\frac{\bar c_{n+1}}{2}s\right)^2\beta_n(T)^{3/2} \\
&& \qquad\qquad \exp\left\{\left(\frac{\bar c_{n+1}^2\beta_n(T)}2
+\frac 1{M_n^2(T)}\right)s^2\right\}
\frac1{1+\alpha_n(T)t^2}
\end{eqnarray*}
and
\begin{eqnarray*}
&&\left|\tilde{\mathbf U}_{n}^{(2)}\tilde\varphi_n(f_n(t+is,T))\right| \\
&&\qquad \le \frac{\bar c_{n+1}^2|M_n'(T)|}{8M_n(T)^3}(s^2+t^2) 
\exp\left\{\left(\frac{\bar c_{n+1}^2\beta_n(T)}2+\frac
1{M_n^2(T)}\right)s^2\right\} \\
&&\qquad\qquad\qquad
\frac1{(1+\alpha_n(T)t^2)^2\left(1-\frac{\bar c_{n+1}}{2M_n(T)}s\right)}
\end{eqnarray*}
for $|s|<\frac{4}{\bar c_{n+1}\sqrt{\beta_{n+1}(T)}}$.

The function $\varphi_n(g_n(x,T))$ can be computed by means of the
application of the inverse Fourier transformation and by replacement of
the domain of integration from the real line to the line
$$
\left\{
z=i\,\textrm{sign}\,x\,\frac2{\sqrt{\beta_{n+1}(T)}}+t,
\;t\in{\mathbb R}^1\right\}.
$$
We get, by applying the above estimates for the Fourier transforms 
$\tilde{\mathbf U}_{n}^{(1)}$ and $\tilde{\mathbf U}_{n}^{(2)}$ and 
exploiting the relation $\frac{M_n'(T)}{2M_n(T)^3}\le\frac1{200}
\beta_{n+1}(T)^2\frac{dM_n(T)^2}{dT}$ together with the fact that the
constants $\alpha_n(T)$ and $\beta_n(T)$ introduced in the definition
of Properties $I(n)$ and $J(n)$ have the same order of magnitude that
\begin{eqnarray}
&&|{\mathbf U}_n\varphi_n(f_n(x,T))|\le -K_1
\frac{dM_n(T)}{dT}e^{-2|x|\beta_{n+1}(T)^{-1/2}}  \label{A9}  \\  
&& \qquad\qquad \le -K_2
\frac{dM_n(T)}{dT}\exp\left\{-\frac1{\sqrt{\beta_{n+1}(T)}}
\left|2x+\frac{x^2}{c^{(n+1)}M_{n+1}(T)}\right|\right\}. \nonumber 
\end{eqnarray}
The estimates obtained for $\tilde{\mathbf U}_n^{(1)}$ and 
$\tilde{\mathbf U}_n^{(2)}$ yield, with the choice $t=0$ and 
some calculation that
\begin{eqnarray}
\|\tilde{\mathbf U}_{n}\tilde\varphi_n(f_n(-is,T))|
&\le&-\frac{9}{10}\frac{dM_n(T)}{dT}\beta_{n+1}(T)^{3/2}s^2
e^{\beta_{n+1}(T)s^2}  \nonumber \\
&&\qquad\qquad\qquad\textrm{if }|s|<\frac{2}{\sqrt{\beta_{n+2}(T)}}.
\label{A10}
\end{eqnarray}
(In the proof of Property $K_4(n+1)$ it will be important that the
right-hand side of (\ref{A10}) is less than the expression at the
right-hand side of the formula which defines Property~$K_4(n+1)$.)

We need a good estimate on the difference of 
${\mathbf R}_nf_n(x,T)-{\mathbf U}_n\varphi_n(f_n(x,T))$ and its 
Fourier transform. These expressions can be bounded similarly 
to the proof of the corresponding inequalities in the proof of 
Proposition~3 in paper~\cite{BM3}. One has to compare the
difference of the corresponding terms in the expressions
${\mathbf Q}_n\varphi_n(f_n(x,T))$ and ${\mathbf R}_n\varphi_n(f_n(x,T))$.
Some calculation yields that
\begin{equation}
\left|Z_n(T)-\frac{\bar c_{n+1}\sqrt\pi}2\right|\le
\frac{\beta_n(T)}{c^{(n)}},\quad
\left|m_n(T)+\frac1{4M_n(T)}\right|\le
\frac{\beta_n(T)}{c^{(n)}}\sqrt{\beta_n(T)}, \label{A11}
\end{equation}
\begin{eqnarray}
\left|\frac{dZ_n(T)}{dT}\right|
&\le& -K\frac{\beta_n(T)}{c^{(n)}}\beta_n^{1/2}(T)\frac{dM_n(T)}{dT},
\label{A12} \\
\left|\frac{d}{dT}\left(m_n(T)+\frac1{4M_n(T)}\right)\right|
&\le& -K\frac{\beta_{n+1}(T)}{c^{(n+1)}}\beta_{n+1}(T)\frac{dM_n(T)}{dT}.
\nonumber
\end{eqnarray}
Relation (\ref{A8}) is a consequence of (\ref{A12}). 
Property $K_1(n+1)$ can be deduced from the above inequalities, since
\begin{eqnarray*}
-\frac{dM_{n+1}(T)}{dT}
&=& -\frac{dM_n(T)}{dT}+\frac1{c^{(n+1)}}\frac{dm_n(T)}{dT}\\
&\ge& -\frac{dM_n(T)}{dT}\left(1-\frac1{c^{(n+1)}}\left(\frac1{4M_n^2(T)}+
K\frac{\beta_n^2(T)}{c^{(n)}}\right)\right) \\
&\ge&-\frac12 \frac{dM_n(T)}{dT}.
\end{eqnarray*}

Now we turn to the proof of Property $K_3(n+1)$. We do it
by estimating the errors we make by replacing the terms
in the sum at the right-hand side of (\ref{A3a}) by their natural
approximation if we replace ${\mathbf R}_nf(x,T)$ by 
${\mathbf U}_n\varphi_n(f_n(x,T))$. (We also use formula (\ref{A2a}) 
in that calculation.) We get, by applying again 
inequalities (\ref{A11}) and (\ref{A12}) together with the estimates 
obtained for $f_n(x,T)$, similarly to the proof of the estimates 
in the lemmas needed for the proof of Lemma~3 in~\cite{BM1} that
\begin{eqnarray*}
&&\left|\frac{\bar{\mathbf Q}_n
f_n(x+m_n(T),T)}{Z_{n}^2(T)}\frac{dZ_n(T)}{dT}\right|\\
&&\qquad\le
K\frac{\beta_n(T)}{c^{(n)}}\left|\frac {dM_n(T)}{dT}\right|
\exp \left\{\frac{-1.5}{\sqrt{\beta_n(T)}}\left|2x+\frac{
x^2}{c^{(n)}M_n(T)}\right|\right\} \\
&& \qquad\qquad \qquad\textrm{if }x\ge -c^{(n+1)}M_{n+1}(T), 
\end{eqnarray*}

\begin{eqnarray*}
&&  \!\!\!\!\!\!\!
\left|\frac{\frac{\partial}{\partial x}
\bar{\mathbf Q}_n f_n(x+m_n(T),T)}{Z_n(T)}\frac{dm_n(T)}{dT}\right.
-\frac{8(r-1)}{\bar c_{n+1}\Gamma(\frac{r-1}2)V(S^{r-2})}
\frac{M'_n(T)}{4M_n^2(T)} \times \\ 
&&\qquad \int_{u\in{\mathbb R}^1,\, \bv\in{\mathbb R}^{r-1}}  
e^{-\bv^2} \varphi_n\left(f_n\left(\frac
x{\bar c_{n+1}}+u-\frac{r-1}{4M_n(T)}+\frac{\bv^2}{2M_n(T)},T\right)\right)\\
&&\qquad\quad \frac\partial{\partial x}\varphi_n
\left(f_n\left(\frac
x{\bar c_{n+1}}-u-\frac{r-1}{4M_n(T)}+\frac{\bv^2}{2M_n(T)},T\right)\right)
\,du\,d\bv \Biggr| \\
&&\le K\frac{\beta_n(T)}{c^{(n)}}\left|\frac{dM_n(T)}{dT}\right|
\exp \left\{\frac{-1.5}{\sqrt{\beta_n(T)}}\left|2x+\frac{
x^2}{c^{(n)}M_n(T)}\right|\right\} \\
&&\qquad\qquad\qquad\qquad \textrm{if }x\ge -c^{(n+1)}M_{n+1}(T) \\
\end{eqnarray*}
and
\begin{eqnarray*}
&& \!\!\!\!\!\!\!
\Biggl|\frac{\bar{\mathbf R}_{n}^{(2)}
f_n(x+m_n(T),T)}{Z_n(T)}+
\frac{8}{\bar c_{n+1}\Gamma(\frac{r-1}2)V(S^{r-2})} 
\frac{M_n'(T)}{2M_n^2(T)}\times \\
&&\quad \int_{u\in{\mathbb R}^1,\,\bv\in{\mathbb R}^{r-1}} 
\bv^2e^{-\bv^2} \varphi_n\left(f_n\left(\frac x{\bar c_{n+1}}+u
-\frac{r-1}{4M_n(T)}+\frac{\bv^2}{2M_n(T)},T\right)\right)\\
&&\quad\qquad\frac\partial{\partial x}\varphi_n\left(f_n\left(\frac
x{\bar c_{n+1}}-u-\frac{r-1}{4M_n(T)}+\frac{\bv^2}{2M_n(T)},T\right)\right)
\,du\,d\bv\Biggr|        \\
&&\le K\frac{\beta_n(T)}{c^{(n)}}\left|\frac{dM_n(T)}{dT}\right|
\exp \left\{-\frac{1.5}{\sqrt{\beta_n(T)}}\left|2x+
\frac{x^2}{c^{(n)}M_n(T)}\right|\right\}\\
&&\qquad\qquad\qquad\qquad \quad\textrm{if }x\ge -c^{(n+1)}M_{n+1}(T).
\end{eqnarray*}
To prove Property~$K_3(n+1)$ we still need an estimate which 
compares the terms
$$
\frac{\bar{\mathbf R}_{n}^{(1)}f_n(x+m_n(T),T)}{Z_n(T)}\quad
\textrm{and}\quad {\mathbf U}^{(1)}_{n}\varphi_n(f_n(x,T)).
$$
We claim that
\begin{eqnarray*}
&&\left|\frac{\bar{\mathbf R}_{n}^{(1)}
f_n(x+m_n(T),T)}{Z_n(T)}
-{\mathbf U}^{(1)}_{n}\varphi_n(f_n(x,T))\right|\le
K\frac{\beta_n(T)}{c^{(n)}}\left|\frac{dM_n(T)}{dT}\right| \\
&&\qquad\exp \left\{-\frac{1.5}{\sqrt{\beta_n(T)}}\left|2x
+\frac{x^2}{c^{(n)}M_n(T)}\right|\right\}
\quad\textrm{if }x\ge -c^{(n+1)}M_{n+1}(T).
\end{eqnarray*}
This estimate can be proved by means of Property $K_3(n)$. With
the help of this relation it can be shown that a negligible error is
committed if in the integrals defining $\bar  {\mathbf R}_{n}^{(1)}
f_n(x+m_n(T),T)$ and ${\mathbf U}^1_{n}\varphi_n(f_n(x,T))$ the
functions $g_n$ and $\varphi_n(g_n)$ are replaced by the function
${\mathbf U}_n\varphi_{n-1}(f_{n-1})$.  After this replacement the 
proof of Theorem~3.2 can be adapted, since we can bound not only 
the function ${\mathbf U}_n\varphi_{n-1}( f_{n-1})$, but also its 
partial derivative with respect to the variable~$x$.

These estimates together imply Property $K_3(n+1)$, and some
calculation shows that a version of Property $K_3(n+1)$, where the
function $g_{n+1}(x,T)$ is replaced by its regularization
$\varphi_{n+1}(g_{n+1}(x,T))$ is also valid. Since we gave
a good estimate on ${\mathbf U}_n\varphi_n(f_n(x,T))$ in (\ref{A9}), some
calculation yields the proof of Property $K_2(n+1)$. It remained to
prove Property $K_4(n+1)$.

Because of (\ref{A10}) and (\ref{A12}) (The latter formula together 
with (\ref{2.22}) and (\ref{2.24}) imply that formula (\ref{A10}) 
remain valid with a slightly bigger coefficient if the term 
$\frac{dM_n(T)}{dT}$ is replaced by $\frac{dM_{n+1}(T)}{dT}$ in it), 
it is enough to give a good bound on the difference 
$\tilde\varphi_{n+1}(g_{n+1}(-is))-\tilde{\mathbf U}_n
\tilde\varphi_n (f_n(-is))$ to prove property $K_4(n+1)$. This can
be done in the following way:

By applying the modified property of $K_3(n+1)$, where the function
$g_{n+1}(x)$ is replaced by $\varphi_{n+1}g_{n+1}(x)$ we get that
\begin{eqnarray*}
&&\left|\frac{\partial^2}{\partial s^2}\left[\tilde\varphi_{n+1}(
g_{n+1}(-is,T))-\tilde{\mathbf
U}_n\tilde\varphi_n(f_n(-is,T))\right]\right|\\
&&\qquad\le-\int
x^2\frac{dM_n}{dT}\frac{\beta_{n+1}(T)}{c^{(n+1)}}
\exp\left\{\left(|t|-\frac{2.8}{\sqrt{\beta_{n+1}(T)}}\right)x\right\}\,dx \\
&&\qquad \le K\frac{\beta_{n+1}^{5/2}(T)}{c^{(n+1)}}\frac{dM_n^2}{dT}
\end{eqnarray*}
if $|s|\le\frac2{\sqrt{\beta_{n+2}(T)}}$.

Since
\begin{eqnarray*}
&&\tilde\varphi_{n+1}(g_{n+1}(0,T))-\tilde{\mathbf
U}_n\tilde\varphi_n(f_n(0,T)) \\
&&\qquad=\left.\frac{\partial}{\partial s}\left(\tilde\varphi_{n+1}(\tilde
g_{n+1}(-is,T)-\tilde{\mathbf
U}_n\tilde\varphi_n(f_n(-is,T)\right)\right|_{s=0}=0,
\end{eqnarray*}
the last relation implies that
$$
\left|\tilde\varphi_{n+1}(\tilde
g_{n+1}(-is,T)-\tilde{\mathbf
U}_n\tilde\varphi_n(f_n(-is,T)\right|
\le
-K\frac{\beta_{n+1}(T)}{c^{(n+1)}}\beta_{n+1}^{3/2}\frac{dM_n(T)}{dT}s^2
$$
if $|s|\le\frac2{\sqrt{\beta_{n+2}(T)}}$. This estimate together
with relation (\ref{A10}) imply Property $K_4(n+1)$ if the number
$\eta$ which is an upper bound for ${\beta_{n+1}(T)}/{c^{(n+1)}}$ is
chosen sufficiently small. Proposition~A is proved. \qed

\section{The Proof of Proposition 1.2.}
\label{introB}

{\it Condition 1}. We have that for $n\ge 1$,
$$
1<c_n=\left(\frac{1+an}{1+a(n-1)}\right)^\lambda
$$
Observe that $c_n$ is decreasing and
$$
\lim_{n\to\infty}c_n=1,\qquad c_n\le c_1=(1+a)^\lambda
$$
This implies Condition 1.

\medskip\noindent
{\it Condition 2.} We have that
$$
(1+an)^{\lambda}\sum_{j=n}^{n+K} (1+aj)^{-\lambda}\ge\frac {K(1+an)^{\lambda}}
{(1+a(n+K))^\lambda} \to K
$$
as $n\to \infty$. This implies Condition 2.

\medskip\noindent
{\it Condition 3.} For $k\le n/2$ we estimate
$$
l_k\sum_{j=k}^n l_j^{-1}=(1+ak)^\lambda\sum_{j=k}^n (1+aj)^{-\lambda}
\ge C (1+ak)^\lambda (1+ak)^{-\lambda+1}=C (1+ak)^{-1}
$$
and for $k>n/2$ and $n\ge j\ge k$ we estimate
$$
l_kl_j^{-1}\ge C_0>0
$$
hence
$$
l_k\sum_{j=k}^n l_j^{-1}\ge C_0(n-k+1)
$$
Thus,
$$
\sum_{k=1}^n\left(l_k\sum_{j=k}^n l_j^{-1}\right)^{-2}\le
C^{-2}\sum_{k=1}^{n/2} (1+ak)^{-2}
+C_0^{-2}\sum_{k=n/2}^n(n-k+1)^{-2}\le C_1
$$
Condition 3 is checked.

\medskip\noindent
Conditions 4 and 5 are obvious. \qed

\bigskip\noindent
{\it Acknowledgements.} An essential part of this work was done at the
Mathematisches Forschungsinstitut Oberwolfach, where the authors enjoyed
their participation in the program ``Research in Pairs''. They
are thankful to the Mathematisches Forschungsinstitut for kind
hospitality and the Volkswagen--Stiftung for support of their stay
at Oberwolfach. The research of the first author (P.B.)
was supported in part by the National
Science Foundation, Grant No. DMS--1565602, the research of the
second author was supported by the Hungarian Foundation 
NKFI--EPR No. K-125569. These supports are gratefully acknowledged.

\end{document}